\documentclass[12pt]{iopart}
\usepackage{graphicx}
\newcommand\varpm{\mathbin{\vcenter{\hbox{%
  \oalign{\hfil$\scriptstyle+$\hfil\cr
          \noalign{\kern-.3ex}
          $\scriptscriptstyle({-})$\cr}%
}}}}

\newcommand\varmp{\mathbin{\vcenter{\hbox{%
  \oalign{$\scriptstyle({+})$\cr
          \noalign{\kern-.3ex}
          \hfil$\scriptscriptstyle-$\hfil\cr}%
}}}}

\begin{document}

\title{Gradient metasurfaces: a review of fundamentals and applications} %Space-variant

\author{Fei Ding, Anders Pors and Sergey I. Bozhevolnyi}
\address{SDU Nano Optics, University of Southern Denmark, Campusvej 55, DK-5230 Odense M, Denmark}
\ead{feid@mci.sdu.dk and seib@mci.sdu.dk}
\vspace{10pt}

\begin{abstract}
In the wake of intense research on metamaterials the two-dimensional analogue, known as metasurfaces, has attracted progressively increasing attention in recent years due to the ease of fabrication and smaller insertion losses, while enabling an unprecedented control over spatial distributions of transmitted and reflected optical fields. Metasurfaces represent optically thin planar arrays of resonant subwavelength elements that can be arranged in a strictly or quasi periodic fashion, or even in an aperiodic manner, depending on targeted optical wavefronts to be molded with their help. This paper reviews a broad subclass of metasurfaces, viz. gradient metasurfaces, which are devised to exhibit spatially varying optical responses resulting in spatially varying amplitudes, phases and polarizations of scattered fields. Starting with introducing the concept of gradient metasurfaces, we present classification of different metasurfaces from the viewpoint of their responses, differentiating electrical-dipole, geometric, reflective and Huygens' metasurfaces. The fundamental building blocks essential for the realization of metasurfaces are then discussed in order to elucidate the underlying physics of various physical realizations of both plasmonic and purely dielectric metasurfaces. We then overview the main applications of gradient metasurfaces, including waveplates, flat lenses, spiral phase plates, broadband absorbers, color printing, holograms, polarimeters and surface wave couplers. The review is terminated with a short section on recently developed nonlinear metasurfaces, followed by the outlook presenting our view on possible future developments and perspectives for future applications.
\end{abstract}

\vspace{2pc}
\noindent{\it Keywords}: gradient metasurfaces, meta-reflectarrays, dielectric metasurfaces, flat optical elements, wavefront shaping, surface wave couplers, nonlinearity
%\vspace{1pc}
%\\
%\submitto{\RPP}
%\maketitle
\section{Introduction}
With the progress and prevalence of nano-fabrication, -characterization, and commercial design tools during the last couple of decades, the early thoughts on new degrees of light control, particularly visualized by Versalago's work on electromagnetic waves propagating in negative index media \cite{Veselago_1968}, have become approachable for the wide optics community, with the accumulated knowledge and experience now being focused towards the design and realization of novel metadevices at practically any frequency regime of interest. It is difficult to precisely define the start of the meta-era, since work already from the 1960s (see, e.g., \cite{Vogel_1964}) may, in today's terminology, be considered as metasurfaces. Nevertheless, it seems safe to say that the intriguing work by T. W. Ebbesen \emph{et al.} in 1998 on extraordinary optical transmission through arrays of subwavelength holes in metal films \cite{Ebbesen_1998} and the visionary concept of perfect lensing by Sir J. B. Pendry in the year 2000 \cite{Pendry_2000} infused a new way of thinking and controlling light. Shortly after, and strongly inspired by the structure of matter, the concept of \emph{metamaterials} was born, which is artificial materials with \emph{effective} optical parameters that are (most often) unattainable with conventional materials \cite{Sihvola_2007,Cai_2010,Simovski_2011}. The unusual optical properties, comprising negative permeability \cite{Pendry_1999,Linden_2004,Dolling_2005} and refractive index \cite{Smith_2000,Shelby_2001,Shalaev_2005}, arise from structuring (conventional) matter on a \emph{subwavelength} scale in which each, so-called, meta-atom features an electric and/or magnetic response that on the scale of the wavelength transpires into effective values of the permittivity and permeability. Metamaterials have been successfully applied in realizing proof-of-concept invisibility cloaks \cite{Pendry_2006,Schurig_2006}, super-resolution imaging with hyperlenses \cite{Zubin_2006,Salandrino_2006,Liu_2007}, optical nanocircuits \cite{Engheta_2005,Engheta_2007}, biosensors \cite{Henzie_2007,Kabashin_2009}, and media with extreme chirality \cite{Rogacheva_2006,Decker_2007}, just to mention a few. That said, the true breakthrough and commercialization of metamaterials have yet failed to happen, which we ascribe to the increasing difficulty in structuring matter in three dimensions with increasing operation frequency and the fact that metals, an often used component of metamaterials, become lossy (i.e., plasmonic) as we approach the optical regime. Furthermore, plasmonic materials, like silver and gold, are not compatible with the widespread CMOS technology, hence hindering easy implementation in present large-scale fabrication facilities. We note that current research efforts explore alternative materials for low-loss, tunable, refractory, and CMOS compatible metamaterials \cite{West_2010,Naik_2013}.

\begin{figure}[tb]
	\centering
		\includegraphics[width=6cm]{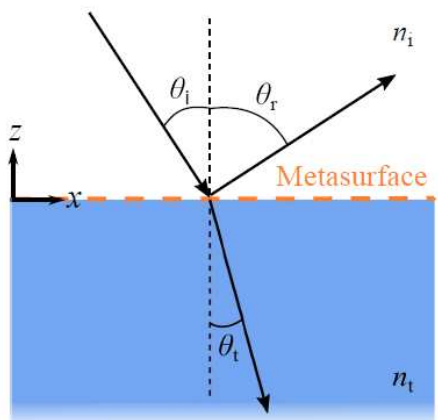}
	\caption{Sketch of the 1D system considered by the group of F. Cappasso in \cite{Yu_2011}, where a metasurface positioned at the interface between to ordinary media (characterized by refractive indexes $n_i$ and $n_t$) necessitates a generalization of the laws of reflection and refraction. Here, $\theta_i$, $\theta_r$, and $\theta_t$ are the incident, reflected, and refracted angles, respectively, that are related to the in-plane wave vectors by $k_{x}^{(i)}=k_0n_i\sin\theta_i$, $k_{x}^{(r)}=k_0n_i\sin\theta_r$, and $k_{x}^{(t)}=k_0n_t\sin\theta_t$.}
	\label{fig:Capasso}
\end{figure}

As concerns started to appear regarding the future of metamaterials, the group of F. Capasso introduced in 2011 the concept of \emph{interface discontinuities}, nowadays referred to as \emph{metasurfaces}, and generalized laws of reflection and refraction \cite{Yu_2011}. The main results of that paper will be discussed in more detail shortly, but first we clarify that metasurfaces are a two-dimensional (2D) array of meta-atoms with subwavelength periodicity and, for this reason, can be considered the planar analog of metamaterials with a thickness much smaller than the wavelength of operation. Interestingly, a considerable amount of experimentally realized metamaterials prior to the year 2011 are planar (see, e.g., \cite{Linden_2004,Dolling_2005,Shalaev_2005,Henzie_2007,Decker_2007}) due to fabrication challenges in entering the third dimension. The conceptual difference in planar metamaterials and metasurfaces, however, owes to the description, where the former is characterized by bulk effective parameters and the latter by induced (electric and magnetic) surface currents. The one-dimensional (1D) system considered by F. Capasso and coworkers is sketched in figure \ref{fig:Capasso},
where an incident plane wave impinges at an angle of $\theta_i$ relative to the surface normal of a metasurface, thus leading to reflection and refraction of the wave at angles $\theta_r$ and $\theta_t$, respectively.
The generalized laws of reflection and refraction can be written as
\begin{eqnarray}
\label{eq:Refl}
& k_{x}^{(r)}-k_{x}^{(i)}=\frac{\mathrm{d}\Phi}{\mathrm{d}x},\\
& k_{x}^{(t)}-k_{x}^{(i)}=\frac{\mathrm{d}\Phi}{\mathrm{d}x},
\label{eq:Refr}
\end{eqnarray}
where $k_{x}^{(i,r,t)}=k_0n_{i,i,t}\sin\theta_{i,r,t}$ is the in-plane wave vector component, $k_0$ is the free-space wave number, and $\Phi=\Phi(x)$ is a position-dependent phase abruptly imprinted on the incident wave by the metasurface. It is evident that in the case of $\mathrm{d}\Phi/\mathrm{d}x=0$, we recover the usual laws of reflection of refraction that imply continuity of the in-plane wave vector. On the other hand, when $\mathrm{d}\Phi/\mathrm{d}x\neq 0$ we are considering a \emph{phase-gradient} metasurface with the possibility to decouple the angle of incidence and the reflected/refracted angle. The most illustrative example of this anomalous behavior is a metasurface featuring a linearly varying phase of the form $\Phi=\pm2\pi x/\Lambda$, where $\Lambda$ is the distance of which the phase has changed by $2\pi$, hereby imposing an additional in-plane wave number of $\mathrm{d}\Phi/\mathrm{d}x=\pm 2\pi/\Lambda$ on the reflected and refracted light. Overall, we believe that one should appreciate the simplicity and visual appeal of the generalized laws of reflection and refraction, particularly with respect to the impact of those equations on the subsequent development of increasingly complex flat optical components. However, the work in \cite{Yu_2011} has also received critical comments due to lack of novelty. The critic is twofold; first, it is easy to show that the response of phase-gradient metasurfaces directly follows well-known Fraunhofer diffraction \cite{Larouche_2012}. For example, the (at first sight) anomalous behavior of light in metasurfaces with a constant phase gradient of $\pm 2\pi/\Lambda$ is equivalent to the functionality of blazed gratings of period $\Lambda$, the only difference being that in blazed gratings the linearly varying phase is achieved through a triangular, sawtooth-shaped profile. The second point of critic comes from the microwave community, practically stating that phase-gradient metasurfaces are merely a scaling of frequency selective surfaces (FSSs) to higher frequencies \cite{Bansal_2011}. As FSSs, including the derived concepts of reflect- and transmitarrays, are made of subwavelength arrays of metal patches or apertures in a metal film and conventionally function as filters \cite{Mittra_1988}, flat parabolic reflectors \cite{Kelkar_1991,Pozar_1993,Pozar_1997}, and flat lenses \cite{McGrath_1986,Pozar_1996,Joumayly_2011}, respectively, it is evident that FSSs and metasurfaces can be considered two sides of the same coin. We, however, argue that the focus on reflect- and transmitarrays for collimation of milli- and micrometer waves for wireless communication has limited the full exploration of FSSs for wave manipulation---a viewpoint that is substantiated by recent years progress in design and applications of reflect- and transmit-arrays by, for example, the groups of A. Grbic \cite{Pfeiffer_2013,Pfeiffer_2013_2,Pfeiffer_2014} and L. Zhou \cite{Sun_2012,Luo_2015}. Coming full circle, we would also like to recognize work from the beginning of this century by the group of E. Hasman on, at that time termed, space-variant subwavelength gratings \cite{Bomzon_2001,Bomzon_2002,Biener_2002,Hasman_2003}, which in today's terminology are known as geometric metasurfaces.

From the above discussion, we hope it is clear that the manipulation of electromagnetic waves by flat and subwavelength-thin devices has been an ongoing effort for several decades, yet only recently experiencing a significant advancement in practically all regimes of the electromagnetic spectrum. In this review, we will attempt to unify the vast amount of work published on metasurfaces and metadevices, hopefully making it the most introductory and extensive overview of this thriving field of research. The paper is organized into three main sections, with the first section introducing a classification of metasurfaces by the properties of the meta-atoms and a discussion of the associated possibilities of efficient wave manipulation. The second section presents realizations of the different classes of metasurfaces that, irrespective of the frequency regime of operation, are exemplified by phase-gradient metasurfaces functioning as blazed gratings. The last section discusses the many and exciting applications of metasurfaces, ranging from flat realizations of conventional optical elements to exotic devices exploiting the new degrees of light control. As a final comment, we would like to acknowledge the many and excellent reviews that already exist, some presenting a comprehensive overview of the field \cite{Yu_2014,Luo_2015_2,Minovich_2015,Zhang_2016,Glybovski_2016,chen2016review,monticone2017metamaterial}, while others are more compact \cite{Meinzer_2014,Koenderink_2015,Tretyakov_2015,Shaltout_2016}, mainly focuses on a certain type of metasurface \cite{Yu_2013,Hum_2014,Guanghao_2015,Estakhri_2016,Jahani_2016}, or a few specific branches of applications \cite{Zheludev_2012,Yu_2015_2,Genevet_2015,Yinghong_2015}.

\section{Classification of metasurfaces}
In the following section, we will discuss different classes of metasurfaces and present simple formulas for their reflection and transmission coefficients, which allow us to emphasize their peculiar properties. In all cases, the metasurfaces will be considered infinitely thin, with the responses related to the (electric, magnetic, and magnetoelectric) polarizabilities of the meta-atoms comprising the metasurface. As an illustration of this idealized situation, it is interesting to note that an induced magnetic response in the plane of the metasurface requires a circulating current in the orthogonal direction, thus exemplifying the apparent paradox of the assumption of zero-thickness metasurfaces. Finally, we assume the incident light to be a plane wave propagating normal to the metasurface and that the induced polarizabilities only have in-plane components.

\subsection{Metasurfaces featuring electric dipole response} \label{sec:EDresponse}
\begin{figure}[tb]
	\centering
		\includegraphics[width=7cm]{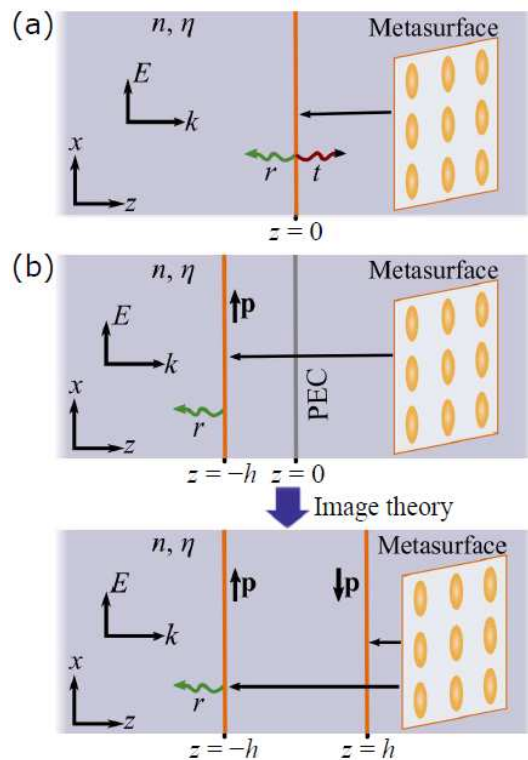}
	\caption{(a) Sketch of configuration consisting of a metasurface extending the $xy$-plane and immersed in a homogeneous lossless medium. The incident plane wave is $x$-polarized and propagates normal to the metasurface. (b) Sketch of a metasurface in front of a metallic backreflector, including the equivalent configuration (bottom sketch) within the approximations of electrostatic image theory.}
	\label{fig:ms1}
\end{figure}

The simplest class of metasurfaces constitutes a subwavelength periodic array of dielectric or metallic inclusions featuring a pure electric dipole (ED) response. For simplicity, here we assume that the metasurface has an isotropic response and is immersed in a homogeneous medium described by the refractive index $n$ and wave impedance $\eta$. The configuration is sketched in \fref{fig:ms1}(a), where the incident $x$-polarized plane wave propagates along the $z$-direction (i.e., normal to the metasurface).
As the metasurface is positioned at $z=0$, the electromagnetic wave in the two half-spaces takes the form
\begin{eqnarray}
\label{eq:Em}
& \mathbf{E}_-(z)=E_0\left(e^{ikz}+re^{-ikz}\right)\hat{\mathbf{x}} ,\\
& \mathbf{H}_-(z)=E_0\eta^{-1}\left(e^{ikz}-re^{-ikz}\right)\hat{\mathbf{y}} , \label{eq:Hm}\\
& \mathbf{E}_+(z)=tE_0e^{ikz}\hat{\mathbf{x}} ,\\
& \mathbf{H}_+(z)=tE_0\eta^{-1} e^{ikz}\hat{\mathbf{y}} \label{eq:Hp},
\label{eq:ms1Field}
\end{eqnarray}
where $E_0$ is the amplitude of the incident electric field, $k=nk_0$ is the wave number, $k_0$ is the wave number in vacuum, and $r$ and $t$ are the reflection and transmission coefficients, respectively. In the above equations, we have used the harmonic-time dependence $\exp(-i\omega t)$, where $\omega$ is the angular frequency. Since the metasurface only features an electric response, which leads to the presence of electric surface currents, the appropriate boundary conditions at the metasurface amount to
\begin{equation}
\hat{\mathbf{z}}\times\left( \mathbf{E}_+-\mathbf{E}_- \right)=0 \quad, \quad \hat{\mathbf{z}}\times\left( \mathbf{H}_+-\mathbf{H}_- \right)=\mathbf{J}_s,
\label{eq:ms1BC}
\end{equation}
where $\mathbf{J}_s$ is the induced electric surface current. If we approximate the optical response from the meta-atom(s) of each unit cell with an effective point dipole moment $\mathbf{p}$, the average surface current is $\mathbf{J}_s=-i\omega \mathbf{p}/A$, where $A$ is the area of the unit cell. Following the discussion in \cite{Kuester_2003}, the electric dipole moment may also be written as $\mathbf{p}=\varepsilon\alpha\mathbf{E}_\textrm{av}$, where $\varepsilon=\varepsilon_0n^2$ is the permittivity of the surrounding medium, $\alpha$ is the electric polarizability, which might incorporate knowledge about coupling between neighboring elements and spatial dispersion effects \cite{Yatsenko_2003,Zhao_2010}, and $\mathbf{E}_{\mathrm{av}}=\frac{1}{2}\left( \mathbf{E}_+ + \mathbf{E}_-\right)$ is the average field acting on the metasurface, hereby leading to the surface current
\begin{equation}
\mathbf{J}_s=-i\omega A^{-1}\varepsilon\alpha \mathbf{E}_{\mathrm{av}}.
\label{eq:ms1J}
\end{equation}
By applying the boundary conditions in equation \eref{eq:ms1BC} on the fields in equations \eref{eq:Em}-\eref{eq:Hp}, we obtain the following reflection and transmission coefficients
\begin{equation}
r=\frac{iC}{1-iC} \quad, \quad t=\frac{1}{1-iC},
\label{eq:ms1:RT}
\end{equation}
where $C=\omega\alpha/(2Av)$ is a dimensionless parameter, and $v$ is the speed of light in the surrounding medium. It is worth noting that the above coefficients have the same functional form as derived for subwavelength thin periodic metal grids, plates, and perforated films \cite{Chen_1973,Lee_1982}. Moreover, it is evident that interaction between the incident wave and the metasurface requires $C\neq 0$, meaning that it is not possible simultaneously to manipulate the incoming wave while having perfect transmission (i.e., $|t|=1$). In order to better visualize the limits of metasurfaces with a purely electric response, we note that the continuity of the transverse electric field across the infinitely thin metasurface transpires to $1+r=t$, while (for the sake of argument) the additional simplifying assumption of a lossless metasurface (i.e., $C$ is a real-valued function) amounts in conservation of the electromagnetic energy, i.e., $|r|^2+|t|^2=1$. In order to satisfy both equalities, it is straightforward to show that the complex transmission coefficient, which we write as $t=|t|e^{i\phi_t}$, must obey the condition
\begin{equation}
|t|=\cos(\phi_t),
\label{eq:ms1:tCond}
\end{equation}
hereby underlining that the amplitude and phase of the transmitted light (and, hence, also the reflected light) are inherently connected, with the phase concurrently limited to the interval $-\pi/2 \leq \phi_t \leq \pi/2$. In other words, isotropic ED-based metasurfaces do not allow for independent engineering of the amplitude and phase of the transmitted or reflected light, while only $\pi$ phase-space can be reached.

Having clarified the limitations of the simplest form of metasurfaces, we now concentrate our attention on the slightly more complex problem of \emph{anisotropic} metasurfaces for which the polarizability and, for this reason, also the reflection and transmission coefficients, turn into $2\times2$ tensors (we ignore the possibility of having an out-of-plane polarizability).
In this context, it is important to remember that a matrix $\overline{\overline{M}}$, representing one of the three quantities, that is rotated by $90^\circ$ with respect to the original $xy$-coordinate system undergoes the following transformation
\begin{equation}
\left(
\begin{array}{cc}
M_{xx} & M_{xy} \\
M_{yx} & M_{yy}	
\end{array}
\right)
\stackrel{90^\circ \mathrm{rot.}}{\longrightarrow}
\left(
\begin{array}{cc}
M_{yy} & -M_{yx} \\
-M_{xy} & M_{xx}	
\end{array}
\right),
\label{eq:rot90}
\end{equation}
which signifies that for passive and reciprocal materials where $M_{xy}=M_{yx}$, the \emph{cross-polarized} light, represented by the off-diagonal tensor elements, may gain an additional $\pi$-phase by rotating the anisotropic metasurface constituents by $90^\circ$. As such, metasurfaces with a pure ED-response, inherently only allowing a phase change of up to $\pi$, may utilize the above geometrical trick to reach the control of the full $2\pi$ phase space for the cross-polarized component of the light. In order for such a metasurface to be effective, however, it is important to maximize the conversion from co- to cross-polarized light as the incident light interacts with the metasurface. We now gauge the limit of the conversion efficiency by studying the reflection and transmission of $x$-polarized normal incident light on anisotropic metasurfaces, where the fields in the two half-spaces are
\begin{eqnarray}
\label{eq:Em2}
& \mathbf{E}_-(z)=E_0e^{ikz}\hat{\mathbf{x}}+E_0e^{-ikz}\left(r_{xx}\hat{\mathbf{x}} + r_{yx}\hat{\mathbf{y}} \right),\\
& \mathbf{H}_-(z)=\frac{E_0}{\eta}e^{ikz}\hat{\mathbf{y}}+\frac{E_0}{\eta}e^{-ikz} \left(-r_{xx}\hat{\mathbf{y}}+ r_{yx}\hat{\mathbf{x}} \right) ,\\
& \mathbf{E}_+(z)=E_0e^{ikz}\left(t_{xx}\hat{\mathbf{x}} + t_{yx}\hat{\mathbf{y}} \right),\\
& \mathbf{H}_+(z)=\frac{E_0}{\eta} e^{ikz}\left(t_{xx}\hat{\mathbf{y}} -t_{yx}\hat{\mathbf{x}} \right) \label{eq:Hp2}.
\label{eq:ms2Field}
\end{eqnarray}
The transmission coefficients can then be derived by applying the boundary conditions in equations \eref{eq:ms1BC} and \eref{eq:ms1J} (together with the assumption of a symmetric polarizability tensor), hereby yielding
\begin{eqnarray}
\label{eq:ms2T_1}
& t_{xx}=\frac{1-iC_{yy}}{(1-iC_{xx})(1-iC_{yy})+C_{yx}^2} ,\\
& t_{yx}=\frac{iC_{yx}}{(1-iC_{xx})(1-iC_{yy})+C_{yx}^2},
\label{eq:ms2T_2}
\end{eqnarray}

where $C_{ij}=\omega\alpha_{ij}/(2Av)$. It is important to note that a specific value of $t_{xx}$ should, \emph{in principle}, be attainable for different sets of ($C_{xx},C_{yy},C_{yx}$), meaning that the cross-polarized transmission, which can be written as $t_{yx}=iC_{yx}t_{xx}/(1-iC_{yy})$, can be controlled rather independently.
It is also worth noting that the continuity of the tangential electric field across the metasurface implies $r_{xx}=t_{xx}-1$ and $r_{yx}=t_{yx}$. In fact, by using these two relations and the simplifying assumption of a lossless metasurface, i.e., $|r_{xx}|^2+|r_{yx}|^2+|t_{xx}|^2+|t_{yx}|^2=1$, we can derive the following relation
\begin{equation}
|t_{yx}|^2=\mathrm{Re}\{t_{xx}\}-|t_{xx}|^2,
\label{eq:ms2Cond}
\end{equation}
where the right hand-side can be seen as a function of two variables $f(\mathrm{Re}\{t_{xx}\},\mathrm{Im}\{t_{xx}\})$, which has a maximum of $|t_{yx}|^2=1/4$ for $t_{xx}=1/2$. In other words, the theoretically maximum conversion efficiency is 25\%, but achieving this ultimate efficiency simultaneously dictates, c.f. equation \eref{eq:ms2T_1}, the condition $(1+iC_{xx})(1-iC_{yy})=C_{yx}^2$. Although this equation has multiple solutions, thus allowing for engineering the phase of $t_{yx}$, we note that most metasurface designs only feature efficiencies of a few percent, with the exception of a recent bi-layer configuration demonstrating conversion efficiency as high as 17\% for linearly polarized incident light at optical wavelengths \cite{Qin_2016}, and another configuration probing the limit with efficiency of 24.7\% for circularly polarized (CP) incident light at microwave frequencies \cite{Ding_2015}.

For linearly polarized light, the control of the cross-polarized transmission is typically realized by utilizing a system of high enough symmetry that a proper alignment of metasurface constituents yields a polarizability tensor that is diagonal, i.e., $\overline{\overline{\alpha}}=\textrm{diag}(\alpha_{xx},\alpha_{yy})$. If we now rotate the metasurface constituents by an angle $\theta$ in the $xy$-plane, the resulting polarizability tensor takes the form
\begin{equation}
\overline{\overline{\alpha}}_\theta =
\left(
\begin{array}{cc}
\alpha_{xx}\cos^2\theta+\alpha_{yy}\sin^2\theta & (\alpha_{xx}-\alpha_{yy})\cos\theta\sin\theta \\
(\alpha_{xx}-\alpha_{yy})\cos\theta\sin\theta & \alpha_{xx}\sin^2\theta+\alpha_{yy}\cos^2\theta	
\end{array}
\right) ,
\label{eq:ms2AlphaRot}
\end{equation}
where it is readily seen that the optimum rotation angle for maximizing the off-diagonal components is $\pi/4$ for which $\cos\theta\sin\theta$ takes on the maximum value of $\frac{1}{2}$. We note that the possibility to engineer both $\alpha_{xx}$ and $\alpha_{yy}$ allows for improved control of both the amplitude and phase of $t_{yx}$. This is crucial in phase-gradient metasurfaces, where one would like to keep the amplitude of $t_{yx}$ constant, while the phase changes along the metasurface in some predefined manner.

\subsection{Geometric metasurfaces} \label{sec:gm}
The extrinsic spin Hall effect of electrons is related to the spatial separation of electrons with opposite spin on interaction with certain scatterers. A similar phenomenon can occur for \emph{circularly} polarized light, where photons of opposite handedness (i.e., spin) are transversely separated by the interaction with a certain class of metasurfaces. The separation of opposite-spin photons occurs due to metasurfaces featuring a  \emph{geometrically-induced} phase-gradient that has an opposite slope for the two spin states. The geometric phase is also known as the Pancharatnam-Berry phase due to the pioneering work on this subject by S. Pancharatnam \cite{Pancharatnam_1956} and M. V. Berry \cite{Berry_1984}.

In order to illustrate the possibility of inducing a geometric phase factor on CP light, we consider, similarly to the previous subsection, a metasurface consisting of mirror-symmetric constituents so that for a proper choice of coordinate system the reflection and transmission tensors are diagonal, here represented by the matrix $\overline{\overline{M}}=\mathrm{diag}\left(M_{xx},M_{yy}\right)$. If the metasurface is rotated by an angle $\theta$ in the $xy$-plane, the resulting tensor $\overline{\overline{M}}_\theta$ takes the form of equation \eref{eq:ms2AlphaRot}. We now define the amplitude of the incident wave
\begin{equation}
\mathbf{E}_0^\pm=\frac{1}{\sqrt{2}}\left(\hat{\mathbf{x}}\pm i\hat{\mathbf{y}}\right),
\label{eq:ms3Circ}
\end{equation}
where $\pm$ stands for right-handed CP (RCP) and left-handed CP (LCP) light, respectively. The transmitted or reflected light can then be written as
\begin{equation}
\overline{\overline{M}}_\theta\cdot \mathbf{E}_0^\pm = \frac{1}{2}\left(M_{xx}+M_{yy}\right)\mathbf{E}_0^\pm + \frac{1}{2}\left(M_{xx}-M_{yy}\right)e^{\pm i2\theta}\mathbf{E}_0^\mp,
\label{eq:ms3RT}
\end{equation}
where the first term represents light of the same handedness as the incident wave, while the second term signifies light of opposite handedness that gains an additional geometric phase of twice the rotation angle $\theta$. More importantly, the sign of the phase term depends on the handedness, meaning that if the rotation angle varies along the metasurface, like $\theta(x)=\pi x/\Lambda$ where $\Lambda$ represents the periodicity, the associated plane wave in the far-field will gain an in-plane wave vector component of $k_x=\pm 2\pi/\Lambda$, i.e., light of opposite handedness will propagate in opposite directions with respect to the normal of the metasurface.

Having verified the presence of a geometrically induced phase term for CP incident light, we now discuss conditions for 100\% conversion efficiency into the second term in equation \eref{eq:ms3RT}. It is clear that the required condition is $M_{yy}=-M_{xx}$, which represents the functionality of a half-wave plate and will make the first term in equation \eref{eq:ms3RT} vanish. The 100\% conversion efficiency, however, is only reached for lossless metasurfaces that either \emph{fully} transmit or reflect light so that $|M_{xx}|=|M_{yy}|=1$ \cite{Luo_2015}. As such, it is evident that simple ED-based metasurfaces cannot satisfy these conditions, since a non-negligible polarizability tensor always leads to both transmission and reflection.

\subsection{Metasurfaces near a metal screen} \label{sec:screen}
In an attempt to reach the 100\% efficiency of geometric metasurfaces, the straightforward implementation amounts in placing the metasurface near an optically thick metal film, hereby ensuring that transmission is zero \cite{Luo_2015}. This type of metasurface, however, is not only limited to controlling circular polarization states, but may also fully control reflected light of linear polarization. We note that metal-backed metasurfaces have different names, typically referenced as reflectarrays \cite{Kelkar_1991,Pozar_1993,Pozar_1997}  or high-impedance surfaces \cite{Sievenpiper_1999} in the milli- and micrometer wave community, while at optical wavelengths they are known as either meta-reflectarrays \cite{Yang_2014}, film-coupled nanoantenna metasurfaces \cite{Moreau_2012}, or gap surface plasmon-based (GSP-based) metasurfaces due to the crucial role of GSPs in molding the response of the reflected light \cite{Pors_2013,Pors_2013_2}.

As a way to exemplify the full control of the reflected light for linearly polarized incident light, we consider the simplest situation of an isotropic ED-based metasurface that is placed in front of a metal screen which is treated as a perfect electric conductor (PEC), with optical properties of the spacer layer being the same as the surrounding medium [see \fref{fig:ms1}(b)]. In the limit of a spacer thickness $h$ that is much smaller than the wavelength of the incident light, we can apply the electrostatic image theory to draw an equivalent configuration, as shown in the lower part of \fref{fig:ms1}(b), which consists of two identical metasurfaces that are separated by the distance $2h$ and feature equal but out-of-phase ED responses. Within the limits of quasistatic theory (i.e., currents still oscillate in time) it is clear that the out-of-phase (i.e., circulating) electric currents generate \emph{magnetic} dipole (MD) radiation at the expense of the total ED response being suppressed. As such, an ED-based metasurface in close proximity of a metal screen can to a first approximation be considered an interface discontinuity featuring magnetic currents and zero transmission. The reflection coefficient can be derived by considering the configuration in \fref{fig:ms1}(a), where the field in the left half-space ($z<0$) follows equations \eref{eq:Em} and \eref{eq:Hm}, and the field in the right half-space ($z>0$) is zero. The proper boundary condition at the metasurface needs to handle the discontinuity of the electric field due to the presence of a magnetic surface current, and it is given by
\begin{equation}
\hat{\mathbf{z}}\times \mathbf{E}_-=\mathbf{M}_s,
\label{eq:ms4:BC}
\end{equation}
where $\mathbf{M}_s$ is the magnetic surface current in analogy to the electric counterpart can be written as
\begin{equation}
\mathbf{M}_s=-i\omega A^{-1}\mu_0\beta \mathbf{H}_\mathrm{av}.
\label{eq:ms4M}
\end{equation}
Here, $\mathbf{m}=\beta\mathbf{H}_\mathrm{av}$  is the magnetic dipole moment, with $\beta$ being the magnetic polarizability of the unit cell constituent and $\mathbf{H}_\mathrm{av}=\frac{1}{2}\left(\mathbf{H}_+ + \mathbf{H}_-\right)$ is the average magnetic field at the metasurface, which in the present configuration amounts to $\mathbf{H}_\mathrm{av}=\frac{1}{2}\mathbf{H}_-$.
By applying the above boundary condition on the electric field in equation \eref{eq:Em}, we obtain the following reflection coefficient
\begin{equation}
r=-\frac{1+iD}{1-iD},
\label{eq:ms4R}
\end{equation}
where $D=\omega\beta/(2 Av)$. If we for simplicity consider the lossless situation (i.e., $D$ is a real-valued function) and write the reflection coefficient as $r=|r|e^{i\phi_r}$, it is easily shown that $|r|=1$ and
\begin{equation}
\phi_r=2\tan^{-1}\left(D\right)+\pi,
\label{eq:ms4:Phase}
\end{equation}

which signifies that the reflection phase spans the entire $2\pi$ phase space when $D$ varies from $-\infty$ to $\infty$. In real configurations, we note that losses and retardation effects do decrease the attainable phase space somewhat, but it is in most cases considerably larger than $\pi$. Moreover, we emphasize that the Ohmic losses, which are typically considered decremental for the metasurface performance, may actually also be exploited to take spatial control of the amplitude of the reflected light. Finally, we point out that when $|D|\gg 1$, which might occur near a resonance in $\beta$, the reflection coefficient is $r\simeq 1$, hereby verifying that the metasurface may behave as an artificial magnetic conductor, also known as an high-impedance surface \cite{Sievenpiper_1999}.

In the above derivation of the functional form of the reflection coefficient for ED-based metasurfaces near a metal screen [equation \eref{eq:ms4R}], we utilized image theory to justify the presence of a magnetic response, hereby implicitly assuming that the reference geometry (i.e., without metasurface), is the free space, as also evident from \fref{fig:ms1}(b). It should, however, be noted that the metal screen can alternatively be viewed as part of the reference geometry, thus entailing an ED-only response of the nearby metasurface, with the interaction with the screen being taken explicitly into account \cite{Estakhri_2014,Estakhri_2015}. Regardless of the view, we emphasize that equation \eref{eq:ms4R} does describe the behaviour of the reflection coefficient when the configuration features a resonance.

As a final comment, it ought to be mentioned that the meta-reflectarrays can be well-described by the coupled-mode theory (CMT) \cite{Fan2003,Wu2011,Che_2015}, which provides a general guidance for designing reflective metasurfaces with tailored functionalities. More specifically, the properties of meta-atoms are fully controlled by the two simple parameters (i.e., the intrinsic and radiation losses), which are, in turn, dictated by the geometrical or material properties of the underlying structures \cite{Che_2015}.

\subsection{Metasurfaces featuring electric and magnetic dipole responses} \label{sec:Huygens}
So far, we have only considered metasurfaces with inclusions featuring ED responses. A second class of metasurfaces, which are typically known as Huygens' metasurfaces and allow for greater control of the light, encompass unit cells with both ED and MD responses \cite{Pfeiffer_2013}. In the study of the fundamental properties of this class of metasurfaces, we again consider an isotropic metasurface in the simple configuration of \fref{fig:ms1}(a). For a $x$-polarized incident wave, the electromagnetic fields are described by equations \eref{eq:Em}-\eref{eq:Hp}, but the boundary conditions that must be satisfied are now
\begin{equation}
\hat{\mathbf{z}}\times\left( \mathbf{E}_+-\mathbf{E}_- \right)=-\mathbf{M}_s \quad, \quad \hat{\mathbf{z}}\times\left( \mathbf{H}_+-\mathbf{H}_- \right)=\mathbf{J}_s,
\label{eq:ms5BC}
\end{equation}
with the surface currents, as defined in equations \eref{eq:ms1J} and \eref{eq:ms4M}, ensuring a discontinuity in both the electric and magnetic field. By applying the boundary conditions to the electromagnetic field, we obtain the following reflection and transmission coefficients
\begin{eqnarray}
\label{eq:ms5R}
& r=\frac{i\left(C-D\right)}{\left(1-iC\right) \left(1-iD\right)} , \\
& t=\frac{1+CD}{\left(1-iC\right) \left(1-iD\right)},
\label{eq:ms5T}
\end{eqnarray}
where $C=\omega\alpha/(2Av)$ and $D=\omega\beta/(2Av)$. It is readily seen that zero transmission and reflection can occur when $C=-D^{-1}$ and $C=D$,
respectively. Interestingly, it should be noted that the latter condition is equivalent to requiring $\alpha=\beta$, which is actually the Kerker condition for dominant forward scattering (relative to the incident light) of light by a single nanoparticle \cite{Kerker_1983,Camara_2011,Pors_2015}. As such, zero-reflection metasurfaces, also conventionally just referred to as Huygens' metasurfaces, are the result of forward-scattering meta-atoms, with the transmission coefficient taking on the simplified expression
\begin{equation}
t=\frac{1+iC}{1-iC}.
\label{eq:ms5T2}
\end{equation}
It is evident that the function is (ignoring the sign difference) equivalent to the reflection coefficient in equation \eref{eq:ms4R} from a metal-backed metasurface, meaning that in the lossless case one can simultaneously achieve perfect transmission and full control of the transmission phase by proper choice of $C$ and $D$, still keeping $C=D$ at all times. The condition $C=D$ might in fact also be purposely violated in order to induce an amplitude modulation on the transmitted fields.

As an alternative to utilize metasurface constituents with both an ED and MD response, it is also possible to stack ED-based metasurfaces that feature out-of-phase electric currents, hereby creating an effective MD response that can be engineered to achieve zero reflection. Such a stacking of metasurfaces is often referred to as transmitarrays \cite{Pozar_1996,Colan_2010} or meta-transmitarrays \cite{Monticone_2013}, depending on the frequency range of operation.

\subsection{Metasurfaces featuring magnetoelectric coupling}
To this end, we have demonstrated the possibility to fully control the reflection and transmission of light by utilizing metasurfaces with electric and magnetic responses. The linear response of metasurfaces, however, can be further generalized by including \emph{magnetoelectric} coupling, which takes into account that the magnetic field may induce an electric response and vice versa. The most general expression of the induced surface currents on the metasurface take on the form
\begin{eqnarray}
\mathbf{J}_s=-i\omega A^{-1}\left[\varepsilon \overline{\overline{\alpha}}\mathbf{E}_\mathrm{av}+v^{-1}\overline{\overline{\gamma}}\mathbf{H}_\mathrm{av}\right], \\
\mathbf{M}_s=-i\omega A^{-1}\left[v^{-1}\overline{\overline{\kappa}}\mathbf{E}_\mathrm{av}+\mu_0\overline{\overline{\beta}}\mathbf{H}_\mathrm{av}\right],
\label{eq:ms6JM}
\end{eqnarray}
where $\overline{\overline{\alpha}}$ and $\overline{\overline{\beta}}$ are the $2\times2$ electric and magnetic polarizability tensors, respectively, and $\overline{\overline{\gamma}}$ and $\overline{\overline{\kappa}}$ describe the magnetoelectric coupling. It should be noted that the above relations define a \emph{bianisotropic} metasurface, with the four tensors obeying the relations $\overline{\overline{\alpha}}=\overline{\overline{\alpha}}^T$, $\overline{\overline{\beta}}=\overline{\overline{\beta}}^T$, and $\overline{\overline{\kappa}}=-\overline{\overline{\gamma}}^T$ for reciprocal metasurfaces \cite{Kong_1972}, hereby illustrating that this class of metasurfaces may be characterized by up to 10 independent parameters. In most cases, however, symmetry of the unit cell and its constituents reduce the number of surface parameters, possibly reaching the lower bound of three independent parameters for bi-isotropic metasurfaces.

In order to exemplify the peculiar optical properties of bianisotropic metasurfaces, we again limit ourselves to $x$-polarized incident light and demand the metasurface to conserve the polarization of the light (i.e., no cross-polarized light), meaning that the surface currents can be written as \cite{Pfeiffer_2014}
\begin{eqnarray}
\mathbf{J}_s=-i\omega A^{-1}\left[\varepsilon \alpha\mathbf{E}_\mathrm{av}+v^{-1}\gamma\mathbf{H}_\mathrm{av}\right], \\
\mathbf{M}_s=-i\omega A^{-1}\left[-v^{-1}\gamma\mathbf{E}_\mathrm{av}+\mu_0\beta\mathbf{H}_\mathrm{av}\right],
\label{eq:ms6JM2}
\end{eqnarray}
where the magnetoelectric coupling is described by a single parameter $\gamma$. Moreover, we consider a configuration similar to \fref{fig:ms1}(a) and calculate the reflection and transmission coefficients for light propagating in both the $\pm z$-directions. In both cases, the boundary conditions at the metasurface are given by equation \eref{eq:ms5BC}, yielding
\begin{eqnarray}
\label{eq:ms6R}
r_\pm=\frac{i(C-D)\pm 2iF}{(1-iC)(1-iD)-F^2}, \\
t_\pm=\frac{1+CD+F^2}{(1-iC)(1-iD)-F^2},
\label{eq:ms6T}
\end{eqnarray}
where $C=\omega\alpha/(2Av)$, $D=\omega\beta/(2Av)$, and $F=\omega\gamma/(2Av)$. It is readily seen that the reflection coefficient is depending on the direction of incident light, which opens up the possibility of designing metasurfaces with markedly different responses in reflection for the two directions of wave propagation. Metasurfaces utilizing this property together with suppression of transmission (i.e., $CD+F^2=-1$) are known as metamirrors \cite{Radi_2014,Asadchy_2015}. Another interesting property of metamirrors is the possibility to only reflect light in a narrow frequency band, related to a resonant behavior of the metasurface, while rendering transparent at other frequencies. In general, when comparing the reflection and transmission coefficients of conventional Huygens' metasurfaces [equations \eref{eq:ms5R} and \eref{eq:ms5T}] with metasurfaces also featuring magnetoelectric coupling [equations \eref{eq:ms6R} and \eref{eq:ms6T}], it is evident that it might be easier to reach either zero reflection or transmission due to the possibility to engineer not only the electric and magnetic response but also the magnetoelectric contribution. For example, zero reflection or transmission can (in principle) occur without a magnetic response (i.e., $D=0$).

As a final comment, we emphasize that the independence of the transmission coefficient in equation \eref{eq:ms6T} on the direction of wave propagation is a direct consequence of the assumption of reciprocity. In this regard, it is important to emphasize that asymmetric transmission can occur for a reciprocal metasurface that converts light to the cross-polarized component. In order to illustrate this fact, we note that the transmission coefficient tensors for $\pm z$-propagating incident light are related by \cite{Menzel_2010}
\begin{equation}
\overline{\overline{t}}_+=\left(
\begin{array}{cc}
t_{xx} & t_{xy} \\
t_{yx} & t_{yy}	
\end{array}
\right)
\quad , \quad
\overline{\overline{t}}_-=\left(
\begin{array}{cc}
t_{xx} & -t_{yx} \\
-t_{xy} & t_{yy}	
\end{array}
\right) ,
\label{eq:Tpm}
\end{equation}
which has the consequence that transmission coefficients like
\begin{equation}
\overline{\overline{t}}_+=\left(
\begin{array}{cc}
0 & 0 \\
1 & 0	
\end{array}
\right)
\quad , \quad
\overline{\overline{t}}_+=\frac{1}{2}\left(
\begin{array}{cc}
1 & -i \\
-i & -1	
\end{array}
\right)
\label{eq:Tlpcp}
\end{equation}
perfectly transmit $x$- and RCP light for propagation along the $+z$-axis, respectively, while these polarization states are completely blocked for $-z$-propagation \cite{Pfeiffer_2014,Niemi_2013}. We note that it is the reciprocity condition in equation \eref{eq:Tpm} that also leads to asymmetric transmission for CP light in certain chiral metasurfaces \cite{Fedotov_2006,Singh_2009}.

\section{Realization of metasurfaces}
With the attempt in previous section to classify the broad range of existing metasurfaces, including a conceptual description of the connection between polarizabilities of meta-atoms and the reflection and transmission coefficients, we now discuss practical realizations. The illustration of the many approaches towards flat optical components will be limited to the ubiquitous case of a blazed diffraction grating, postponing the wide range of metasurface applications to the subsequent section. The blazed grating functionality, as discussed in the Introduction, is realized by incorporating a constant phase gradient on the metasurface so that reflected/transmitted light will gain an additional in-plane momentum, hereby resulting in the famous anomalous propagation of light. However, before discussing the many different implementations of blazed grating functionality, we would like to pay attention to the meta-atoms that occupy the metasurface unit cells.

\subsection{Fundamental building blocks} \label{sec:BuildingBlocks}
In the previous section, we assumed metasurfaces featuring electric, magnetic, and magnetoelectric responses without specifying the origin of this behavior. Here, we would like to highlight the optical properties of four nanostructures that are indispensable building blocks for optical and near-infrared metamaterials and metasurfaces; that is, elongated metallic nanorods, metal-insulator-metal (MIM) resonators, metallic split ring resonators (SRRs), and silicon nanodisks. We note that the metallic nanorod and SRR also have low-frequency regime counterparts, though the relaxed fabrication constraints at these frequencies sometimes result in more complex meta-atoms, like fractal-shaped configurations \cite{Zubir_2009}.

\begin{figure}[tb]
	\centering
		\includegraphics[width=15.5cm]{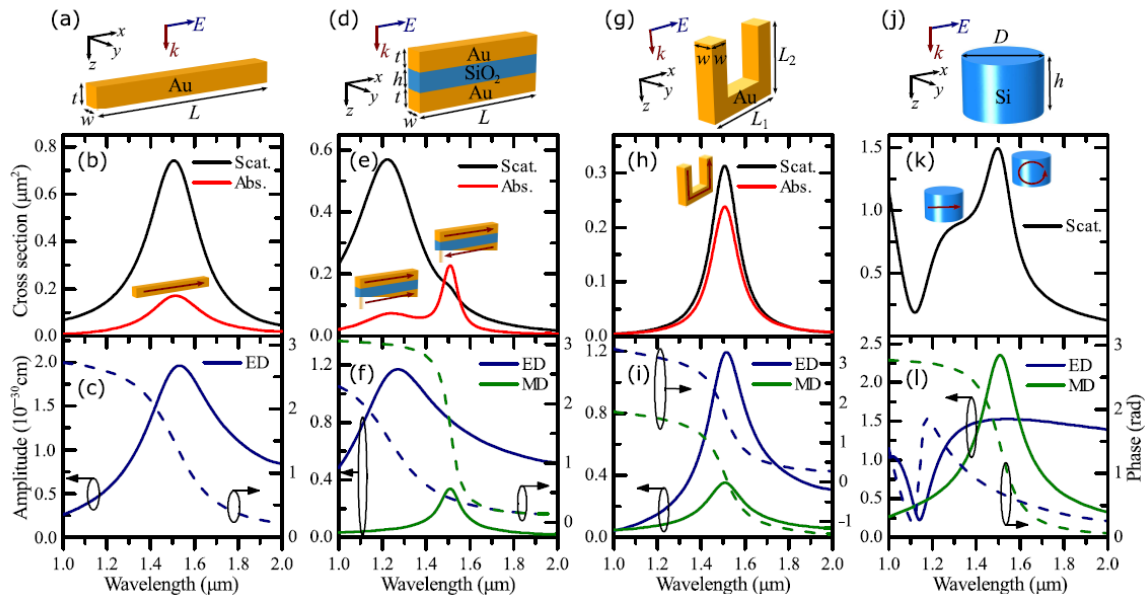}
	\caption{Optical properties in the near-infrared regime of (a-c) gold nanorod, (d-f) MIM-resonator, (g-i) split ring resonator, and (j-l) silicon nanodisk. (a,d,g,j) show sketches of the four different configurations, with (b,e,h,k) presenting the calculated absorption (red) and scattering (black) cross sections as a function of wavelength for $x$-polarized plane wave incident light propagating along the $z$-axis. The dimensions are $L=450$\,nm and $w=t=50$\,nm for the nanorod; $L=350$\,nm and $w=t=h=50$\,nm for the MIM-resonator; $L_1=220$\,nm, $L_2=190$\,nm, and $w=50$\,nm for SRR; $D=430$\,nm and $h=300$\,nm for the silicon nanodisk. The insets represent schematic drawings of the induced (polarization) currents in the nanostructures. (c,f,i,l) show the amplitude and phase of the induced electric (blue) and magnetic (green) dipole moments for incident light with amplitude $1$\,V/m. The MD moment is divided by the speed of light $c$ to have the units of the ED moment.}
	\label{fig:BuildingBlocks}
\end{figure}

In the pursuit of a resonant electric dipole response, the metallic nanorod configuration, as sketched in figure \ref{fig:BuildingBlocks}(a), appears as an attractive candidate due to the simple geometry and elongated shape.
The long axis of the nanorod, which is considerably larger than the subwavelength cross sectional dimensions (i.e., $w,t\ll L$), implies a strongly anisotropic response, with noticeable light-matter interaction being limited to polarizations along this axis. Since metallic nanorods can be considered as plasmonic waveguides of finite length, it follows that the associated resonances can be described as standing-wave resonances of constructive interfering propagating and counter-propagating plasmonic modes \cite{Ditlbacher_2005,Bozhevolnyi_2007,Dorfmuller_2009}. In the view of nanorods as Fabry-P{\'e}rot resonators, the resonance condition can be expressed as
\begin{equation}
\textrm{Re} \{k_{\textrm{spp}}\} L+\Gamma=\pi m,
\label{eq:FPres}
\end{equation}
where $k_{\textrm{spp}}$ is the (complex) wave number of the plasmonic waveguide mode, $L$ is the length of the nanorod, $\Gamma$ is the reflection phase of the mode at nanorod terminations, and $m$ is an integer. It should be noted that the quantity $\Gamma/\textrm{Re} \{k_{\textrm{spp}}\}$ can be considered as an effective increase of the resonator length $L$ due to the plasmonic mode extending beyond the geometrical end facets \cite{Novotny_2007}. We note that this is in stark contrast to conventional linear dipole antennas at radio frequencies where electromagnetic waves induce electric currents in the metal with resonances occurring at integer numbers of half wavelengths \cite{Jackson}. In regard to the symmetry of the standing wave resonances in nanorods, it should be noted that only modes of odd integers $m$ can be excited by normal incident light. At the same time, the requirement that metasurfaces should consist of subwavelength unit cells leads to the utilization of the smallest resonant configuration, corresponding to nanorods with the fundamental resonance ($m=1$) close to the working wavelength. As an illustrative example, figure \ref{fig:BuildingBlocks}(b) shows scattering and absorption cross sections of a gold nanorod in air (with dimensions $w=t=50$\,nm and $L=450$\,nm) for a normal incident $x$-polarized plane wave. The wavelength of the fundamental resonance, as characterized by maximum in the optical cross sections, is $\lambda \simeq 1.5$\,$\mu$m, with scattering being the dominant decay channel for the plasmonic mode. This is in agreement with the theory of multipole expansion of electromagnetic fields, which dictates that systems featuring a dominant ED moment will be strongly radiating \cite{Jackson, Papasimakis_2009,Pors_2011}. As such, it is customary to denote nanostructures with dominant or suppressed ED moments as bright and dark configurations, respectively, since (by reciprocity) it provides the link for external light to interact with the nanostructure \cite{Zhang_2008}. In the present nanorod example, figure \ref{fig:BuildingBlocks}(c) displays the induced ED moment in the nanorod as a function of wavelength. As expected, the ED moment is resonantly enhanced at $\lambda\simeq 1.5$\,$\mu$m, while the phase undergoes a change of close to $\pi$ for wavelengths of opposite sides of the resonance.

Having clarified the optical properties of elongated nanorods, we turn to a slightly more complicated configuration consisting of two metal nanorods separated by a nanometer-thin dielectric spacer, such as silicon dioxide (SiO$_2$) (figure \ref{fig:BuildingBlocks}(d)), also known as MIM resonator. The optical properties of this kind of configuration can be qualitatively explained within the hybridization model \cite{Prodan_2003}, where the near-field interaction between the two (identical) nanorods will lead to two new super modes: a high-energy mode characterized by in-phase polarization currents in the nanorods, thus having a large total ED moment, and a low-energy mode with out-of-phase currents, hereby leading to a suppression of the total ED moment in favor of a magnetic response (see insets in figure \ref{fig:BuildingBlocks}(e)). In order to illustrate this line of thought, figure \ref{fig:BuildingBlocks}(c) presents numerical calculations of the optical cross sections of a gold-SiO$_2$-gold resonator ($L=350$\,nm and $w=t=h=50$\,nm, c.f. figure \ref{fig:BuildingBlocks}(d)) interacting with a $z$-propagating $x$-polarized incident plane wave. In accordance with the hybridization model, the spectra features a broad resonance at $\lambda \simeq 1.2$\,$\mu$m that efficiently scatters light and a narrower resonance at $\lambda \simeq 1.5$\,$\mu$m whose existence is mainly seen in the absorption cross section. The noticeable different optical properties of the two plasmonic modes relate to the resonances being of electric and magnetic origin, as confirmed by calculating the induced ED and MD moments of the MIM resonator (figure \ref{fig:BuildingBlocks}(f)). It is worth noting that the magnetic resonance of MIM structures has been utilized in the past to create metamaterials with negative permeability \cite{Yuan_2007,Cai_2007} and negative refractive index \cite{Shalaev_2005,Dolling_2006,Dolling_2007} at optical and near-infrared wavelengths. Moreover, in the retardation-based regime (i.e., $L \ll \lambda$ is not strictly valid), which is outside the scope of the hybridization model, MIM resonators also function as Fabry-P{\'e}rot resonators with the magnetic resonance being the fundamental mode of standing-wave GSP waveguide modes \cite{Bozhevolnyi_2007,Miyazaki_2006,Jung_2009}.

Metallic split ring resonators, as sketched in figure \ref{fig:BuildingBlocks}(g), are ubiquitous meta-atoms for realizing a magnetic response in practically any frequency regime. For example, SRRs were a key element in the first experimental verification of metamaterials with negative refraction at microwave frequencies \cite{Shelby_2001} and realizing negative permeability at near-infrared wavelengths \cite{Linden_2004}. It is mainly only at optical wavelengths that MIM resonators are used instead of SRRs, which owes to the fact that the finite plasma frequency of real metals leads to a saturation of the magnetic resonance wavelength as the SRR is being scaled down \cite{Zhou_2005,Klein_2006}. In order to change the scaling and electromagnetic response of SRRs, these configurations may consist of several rings \cite{Pendry_1999} or a single ring with multiple cuts \cite{Zhou_2005}, however, here we restrict ourselves to discuss the simplest case of a single ring featuring only one cut (figure \ref{fig:BuildingBlocks}(g)). For such a U-shaped SRR with the dimensions $L_1=220$\,nm, $L_2=190$\,nm, and $w=50$\,nm, figure \ref{fig:BuildingBlocks}(h) demonstrates the significant increase in both scattering and absorption cross sections at the magnetic resonance at $\lambda \simeq 1.5$\,$\mu$m. It should be emphasized that the resonance is equivalent to the fundamental resonance of straight nanorods (figure \ref{fig:BuildingBlocks}(a)-\ref{fig:BuildingBlocks}(c)), where the magnetic response arises from bending the nanorod into a U-shape so that the linear plasmonic current becomes a partial circulating current (see inset of figure \ref{fig:BuildingBlocks}(h)) \cite{Rockstuhl_2006,Pors_2010}. For this reason, the SRR features both an ED and MD response (pointing along the $x$- and $y$-axis, respectively), as depicted in figure \ref{fig:BuildingBlocks}(i), with the possibility to engineer the relative strength between these two dipole moments by the degree of which the nanorod is bent into a one-cut loop. Moreover, it ought to be mentioned that SRRs, though often just referred to as a magnetic meta-atoms, also include a magnetoelectric response that arises from $x$-polarized light inducing an MD along the $y$-axis, and, conversely, a magnetic field along the $y$-axis generating an ED along the $x$-direction \cite{Chen_2005,Zhaofeng_2009}.

The last nanostructure that we would like to highlight is a silicon nanodisk (figure \ref{fig:BuildingBlocks}(j)), which seems to be an essential building block for the realization of near-infrared all-dielectric Huygens' metasurfaces \cite{Decker_2015,Arbabi_2015_2,Yu_2015,Arbabi_2015}. The first point to note is that this type of metasurface utilizes the dielectric properties of silicon, characterized by negligible absorption for $\lambda>0.7$\,$\mu$m and a high refractive index of $n\sim 3.5$ throughout the entire near-infrared regime. The second point has its origin in the Mie theory, stating that spherical particles in general feature both electric and magnetic resonances \cite{Bohren}, though the condition for excitation of magnetic modes can only be reached in magnetodielectric \cite{Holloway_2003} or high-permittivity \cite{Zhao_2009,Evlyukhin_2012} particles. Since the overall conclusions of Mie theory are generally valid for arbitrarily shaped nanoparticles, it follows that nano-disks and -cubes so too feature electric and magnetic resonances \cite{Staude_2013,Sikdar_2015,Shalaev_2015}. Moreover, by proper choice of geometrical parameters it is possible to have spectrally overlapping ED and MD resonances that, in accordance with the Kerker conditions \cite{Kerker_1983}, leads to strongly directional scattering of light \cite{Staude_2013,Sikdar_2015, Krasnok_2012, Fu_2013}, while perfect matching of the ED and MD moments in the arrayed counterpart (i.e., metasurface) results in zero reflection of light (see section \ref{sec:Huygens}), with the phase of the transmitted light being determined by the exact values of the ED and MD moments \cite{Decker_2015,Shalaev_2015}. As an illustration of the optical properties of silicon nanodisks, figure \ref{fig:BuildingBlocks}(k) displays the scattering cross section for a disk with diameter $D=430$\,nm and height $h=300$\,nm. It is seen that the spectrum features a peak at $\lambda\simeq 1.5$\,$\mu$m and a shoulder at $\lambda\simeq 1.3$, which is the signature of the excitation of a second, though weaker, mode. The origin of the spectral peaks is elucidated through the calculation of the induced dipole moments (figure \ref{fig:BuildingBlocks}(l)), which clarifies that the shoulder in the spectrum originates from a broad ED resonance, while the peak at $\lambda\simeq 1.5$\,$\mu$m arises from excitation of a strong MD mode. Interestingly, it is noticed that the ED and MD moments are equal at $\lambda \simeq 1.6$\,$\mu$m, meaning that this nanodisk configuration satisfies the criterion for becoming an element in a phase-gradient Huygens' metasurface working at $1.6$\,$\mu$m. As a final comment, we stress that all-dielectric metasurfaces operating at visible wavelengths can be advantageously designed using titanium dioxide (TiO$_2$), which features  a relatively high refractive index ($n>2.3$) in the entire visible spectrum and, unlike silicon, negligible absorption \cite{Devlin_2016}.

\subsection{V-antenna metasurfaces} \label{sec:Vmetasurface}
\begin{figure}[tb]
	\centering
		\includegraphics[width=7.5cm]{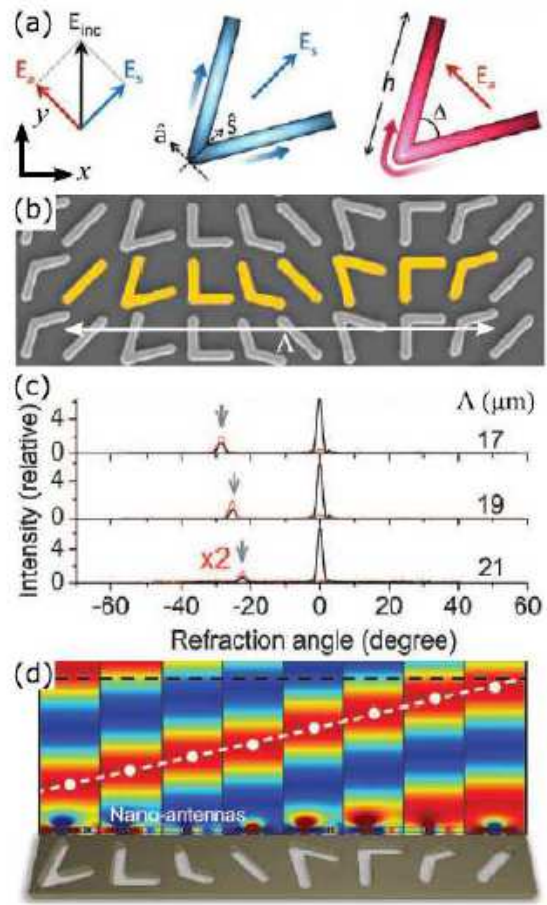}
	\caption{(a) Top-view of metallic V-antennas with colors indicating the current distribution of the symmetric (blue) and antisymmetric (pink) mode. (b) Scanning electron microscopy (SEM) image of fabricated metasurface featuring a linear phase gradient for cross-polarized light along the $x$-axis with period $\Lambda$. (c) Measured far-field intensity as a function of angle and periodicity ($\Lambda=17,19,21$\,$\mu$m) for the total (black) and cross-polarized (red) transmitted light when the incident light is $y$-polarized and normal to the metasurface. The working wavelength is $\lambda=8$\,$\mu$m. (d) Sketch of super cell of Babinet-inverted V-antenna metasurface featuring a linear phase gradient for the cross-polarized light. The image shows simulations of the transmitted cross-polarized electric field of the individual V-apertures in a gold film at the wavelength $\lambda=676$\,nm. (a-c) Adapted with permission from \cite{Yu_2011}; Copyright 2011 American Association for the Advancement of Science. (d) Adapted with permission from \cite{Ni_2013}; Copyright 2013 Changchun Institute of Optics, Fine Mechanics and Physics.}
	\label{fig:Vantenna}
\end{figure}

In connection with the derivation of the generalized laws of reflection and refraction by N. Yu \emph{et al.} \cite{Yu_2011}, the concurrent realization of interface discontinuities was based on ED-only metasurfaces, constructed by a periodic arrangement of V-shaped metallic nanoantennas, as sketched in figure \ref{fig:Vantenna}(a).
This type of nanostructure is characterized by the length $h$ and angle $\Delta$, and in the wavelength range of interest it features two resonances (termed symmetric and antisymmetric with respect to the current flow in the two arms) that can be excited by orthogonally polarized light. In accordance with the discussion in section \ref{sec:EDresponse}, the V-antenna can be described by a diagonal polarizability tensor in the $(\hat{\mathbf{a}}, \hat{\mathbf{s}})$ basis, but in order to maximize the off-diagonal elements the V-antenna is rotated 45$^\circ$ relative to the incident polarization (along the $y$-axis). As a proof-of-concept experiment, the metasurface was realized in the infrared regime  at $\lambda=8$\,$\mu$m (and later extended to the near-infrared range \cite{Ni_2012}) with the linearly varying phase being discretized into eight steps along the $x$-axis, as seen in the top-view image of the metasurface in figure \ref{fig:Vantenna}(b). The V-antennas (or nanorods) are 50\,nm thick, made of gold, and are fabricated on a silicon wafer with dimensions carefully chosen so that the amplitude of cross-polarized scattered light is roughly the same for all the meta-atoms, while the phase increases in steps of $\pi/4$ along the $x$-axis. It is worth noting that the latter four meta-atoms in the super cell are equivalent to the first four elements, just rotated 90$^\circ$ in the $xy$-plane to induce an additional $\pi$-phase shift [see equation \eref{eq:rot90}]. For normal incident $y$-polarized light, figure \ref{fig:Vantenna}(c) displays the transmitted far-field intensity of the total (black) and cross-polarized (red) light as a function of angle and metasurface periodicity $\Lambda$. As seen, it is only the cross polarized light that feels the linear phase gradient and, thus, in accordance with the theory of blazed gratings, is being redirected into the $-1$ diffraction order. Moreover, it is seen that the fraction of light being anomalously refracted compared to directly transmitted increases with decreasing periodicity $\Lambda$, which is a natural consequence of a closer packing of the V-antennas. That said, it should be noted that the absolute conversion efficiency, defined as the power in cross-polarized light to the incident power, is only on the order of a few percent \cite{Aieta_2012}.

In the above example of the performance of V-antenna metasurfaces, it is clear that the cross-polarized component of the transmitted light is noticeably weaker than the co-polarized one. Since the co-polarized component does not feel the phase-gradient, it is typically an undesirably background field that complicates measurements or the practical utility of the designed meta-device. In an attempt to increase the ratio between cross- and co-polarized transmitted light, it has been suggested to utilize Babinet's principle, stating that the complementary configuration of an opaque metal film with V-apertures also work as a phase-gradient metasurface for cross-polarized light \cite{Ni_2013}. The validity of this line of thought is illustrated in figure \ref{fig:Vantenna}(d), which shows juxtaposed transmitted light from eight different apertures in a 30\,nm thick gold film at the wavelength $\lambda=676$\,nm, thus verifying the possibility to induce a linear phase-gradient on the cross-polarized light, with the additional benefit that the ratio of cross- and co-polarized light is roughly one order of magnitude higher than for V-antenna metasurfaces \cite{Ni_2013}. As a final note, we would like to mention that the low conversion efficiency of either V-antenna or -aperture metasurfaces can be considerably enhanced by a unification of the two types of metasurfaces into a single bi-layer structure with strong interlayer coupling. Such a configuration has been realized at visible wavelengths ($\lambda\simeq 770$\,nm), demonstrating an experimental and simulated conversion efficiency of $\sim 17$\% and $\sim 29$\%, respectively \cite{Qin_2016}. The (at first sight) paradox that the simulated result exceeds the theoretical maximum value of 25\% (see section \ref{sec:EDresponse}) is due to the finite, though subwavelength, thickness of the bi-layer of $130$\,nm $\sim\lambda/6$, thus leading to non-negligible retardation effects and a break of the radiation symmetry.

\subsection{Geometric metasurfaces}\label{sec:GMs}
\begin{figure}[tb]
	\centering
		\includegraphics[width=14cm]{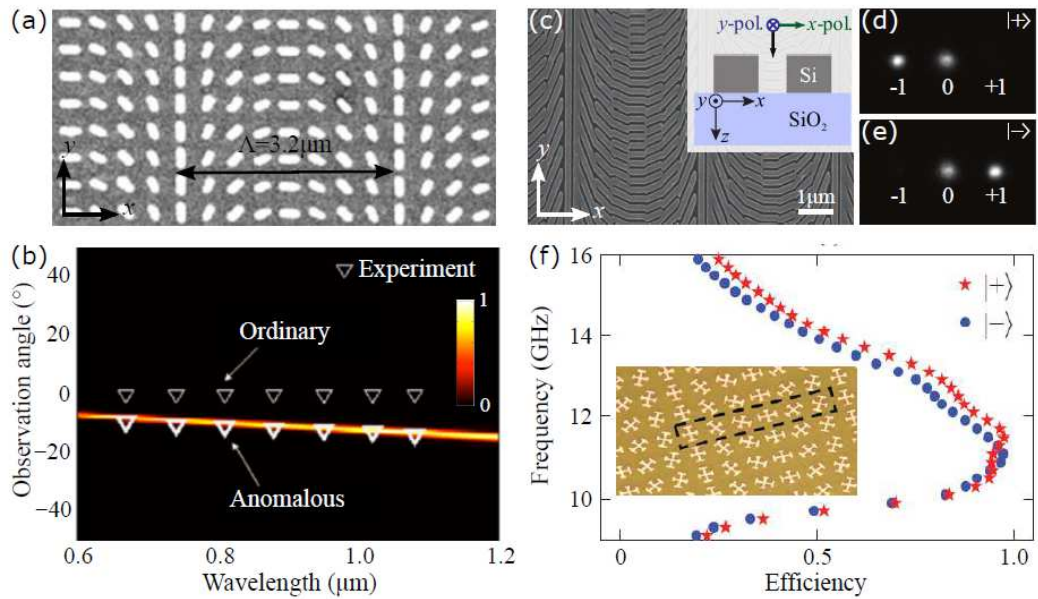}
	\caption{(a) Top-view SEM image of geometric metasurface consisting of 40\,nm thick gold nanorods on an ITO coated glass substrate. The inter-particle spacing is 400\,nm, and the geometrically induced phase gradient is imprinted along the $x$-axis by rotating the nanorods in steps of $\pi/8$. (b) Colormap shows the  calculated (normalized) far-field intensity of RCP light in the substrate as a function of wavelength and refraction angle when the LCP incident light impinges normally to the metasurface. Triangular markers indicate experimentally measured anomalous refraction angles. (c) Top-view SEM image of an all-dielectric geometric metasurface for visible light with the linear phase-gradient along the $x$-axis. Inset shows a side-view drawing of the metasurface. (d,e) Measured far-field intensity at $\lambda=550$\,nm for RCP ($|+\rangle$) and LCP ($|-\rangle$) incident light. (f) Microwave experiment of geometric metasurface placed atop a dielectric spacer and metal back-reflector. Inset shows an image of the fabricated metasurface, with the dashed rectangle highlighting the boundary of the super-cell within the metallic meta-atom is rotated in steps of $\pi/6$. The graph presents the fraction of light that is anomalously reflected for both RCP (stars) and LCP (circles) incident light. (a,b) Adapted with permission from \cite{Huang_2012}; Copyright 2012 American Chemical Society. (c-e) Adapted with permission from \cite{Lin_2014}; Copyright 2014 American Association for the Advancement of Science. (f) Adapted with permission from \cite{Luo_2015}; Copyright 2015 Wiley-VCH.}
	\label{fig:Geometric}
\end{figure}

Whereas V-antenna and -aperture metasurfaces rely on resonant scattering of light, with the consequent result that the working bandwidth is limited due to dissimilar dispersion of the different meta-atoms, geometric metasurfaces, on the other hand, are dispersionless in the sense that the induced phase on the cross-polarized light is of pure geometric origin, cf. equation \eref{eq:ms3RT}. Having said that, it should be noted that the conversion efficiency may be strongly wavelength-dependent, since it relies on the ability of the basic unit cell to function as an efficient half-wave plate, i.e, ideally we aim at $|M_{xx}|=|M_{yy}|=1$ and $\Delta\phi=\phi_{xx}-\phi_{yy}=\pi$ [see equation \eref{eq:ms3RT}] . Within the concept of metasurfaces, the first realization at optical wavelengths consisted of an array of identical gold nanorods on an indium tin oxide(ITO) coated glass substrate that are rotated in steps of $\pi/8$ along the $x$-axis, thus inducing a linear phase variation in this direction \cite{Huang_2012}. Figure \ref{fig:Geometric}(a) shows a SEM image of the fabricated metasurface, while the colormap of figure \ref{fig:Geometric}(b) displays calculation of the transmitted cross-polarized light as a function of wavelength and observation angle when the LCP light impinges normally to the surface.
In accordance with the dispersionless nature of the phase gradient, it is seen that cross-polarized light is exclusively being anomalously redirected in the entire wavelength range, with the angle of refraction being in agreement with experimental measurements (triangular markers). The work in \cite{Huang_2012} does not state the conversion efficiency of the metasurface, but we speculate it is below 10\%, which we base on the relatively low density of nanorods (unit cell size is 400\,nm) with respect to the wavelength region, together with the fact that the nanorod unit cell only features an ED response, thus resulting in both reflection and transmission (one of which is undesirable) and an inability to reach $\Delta\phi=\pi$.

In the strive towards the ultimate goal of real-life meta-devices for efficient manipulation of light by phase-only engineering, plasmonic metasurfaces feature two significant drawbacks, one being the unavoidable loss of electromagnetic energy to Ohmic heating in the metal nanostructures, while the second hurdle relates to typical plasmonic materials, like gold and silver, not being compatible with mature semiconductor fabrication technologies \cite{West_2010}. As a way to (at least partially) circumvent these drawbacks, recent works have focused on all-dielectric metasurfaces consisting of nanostructured silicon \cite{Jahani_2016}. Here, we would like to highlight a silicon-based geometric metasurface that works in transmission at a wavelength of $\lambda\simeq 550$\,nm (figure \ref{fig:Geometric}(c)) \cite{Lin_2014}. The metasurface consists of subwavelength spaced silicon nanostripes for which light of orthogonal polarizations acquires a phase difference of $\Delta\phi\simeq\pi$ after propagation through the metasurface (see inset of figure \ref{fig:Geometric}(c)). As such, the working principle is related to form birefringence \cite{Xu_1995}, but it should be noted that the 120\,nm wide and 100\,nm high nanostripes feature a resonant response for $x$-polarization, hereby ensuring $\pi$-phase retardation over a subwavelength thickness. The blazed grating functionality is achieved by discretizing the geometric phase in eight steps, as shown in figure \ref{fig:Geometric}(c), with the corresponding measured diffraction patterns at $\lambda=550$\,nm displayed in figure \ref{fig:Geometric}(d) and \ref{fig:Geometric}(e) for normal incident RCP and LCP light, respectively. In accordance with theory, the two spin states experience opposite slopes of the constant phase gradient, thus resulting in cross-polarized light being redirected into either $\pm 1$ diffraction order. The noticeable power in the zeroth diffraction order is due to $|t_{xx}|\neq |t_{yy}|$, accordingly implying an imperfect conversion of co- to cross-polarized light. In fact, the conversion efficiency is estimated to $\sim 20$\% \cite{Qin_2016}, with the limiting factors (besides $|t_{xx}|\neq |t_{yy}|$) being reflection and the significant absorption in silicon at the working wavelength.

As the last example of geometric metasurfaces, we would like to pay attention to a metallic metasurface working in reflection and realized in the microwave regime \cite{Luo_2015}. The metasurface is, in principle, a reflectarray consisting of a metal film overlaid by a dielectric spacer and an array of cross-shaped meta-atoms (see inset of figure \ref{fig:Geometric}(f)) that are designed so that $r_{xx}\simeq-r_{yy}$ near the frequency of 12\,GHz. At the same time, the small Ohmic losses in the microwave regime allow for $|r_{ii}|$ to approach one, thereby setting the scene for close to 100\% conversion efficiency. The experimental measurements, as displayed in figure \ref{fig:Geometric}(f) as a function of frequency, readily demonstrate a record-high conversion efficiency of $>90$\,\% close to the design frequency. We note that a similar metasurface design can be realized at optical wavelengths without any major degradation in the conversion efficiency that may still reach 80\,\% \cite{Zheng_2015}.

\subsection{Meta-reflectarrays}\label{sec:RMs}
\begin{figure}[tb]
	\centering
		\includegraphics[width=8.5cm]{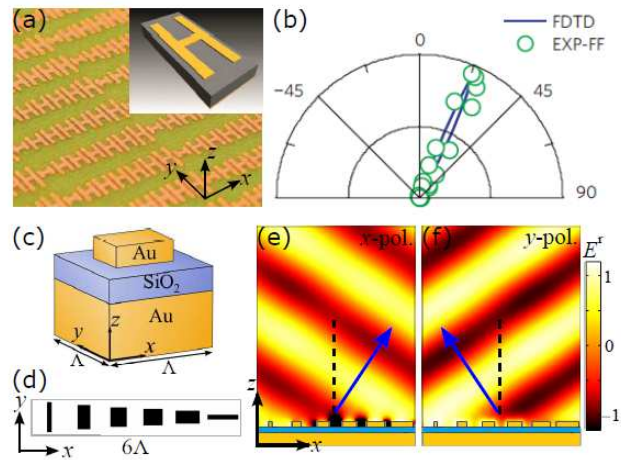}
	\caption{(a) Image of microwave meta-reflectarray that features a linear phase gradient on the reflection coefficient along the $x$-axis, with the inset showing a sketch of the basic unit cell consisting of a metallic H-shaped meta-atom atop a dielectric spacer and metal back-reflector (the unit cell size is $2.5\times 6$\,mm$^2$). (b) Angular resolved far-field intensity of reflected light when incident light is $x$-polarized and impinges normally to the metasurface featuring a super cell periodicity of $50$\,mm. Solid line and open circles represent simulation and experimental results, respectively. The working frequency is 15\,GHz. (c) Unit cell of optical meta-reflectarray consisting of an optically thick gold substrate overlaid by a nanometer-thin glass spacer and a gold nanobrick. (d) Top-view of birefringent super-cell that anomalously reflects incident light to opposite sides of the surface normal for $x$- and $y$-polarizations. The design wavelength is $\lambda=800$\,nm and the unit cell size is $\Lambda=240$\,nm. (e,f) Full-wave simulation of the reflected light for $x$- and $y$-polarized normal incident light, respectively. (a,b) Adapted with permission from \cite{Sun_2012}; Copyright 2012 Nature Publishing Group. (c-f) Adapted with permission from \cite{Pors_2013}; Copyright 2013 Nature Publishing Group.}
	\label{fig:Metareflectarray}
\end{figure}

As evident from the above discussion, meta-reflectarrays pay an important role in realizing highly-efficient geometric metasurfaces for manipulation of CP light. Moreover, meta-reflectarrays are also among the most efficient metasurfaces for control of linearly polarized light and particularly attractive due to the simple design that only requires one step of lithography. As a first experimental verification of the blazed grating functionality, figure \ref{fig:Metareflectarray}(a) displays a top-view image of a metasurface designed to anomalously reflect $x$-polarized incident light at the frequency of $15$\,GHz, with the inset illustrating the composition of the unit cell.
By meticulously designing the H-shaped metallic meta-atoms, the incident light experiences a constant phase gradient along the $x$-axis, thereby leading to diffraction into $+1$ order exclusively, as verified experimentally by angle-resolved far-field measurements from a metasurface featuring a super cell period of 50\,mm (figure \ref{fig:Metareflectarray}(b)). It is worth noting that this type of meta-reflectarrays manipulates the co-polarized light with (ideally) zero conversion to the cross-polarized component, while the limiting factors of the manipulation efficiency are Ohmic losses and (to a lesser extent) a non-ideal phase gradient. For this reason, the metasurface in figure \ref{fig:Metareflectarray}(a) shows close to 100\% efficiency, while the highest efficiency for similar configurations at optical and near-infrared wavelengths amounts to $\sim 80$\,\% \cite{Sun_2012_2}, though replacement of the metallic meta-atoms with silicon counterparts can push the efficiency towards 100\% again \cite{Yang_2014}.

Besides the ease of fabrication and high efficiency in manipulating the reflection phase, meta-reflectarrays may show even greater control of the reflected light by engineering of the reflection amplitude and phase for one polarization or the reflection phases for two orthogonal polarizations simultaneously \cite{Pors_2013,Pors_2013_2,Farahani_2013}. The latter property may lead to the design of birefringent metasurfaces with different optical properties for orthogonal polarizations. As an example, figure \ref{fig:Metareflectarray}(c) shows a sketch of a gold-glass-gold unit cell for a birefringent metasurface working at the wavelength of 800\,nm, where the two widths of the top nanopatch can be used to independently control the reflection phase of $x$- and $y$-polarized light. Figure \ref{fig:Metareflectarray}(d) displays a top-view image of a metasurface super cell that incorporates a linearly varying phase (in steps of 60$^\circ$) along the $x$-axis but the slope is of opposite sign for the two orthogonal polarizations, thus leading to a splitting of $x$- and $y$-polarized light into $\pm 1$ diffraction order, respectively. The polarization splitter functionality is visualized in Figs. \ref{fig:Metareflectarray}(e) and \ref{fig:Metareflectarray}(f), where it is readily seen that reflected light propagates in opposite directions relative to the surface normal for orthogonal polarizations. The theoretical efficiency of the proposed polarization splitter is $\sim 80$\,\%, though proof-of-concept experiments show a reduced efficiency of $\sim 50$\,\% \cite{Pors_2013}.

\subsection{Huygens' metasurfaces}
\begin{figure}[tb]
	\centering
		\includegraphics[width=15.5cm]{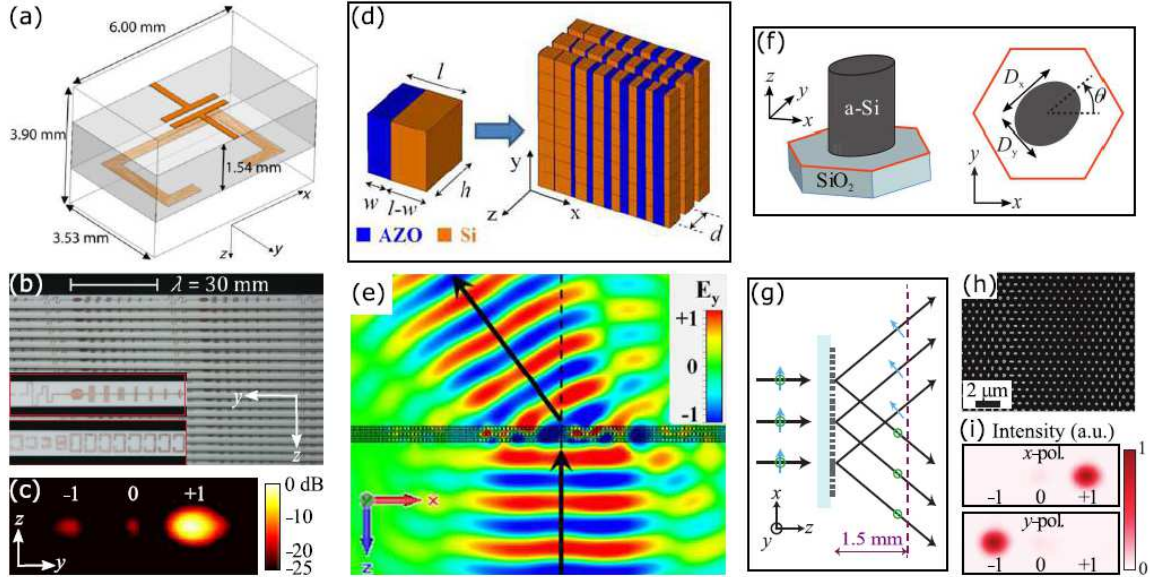}
	\caption{Huygens' metasurfaces realized with (a-c) metallic meta-atoms for the microwave regime, (d,e) three layers of semiconductor-dielectric fillings for the mid-infrared regime, and (f-i) elliptically shaped silicon nanoposts for the optical frequency range. (a) Sketch of unit cell consisting of a dielectric substrate with patterned copper traces on both the top and bottom side. The incident light is $y$-polarized and propagates along the $x$-direction. (b) Photograph of part of the fabricated Huygens' surface featuring a linear phase-gradient along the $y$-direction, with insets showing the super cell on the two sides of the substrate. (c) Measured far-field intensity on a logarithmic scale at the working frequency of 10\,GHz. (d) Sketch of composite metasurface consisting of three layers separated by the distance $d$ in the propagation direction $z$ for $y$-polarized incident light. The unit cell consists of blocks of AZO and silicon. (e) Simulation of refraction of normal incident wave by metasurface featuring a linear phase gradient along the $x$-axis. The geometrical parameters are $l=h=250$\,nm, $d=375$\,nm, and the working wavelength is $\lambda=3$\,$\mu$m. (f) Sketch of hexagonal unit cell, with the amorphous silicon nanoposts being characterized by the parameters $(D_x,D_y, \theta)$. (g) Sketch of working principle of birefringent metasurface where $x$- and $y$-polarized incident light experience phase gradients of opposite sign. (h) SEM image of fabricated metasurface featuring 715\,nm tall nanoposts with diameters varying between $65$ to 455\,nm and lattice constant of $650$\,nm. The working wavelength is $\lambda=915$\,nm. (i) Measured far-field intensity for $x$- and $y$-polarized light. (a-c) Adapted with permission from \cite{Pfeiffer_2013}; Copyright 2013 American Physical Society. (d,e) Adapted with permission from \cite{Monticone_2013}; Copyright 2013 American Physical Society. (f-i) Adapted with permission from \cite{Arbabi_2015}; Copyright 2015 Macmillan Publishers Limited.}
	\label{fig:Huygens}
\end{figure}

Metasurfaces that efficiently manipulate the phase of the transmitted light can be realized in multiple ways, with the preferred implementation being dependent on the frequency regime of interest. Here, we discuss realizations in the microwave, infrared, and optical regimes that are based on metallic, semiconductor, and dielectric meta-atoms, respectively. Assuming a $y$-polarized wave that propagates in the $x$-direction, figure \ref{fig:Huygens}(a) depicts the unit cell for a Huygen's metasurface at microwave frequencies, which consists of a metallic antenna at the top giving rise to an ED response and a bottom SRR that ensures a MD response.
By properly adjusting both layers simultaneously, it is possible to find several unit cells with the zero-reflection condition but different phases for the transmitted light, thereby realizing the necessary ingredients for a phase-gradient metasurface. Figure \ref{fig:Huygens}(b) shows a photograph of a fabricated metasurface that incorporates a constant phase gradient along the $y$-axis, thus resulting in incident light being dominantly refracted into $+1$ diffraction order (figure \ref{fig:Huygens}(c)). The fabricated metasurface features a manipulation efficiency of $86$\,\% at the frequency of 10\,GHz. However, scaling of the design principle to near-infrared and optical wavelengths implies considerable degradation in the efficiency due to the detrimental influence of Ohmic heating in the metallic meta-atoms \cite{Pfeiffer_2014_2}.

As an alternative approach in realizing Huygens' metasurfaces, particularly relevant for the infrared regime, figure \ref{fig:Huygens}(d) sketches a three-layer phase-gradient metasurface for which light manipulation, in continuation of the concept of optical nanoelements \cite{Engheta_2005}, is realized by unit cells with varying portions of positive and negative permittivity material, here realized using silicon and aluminum-doped zinc oxide (AZO). We note that AZO is an attractive material at infrared wavelengths due to its plasmonic response and tunability of the optical properties. The sketched composite metasurface is designed for $y$-polarized incident light at a wavelength of 3\,$\mu$m and implements a constant phase-gradient along the $x$-axis, where the two identical outer layers, separated by the distance $2d=750$\,nm, feature out-of-phase (polarization) currents, hereby creating an effective magnetic response, with the inner layer producing the electric response \cite{Pfeiffer_2013_2}. The functionality of the composite metasurface is verified numerically (figure \ref{fig:Huygens}(e)), where more than 75\% of the incident power is being anomalously refracted.

As the third, yet very different, realization of Huygens' metasurfaces, we discuss an all-dielectric approach that is appealing at optical and near-infrared wavelengths. The metasurface consists of an hexagonal array of elliptically-shaped amorphous silicon nanoposts atop a glass substrate, where the diameters and orientation of the nanoposts (see figure \ref{fig:Huygens}(f)) can be used to independently control the transmission phase of two orthogonal polarizations simultaneously, thus creating a birefringent response at will. It should be noted that the height of the posts is on the scale of the free-space wavelength and, as such, may be considered as Fabry-P{\'e}rot waveguide resonators for which orthogonally polarized modes experience different effective indexes. To be in line with section \ref{sec:BuildingBlocks}, however, we note that each nanopost may also be viewed as a high-permittivity resonator featuring precisely tailored ED and MD responses, so that light is scattered in the forward direction \cite{Arbabi_2015}. Moreover, we emphasize that the coupling between the different nanoposts is weak, hereby underlining that the posts can also be arranged into a more conventional square lattice arrangement \cite{Arbabi_2015_2}. As a way of example, the all-dielectric metasurface is realized at the wavelength of 915\,nm using 715\,nm tall nanoposts in a hexagonal array of lattice constant 650\,nm. Note that the lattice constant is chosen small enough to avoid diffraction and, thus, function as a metasurface for close-to-normal incident light. In connection with the discussed birefringent meta-reflectarrays, we here present the same functionality of splitting $x$- and $y$-polarized light by phase-gradient metasurfaces imprinting a linear phase with opposite slopes for orthogonal polarizations (see working principle in figure \ref{fig:Huygens}(g)). A top-view of part of the fabricated metasurface is shown in figure \ref{fig:Huygens}(h), and the associated far-field diffraction patterns, as displayed in figure \ref{fig:Huygens}(i), verify the splitting of $x$- and $y$-polarized light into $\pm 1$ diffraction order, respectively. We stress that despite slight fabrication imperfections, the experimentally measured manipulation efficiencies are above 70\% for both polarizations. Finally, we note that the same level of manipulation efficiency has also been reached in a, one might say, complementary metasurface design consisting of a square array of nanodisks of the same size, hereby creating a polarization-independent response, where the control of the transmission phase is achieved through variation of the lattice period \cite{Chong_2015,Chong_2016}.

\subsection{Metamirrors}
\begin{figure}[tb]
	\centering
		\includegraphics[width=7.5cm]{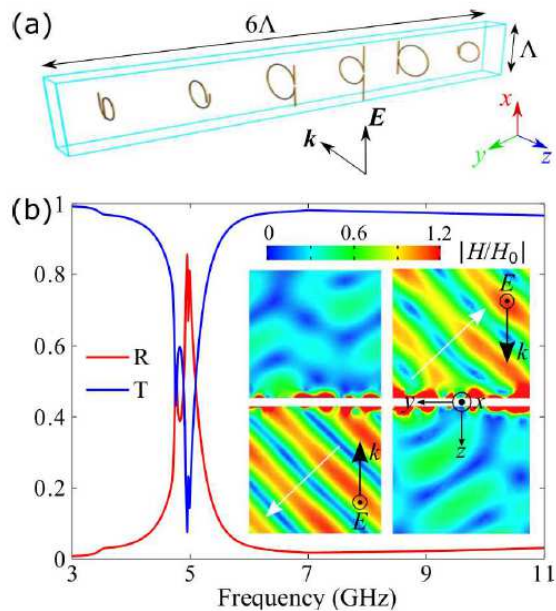}
	\caption{(a) Sketch of metamirror super cell that incorporates a linear phase gradient along the $x$-axis at 5\,GHz. The unit cell size is $\Lambda=14.4$\,mm. (b) Calculated reflectance and transmittance as a function of frequency for the metamirror in (a). Insets show the magnitude of reflected and transmitted magnetic field for incident light propagation along the $\pm z$-axis at 5\,GHz. (a,b) Adapted with permission from \cite{Asadchy_2015}; Copyright 2013 American Physical Society.}
	\label{fig:Metamirror}
\end{figure}

With metamirrors, we come to the class of metasurfaces featuring resonant electric, magnetic, and magnetoelectric responses that are tailored in such a way that transmission is suppressed, while the reflection phase is meticulously controlled. As a way of realization in the microwave regime, figure \ref{fig:Metamirror}(a) displays a drawing of a super cell consisting of six metallic $\Omega$-type meta-atoms that imprint a constant phase gradient on the reflection coefficient along the $y$-axis for $-z$-propagating, $x$-polarized incident light.
The electric and magnetic responses of each unit cell can be engineered by the straight and loop section of the meta-atoms, respectively. The functionality of the proposed metamirror has been verified both numerically and experimentally \cite{Asadchy_2015}, with figure \ref{fig:Metamirror}(b) showing the calculated reflectance and transmittance as a function of frequency. In line with the property of metamirrors, the metasurface is close to being transparent for most frequencies. Only close the the design frequency at 5\,GHz do we observe considerably light-matter interaction, resulting in 6\,\% transmitted light and 86\,\% being anomalously reflected (see inset of figure \ref{fig:Metamirror}(b)). It is worth noting that the presented metasurface shows almost the same behaviour for $\pm z$-propagation of light (inset of figure \ref{fig:Metamirror}(b)), which, c.f. equation \eref{eq:ms6R}, indicates a relatively weak magnetoelectric response relative to the ED and MD contributions. As a final comment, we note that metamirrors can also be realized using all-dielectric meta-atoms \cite{Odit_2016} and, hence, scalable to the near-infrared and optical regimes \cite{Asadchy_2016}.

\subsection{A note on the size of metasurface unit cells}
From the above discussion of realizable metasurfaces, it is evident that the view of metasurfaces as interface discontinuities with effective surface responses is a somewhat idealization, since many realizations, particularly dielectric Huygens' metasurfaces, feature nonnegligible thicknesses and unit cell sizes compared to the operation wavelength. The effect of retardation due to the finite thickness of the unit cell is typically managed automatically, as most metasurface designs are based on full-wave numerical simulations. The influence of the finite width of the unit cell, on the other hand, is the origin of spatial dispersion \cite{Silveirinha_2007,Menzel_2008}, which might degrade the performance of the metasurface for off-normal incident light. Moreover, in order to avoid a square-array metasurface to function as a diffraction grating, the angle of incidence $\theta_i$ (measured from the surface normal) must be smaller than $\theta_{\textrm{max}}=\sin^{-1}\left(\lambda/\Lambda-1\right)$, where $\Lambda$ is the unit cell width and $\lambda$ is the wavelength. In other words, for operation of the metasurface with close-to-normal incident light, it should satisfy $\Lambda<\lambda$, while for grazing incident light the criterion becomes $\Lambda<\lambda/2$.

\section{Applications of metasurfaces}
Metasurfaces have been utilized in a wide range of applications, ranging from flat and compact versions of conventional optical elements, via metadevices tackling specific applications, to control of enhanced nonlinear processes. Here, we try to give an overview of the many, yet very different, applications.

\subsection{Flat optical elements}

\subsubsection{Wave plates}
\begin{figure*}[tb]
	\centering
		\includegraphics[width=15.5cm]{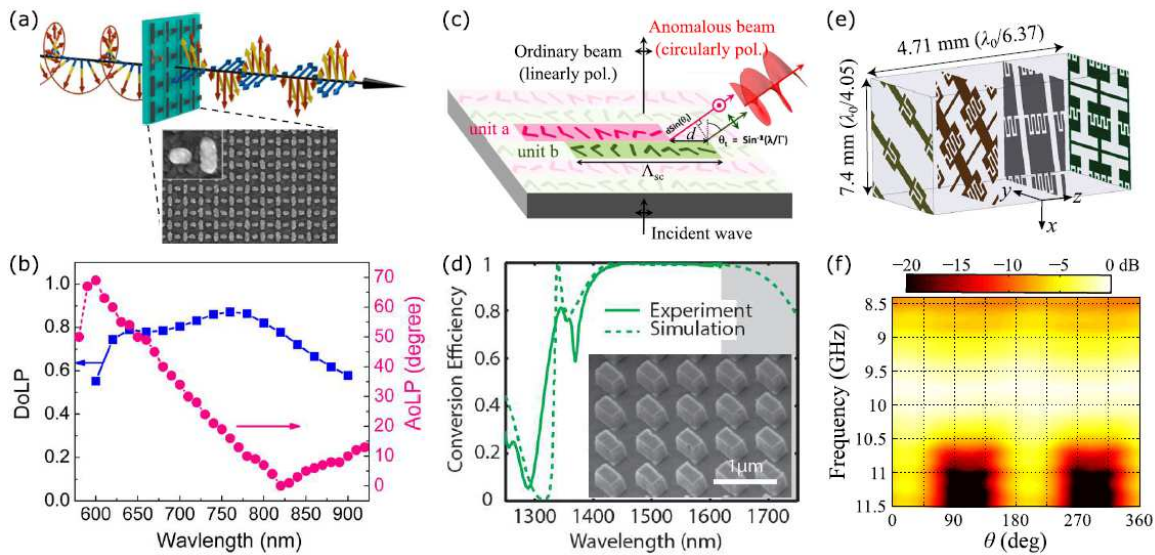}
	\caption{(a) Artistic view of a plasmonic nanorod-based metasurface functioning as a QWP, together with a SEM image of a similar structure with array periodicity of 240\,nm and $190$\,nm along the $x$- and $y$-direction, respectively. (b) Measurements of the DoLP and AoLP as a function of wavelength for CP incident light on the metasurface in (a). (c) Illustration of a metallic V-antenna metasurface that redirects part of the incident light and converts the polarization from linear to circular. (d) Dielectric meta-reflectarray (see inset) that functions as a half-wave plate at near-infrared wavelengths. The graph shows the measured (solid) and simulated (dashed) polarization conversion efficiency, $|r_\mathrm{cr}|^2/(|r_\mathrm{co}|^2+|r_\mathrm{cr}|^2)$, where $r_\mathrm{co}$ and $r_\mathrm{cr}$ are the reflection coefficients of the co- and cross-polarized light, respectively. (e) Sketch of a unit cell of a bianisotropic Huygens' metasurface that is designed to function as a polarization rotator at microwave frequencies. (f) Measured amplitude of cross-polarized transmission coefficient as a function of frequency and angle of incident linear polarization $\theta$ measured from the $x$-axis. (a,b) Adapted with permission from \cite{Zhao_2013}; Copyright 2013 American Chemical Society. (c) Adapted with permission from \cite{Yu_2012}; Copyright 2012 American Chemical Society. (d) Adapted with permission from \cite{Yang_2014}; Copyright 2014 American Chemical Society. (e,f) Adapted with permission from \cite{Pfeiffer_2014}; Copyright 2014 American Physical Society.}
	\label{fig:PolControl}
\end{figure*}

Homogeneous metasurfaces, defined by an array of identical unit cells, represent a subclass of metasurfaces that cannot change the wavefront of the incident light, but may be able to control the polarization by a properly engineered anisotropic response. For example, metasurfaces featuring the reflection or transmission matrix $\overline{\overline{M}}=\mathrm{diag}(M_{xx},M_{yy})$ function as quarter- and half-wave plates when $|M_{xx}|=|M_{yy}|$ and the phase difference $\Delta\phi=\phi_{xx}-\phi_{yy}$ is $\pi/2$ and $\pi$, respectively, and the incident light is circularly or diagonally [i.e., $\mathbf{E}_i=1/\sqrt{2}(1,\pm i)^T$ or $\mathbf{E}_i=1/\sqrt{2}(1,\pm1)^T$] polarized. The quarter-wave plate (QWP) functionality is particularly simple to realize since $\Delta\phi=\pi/2$ can be achieved with ED-only metasurfaces, such as arrays of metallic nanorods \cite{Pors_2011_2,Zhao_2013,Li_2015} or the complementary configuration of slits in a metal film \cite{Ogut_2010,Zhao_2011,Khoo_2011,Baida_2011,Roberts_2012}. The QWP response is achieved by designing nanorods or slits that have oppositely detuned resonances with respect to the operation wavelength for orthogonal polarizations, hereby creating a phase difference between light scattered by the two polarizations. One example is depicted in figure \ref{fig:PolControl}(a), where each unit cell consists of two orthogonally-oriented silver nanorods that are detuned by $\Delta\phi=\pi/2$, but otherwise show similar dispersion, thereby leading to a broadband QWP operation.

The particular configuration in figure \ref{fig:PolControl}(a) is designed for visible light, and the QWP functionality is probed by measuring the degree and angle of the linear polarization (DoLP and AoLP) in transmission for CP incident light (figure \ref{fig:PolControl}(b)). The DoLP, though not perfect, demonstrates the weak wavelength dependence between $\lambda\sim 620-830$\,nm, while the AoLP signifies that $|t_{xx}|\neq |t_{yy}|$ for most wavelengths. In relation to conventional QWPs, we note that the wavelength-dependent AoLP corresponds to varying fast and slow axes, with $\lambda\simeq 670$\,nm representing the situation of optical axes coinciding with the nanorod axes (i.e., AoLP $=45^\circ$).
As a different realization of QWP, figure \ref{fig:PolControl}(c) displays a drawing of a V-antenna metasurface consisting of two super cell units that are offset along the $x$-direction by $d=\Lambda_{sc}/4$, where $\Lambda_{sc}$ is the super cell periodicity. Each super cell features a constant phase gradient, thus leading to cross-polarized light being redirected into the first diffraction order. The difference in rotation by 45$^\circ$ for mutual V-antennas and the offset $d$, however, ensure that cross-polarized light from the two super cells are orthogonally polarized and with a phase difference of $\Delta\phi=\pi/2$. Importantly, since the phase difference $\Delta\phi$ is achieved by geometrical means, this type of QWP is intrinsically broadband. The metasurface has been realized in the infrared regime, demonstrating good QWP functionality in the wavelength range $5-12$\,$\mu$m with a maximum conversion efficiency of $\sim 10$\% near $\lambda=8$\,$\mu$m \cite{Yu_2012}. We emphasize that the two QWP metasurfaces of figure \ref{fig:PolControl}(a) and \ref{fig:PolControl}(c) are (in the lossless and ultrathin approximation) limited to manipulate 50\% and 25\% of the incident light, respectively, due to symmetric radiation into the two half spaces. Only by utilizing more complex metasurfaces, like meta-reflectarrays and Huygens' metasurfaces, can the polarization conversion approach 100\% \cite{Markovich_2013}. For example, plasmonic meta-reflectarrays have shown QWP functionality with theoretical efficiency above 80\% \cite{Pors_2013_3} and $\sim 90$\% \cite{Chen_2014} around $\lambda=800$\,nm and $\lambda=1550$\,nm, respectively, and an efficiency that will steady increase for designs at lower frequencies. Also, we emphasize that meta-reflectarrays may show broadband functionality, which can be achieved by proper compensation of the dispersion of the meta-atoms by the phase accumulation in the dielectric spacer \cite{Jiang_2014,Guo_2015}.

The simplicity, high efficiency, broadband response, and full phase control with meta-reflectarrays have also rendered these metasurfaces attractive for half-wave plate (HWP) functionality. The ability to produce orthogonally polarized light upon reflection was already shown in 2007 at microwave frequencies \cite{Hao_2007} and later extended to the visible \cite{Hao_2009,Dai_2014}, near-infrared \cite{Pors_2013_4,Ding_2015_2}, and mid-infrared regimes \cite{Levesque_2014,Ribaudo_2014}, though accompanied by slightly reduced efficiencies due to Ohmic losses. Here, we would like to highlight a dielectric-based meta-reflectarray in which the meta-atoms are made of silicon [see inset of figure \ref{fig:PolControl}(d)], thereby considerably reducing the level of absorption within the metasurface for near-infrared applications. The metasurface in figure \ref{fig:PolControl}(d), consisting of a silver back-reflector, 200\,nm PMMA spacer, and an array of silicon nanopatches, shows a practically perfect conversion of the incident light to the cross-polarized component within the wavelength range $1420-1620$\,nm, while the associated reflectance is measured to be above $97$\% within this bandwidth \cite{Yang_2014}. We note that similar efficient polarization control can also be reached in transmission using all-dielectric metasurfaces, similar to the one sketched in figure \ref{fig:Huygens}(f).
As a last example of polarization control with metasurfaces, we would like to discuss the unique properties offered by bianisotropic surfaces. Particularly, we focus on the transmission matrix
\begin{equation}
\label{eq:polRotator}
\overline{\overline{t}}_+=\left(
\begin{array}{cc}
0 & -1 \\
1 & 0	
\end{array}
\right),
\end{equation}
which represents a polarization rotator that rotates any incident linear polarization by $90^\circ$---a conventional HWP only achieves the same functionality for diagonally polarized light. Figure \ref{fig:PolControl}(e) depicts a metallic Huygens' metasurface that works as a polarization rotator at microwave frequencies. The functionality can actually be achieved by three sheets only, but the present configuration adds an additional layer to increase the level of transmission and bandwidth \cite{Pfeiffer_2014}. The polarization rotator has been realized for operation at 10\,GHz, and figure \ref{fig:PolControl}(f) displays the measured transmission amplitude of the cross-polarized light as a function of frequency and polarization angle of the incident (linearly polarized) light. In line with equation \eref{eq:polRotator}, the transmission is close to unity (0\,dB) and independent of the angle of the linear polarized light at the design frequency, while the metasurface demonstrates a 3\,dB-bandwidth of $\sim 9$\,\%.

\subsubsection{Lenses and reflectors}
\begin{figure*}[tbh]
	\centering
		\includegraphics[width=16.0cm]{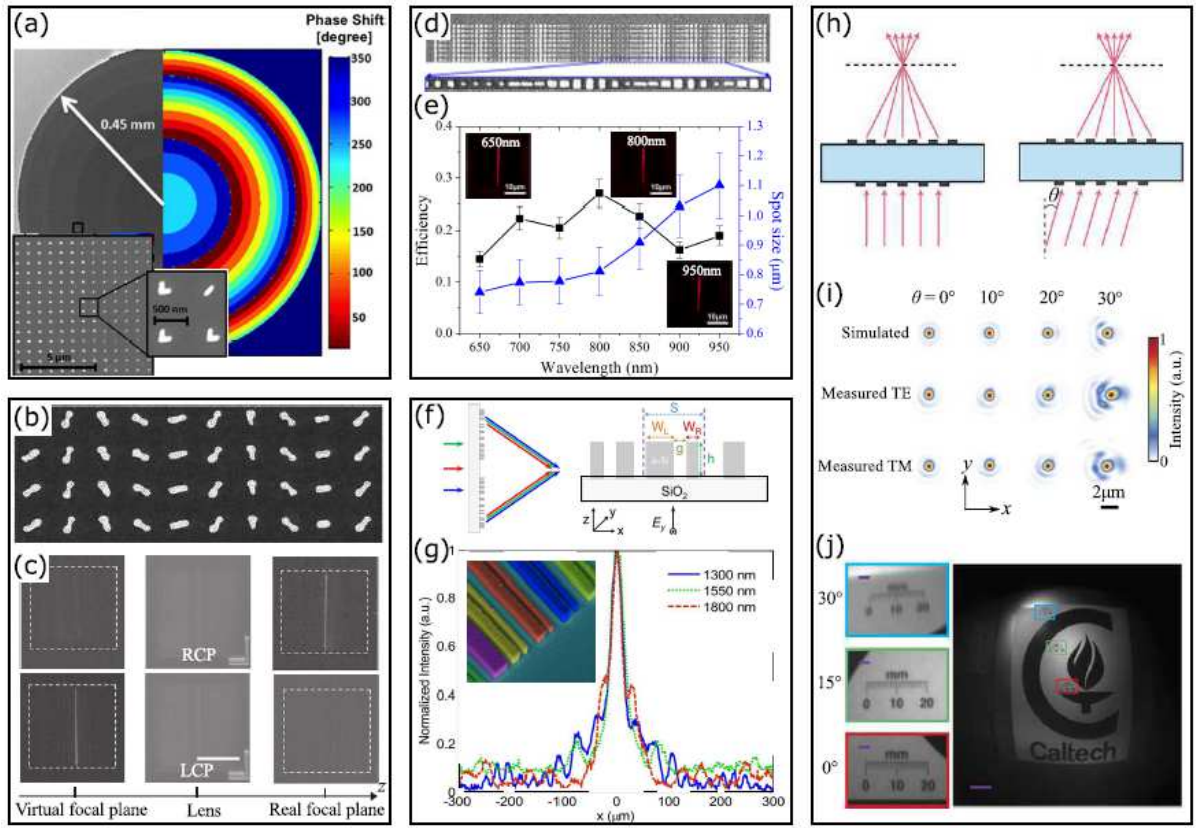}
	\caption{{\scriptsize (a) SEM image of a plasmonic V-antenna metasurface that functions as a lens with focal length of 3\,cm at $\lambda=1.55\mu$m (left), together with the discretized phase profile (right) (b) Representative SEM image of a geometric metasurface consisting of gold nanorods that behaves as a dual-polarity cylindrical lens for CP light. (c) Optical microscope images of the virtual focal plane, the metasurface, and real focal plane for the metasurface in (b) for incident RCP (top row) and LCP (bottom row) light at $\lambda=740$\,nm. The scale bar is 50\,$\mu$m. (d) SEM image of plasmonic meta-reflectarray that is designed to function as a parabolic mirror in one dimension at $\lambda=800$\,nm. (e) Measured efficiency and spot size ($e^{-1}$ full width) for the metasurface mirror in (d). Inset shows the optical images of the focal plane for three different wavelengths. (f) Sketch of the working principle of a achromatic metasurface lens (left) and a 1D realization that uses aperiodic super cells consisting of coupled rectangular silicon meta-atoms on a glass substrate (right). (g) Experimentally measured intensity at the focal length of $\simeq 7.5$\,mm for the three wavelengths $1300$, $1550$, and $1800$\,nm. Inset shows a color-coded SEM image of the fabricated achromatic metasurface lens. (h) Schematic illustration of focusing of on-axis and off-axis light by a metasurface doublet lens corrected for monochromatic aberrations. (i) Simulated and measured focal plane intensity profiles of the metasurface doublet lens for different incident angles. (j) Imaging with the metasurface doublet lens. (a) Adapted with permission from \cite{Aieta_2012}; Copyright 2012 American Chemical Society. (b,c) Adapted with permission from \cite{Chen_2012}; Copyright 2012 Nature Publishing Group. (d,e) Adapted with permission from \cite{Pors_2013_5}; Copyright 2013 American Chemical Society. (f,g) Adapted with permission from \cite{Khorasaninejad_2015}; Copyright 2015 American Chemical Society. (h-j) Adapted with permission from \cite{Arbabi2016doublet}; Copyright 2016 Nature Publishing Group.}}
	\label{fig:Lenses}
\end{figure*}

In conventional optical systems, light is focused by either dielectric lenses or parabolic reflectors that, on the scale of the wavelength, are bulky and curved components, thus preventing straightforward miniaturization. Metasurfaces, on the other hand, represent flat and subwavelength-thin devices that can achieve the same functionality by incorporating a parabolic transmission/reflection phase, i.e.,
\begin{equation}
\phi(x,y)=\frac{2\pi}{\lambda}\left(\sqrt{x^2+y^2+f^2}-f\right),
\label{eq:parPhase}
\end{equation}
where we have assumed the metasurface to lie in the $xy$-plane, and $f$ is the focal length. Since focusing is of paramount importance in practically any setup, it is also a functionality that has been realized with many different types of planar configurations, comprising early work on nanohole arrays in metal films \cite{Huang_2008,Gao_2010} and arrays of plasmonic slit waveguides \cite{Verslegers_2009,Ishii_2011}. However, recent progress is related to wavefront engineering with metasurfaces, which includes focusing by V-antenna/aperture configurations \cite{Aieta_2012,Ni_2013,Jiang_2013}, geometric metasurfaces \cite{Lin_2014,Chen_2012}, meta-reflectarrays \cite{Li_2012,Pors_2013_5}, metallic \cite{Pfeiffer_2013_3,Wong_2014}, semiconductor \cite{Monticone_2013} and dielectric \cite{Arbabi_2015_2,West_2014,Arbabi_2015_3} Huygens' metasurfaces, and metamirrors \cite{Asadchy_2015,Asadchy_2016}. Here, we review some of these works, while also addressing the important topic of achromatic lenses and metasurface doublet lenses corrected for monochromatic aberrations. Starting with V-antenna metasurfaces, figure \ref{fig:Lenses}(a) conveniently illustrates the principle of meta-lenses, where the dimensions of the meta-atoms are varied in the radial direction so to achieve a parabolic phase profile. As a consequence of the difficulty in realizing meta-atoms with fully independent scattering amplitudes and phases, the continuously varying phase profile is conventionally approximated by a step-like function, where the phase takes on a finite number of values, thus making the metasurfaces consist of a series of concentric rings in which the metasurface is locally homogeneous.

Despite the step-wise approximation to the true phase profile of Eq. \eref{eq:parPhase}, we note that the profile is not derived from paraxial approximations and, as such, metasurface lenses do not for normal incident light suffer from monochromatic aberrations. For off-normal incident light, aberrations do exist, but it can be shown that spherical aberrations and image distortion remain suppressed \cite{Aieta_2013}. Meta-lenses based on geometric metasurfaces have the peculiar property that the sign of the geometric phase changes with the handedness of the incident light. This property is shown by the plasmonic metasurface in figure \ref{fig:Lenses}(b), which induces a parabolic phase along the horizontal direction, thus functioning as a cylindrical lens (i.e., light is being focused to a line). The experimental proof of the dual-polarity property is conducted at $\lambda=740$\,nm, and Figure \ref{fig:Lenses}(c) shows optical images for RCP and LCP incident light at the virtual and real focal planes, hereby confirming that the metasurface behaves as a convex and concave lens, respectively. It should be noted that the V-antenna and geometric metasurfaces in figure \ref{fig:Lenses}(a) and \ref{fig:Lenses}(b) demonstrate the principle of meta-lenses, but the associated focusing efficiencies are only $\sim 1$\% and $\sim 5$\%, respectively. Huygens's metasurfaces, however, may feature considerably larger efficiencies, with, for example, an all-dielectric realization demonstrating efficiencies of $\sim 80$\% at near-infrared frequencies \cite{Arbabi_2015_2,Arbabi_2015_3}.

As complementary to conventional focusing of transmitted light by a dielectric lens, reflected light may be focused by a parabolic mirror/reflector. The flat version of a parabolic reflector can be conveniently realized by meta-reflectarrays, one of which is exemplified in figure \ref{fig:Lenses}(d) for focusing around $\lambda=800$\,nm. The metasurface, consisting of an array of gold nanopatches atop a thin dielectric spacer and gold film, imposes a parabolic phase on the incident light along the horizontal direction, thus resulting in 1D focusing of the reflected light. The good focusing ability has been experimentally confirmed for several wavelengths, with the inset of figure \ref{fig:Lenses}(e) displaying optical images at the focal plane for $\lambda=$ 650, 800, and 950\,nm, while a more quantitative evaluation of the focusing reveals an efficiency of up to $\sim 30$\% and a full-width of the focusing spot on the order of the wavelength (see Figure \ref{fig:Lenses}(e)). We emphasize that the broadband behavior originates from the low quality factor associated with plasmonic resonances.

The fact that the far-field response of phase-gradient metasurfaces follows diffraction theory \cite{Larouche_2012} has the important consequence that metasurfaces inherit the strong spectral dispersion of conventional diffractive optical elements. The dispersion may be an advantage in applications related to separation of wavelengths, but for imaging purposes, where the diffractive nature of meta-lenses implies a change in focal length proportional to $1/\lambda$, the dispersion results in strong chromatic aberrations. Recently, several efforts have been dedicated to approach this problem by either dispersive phase compensation using metasurfaces with unit cells consisting of multiple meta-atoms \cite{Aieta_2015,Khorasaninejad_2015,Arbabi_2016}, birefringent unit cells with equal focal length for orthogonally polarized light at two different wavelengths \cite{Eisenbach_2015}, or metasurface designs based on filter circuit theory \cite{Cheng_2015} or the principle of holography \cite{Zhao_2015}. Of these different approaches, the utilization of complex unit cells, thus featuring more geometrical parameters to control the phase response at multiple wavelengths simultaneously, seem most attractive with respect to versatility and quality of the focal spot. The left inset of figure \ref{fig:Lenses}(f) illustrates the working principle of an achromatic meta-lens (in which incident light of different wavelengths is focused to the same point), while the right inset represents a realization of an all-dielectric cylindrical meta-lens that focuses three different near-infrared wavelengths to the same focal point \cite{Khorasaninejad_2015}. Here, the two widths of the rectangular resonators ($W_L$ and $W_R$) and the separation $g$ are the free parameters that allows one to engineer the transmission phase at the wavelengths $1300$, $1550$, and $1800$\,nm simultaneously, keeping the period and resonator height fixed at $S=1000$\,nm and $h=400$\,nm, respectively. Based on numerical simulations of a multitude of ($W_L$,$W_R$,$g$) combinations, a 600\,$\mu$m wide meta-lens with focal length of $f=7.5$\,mm has been realized (see inset of figure \ref{fig:Lenses}(g)), and figure \ref{fig:Lenses}(g) displays the measured intensity at the focal distance at the three wavelengths. Taking into account that the numerical aperture (NA) is 0.04 (i.e., weak focusing), the lens actually produces an almost diffraction-limited spot, with focusing efficiencies of $\sim 15$, 10, and 21\% at the three wavelengths. Overall, the meta-lens demonstrates the possibility to focus spectrally separated wavelengths to the same spot. However, it should be emphasized that wavelengths of light deviating from the designed ones feature varying focal lengths, meaning that no, so far, realized meta-lenses can image spectrally-broad sources without noticeable chromatic aberrations.

Aside from the comparable or even superior performance than conventional lenses in terms of point-to-point interconnection scheme at normal incidence, the meta-lenses mentioned above are suffering from severe monochromatic aberrations for off-normal incident light, in particular for lenses with high-NA, resulting in small field of view and poor imaging capabilities \cite{Lalanne2016arXiv}. To correct the monochromatic aberrations and increase the field of view of meta-lenses, Faraon's group proposed a metasurface doublet lens for the operation wavelength of 850\,nm by cascading two metasurface plates that are patterned on two sides of a single transparent substrate \cite{Arbabi2016doublet}. Here,the first metasurface functions as a corrector plate and the second operates the main portion of focusing, thereby significantly reducing the monochromatic aberrations (figure \ref{fig:Lenses}(h)). The metasurface doublet lens has been realized by using all-dielectric metasurface platform (i.e., amorphous silicon nanoposts with different diameters that are covered by the SU-8 polymer), and figure \ref{fig:Lenses}(i) shows the simulated and measured focal plane intensity profiles of the metasurface doublet lens for different incident angles. The doublet lens has a nearly diffraction limited focal spot for incident angles up to more than $25^\circ$  while the singlet exhibits significant aberrations even at incident angles of a few degrees. Moreover, the metasurface doublet lens shows superior imaging performance over a wide field of view (figure \ref{fig:Lenses}(j)).

\subsubsection{Spiral phase plates}
Vortex beams are a set of beam-like solutions to the paraxial wave equation that are characterized by a doughnut-shaped intensity profile and a spiral wavefront with the phase dependence $\textrm{exp}(il\theta)$, where $\theta$ is the azimuthal angle in the plane perpendicular to the propagation direction and $l$ is an integer, known as the topological charge, that describes the number of twists within one wavelength. We note that vortex beams have found usage in many different applications, which is a consequence of its unique vectorial nature and the fact that the spiral wavefront is a manifestation of the photons carrying orbital angular momentum. For example, the possibility to tightly focus vortex beams and generate unique focal distributions can be advantageously exploited in high-resolution imaging \cite{Dorn_2003,Hell_2003} or in optical particle trapping/manipulation \cite{Padgett_2011}, while the orbital angular momentum property represents a new optical degree of freedom that can increase the bandwidth in optical communication systems, including within quantum information processing \cite{Terriza_2007}. Irrespective of the application, however, one needs to generate the vortex beam, which is typically done directly by the light source (i.e., laser) or by an optical component that converts a Gaussian input beam to a vortex beam \cite{Zhan_2009}. The latter approach is readily realizable with metasurfaces, as exemplified by the design of hologram-based fork-gratings \cite{Chen_2016} and spatially-varying linear polarizers \cite{Zhao_2013_2}. The most sought approach, however, is concerned with the implementation of the azimuthally-varying phase
\begin{equation}
\phi(x,y)=l\theta(x,y)=l\tan^{-1}(y/x),
\label{eq:sps}
\end{equation}
which represents the flat version of a spiral phase plate with topological charge $l$. This type of phase profile has been demonstrated from the visible to the terahertz regime by using V-antenna/aperture metasurfaces \cite{Yu_2011,Genevet_2012,He_2013}, geometric metasurfaces \cite{Biener_2002,Huang_2012,Li_2013,Ma_2015,Pu_2015}, meta-reflectarrays \cite{Yang_2014}, and dielectric Huygens' metasurfaces \cite{Arbabi_2015,Shalaev_2015,Chong_2015,Zhijie_2016}.
\begin{figure*}[tbh]
	\centering
		\includegraphics[width=7.5cm]{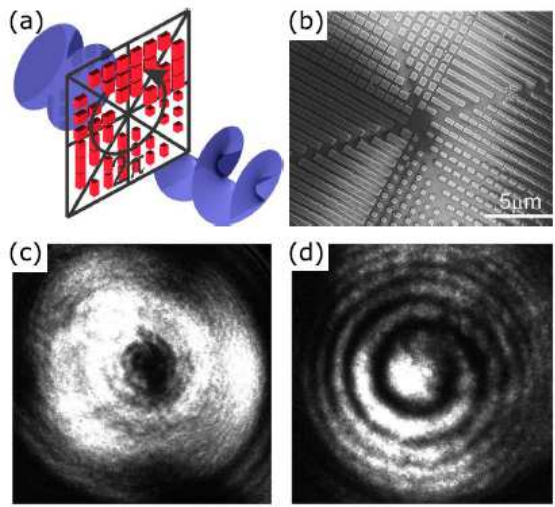}
	\caption{(a) Illustration of a metasurface-based spiral phase plate that converts an incident Gaussian beam into a vortex beam with helical wavefront. (b) SEM image of an all-dielectric metasurface that implements a $2\pi$ phase-variation (along the polar angle) in 8 steps at $\lambda=1.55$\,$\mu$m, with (c) displaying the doughnut-shaped intensity distribution of the transmitted beam. (d) Recorded image of a spiral-shaped intensity pattern arising from interference between the transmitted beam and a co-propagating Gaussian reference beam. (a-d) Adapted with permission from \cite{Shalaev_2015}; Copyright 2015 American Chemical Society.}
	\label{fig:OAM}
\end{figure*}

Figure \ref{fig:OAM}(a) illustrates the basic working principle in which a Gaussian incident beam is transformed to a vortex beam of $l=1$ by a metasurface implementing (in a discrete number of steps) a $2\pi$ phase variation in the azimuthal direction. The corresponding all-dielectric realization is shown in figure \ref{fig:OAM}(b), which consists of eight sectors of silicon nanobricks on top of a glass substrate, hereby introducing an additional $\pi/4$ phase shift between neighboring sections. The intensity of the transmitted beam at a wavelength of $\lambda=1.55$\,$\mu$m is presented in figure \ref{fig:OAM}(c), thus confirming the doughnut-shaped intensity profile, with the characteristic zero intensity in the core of the beam arising from the phase singularity at the center of the spiral phase. The topological charge, on the other hand, is probed by interfering the transmitted beam with a co-propagating Gaussian reference beam, hereby creating a spiral-shaped interference pattern (see figure \ref{fig:OAM}(d)). As a final comment, we note that the metasurface in figure \ref{fig:OAM}(b) features a limited efficiency of $\sim 45$\% due to design and realization imperfections, though related designs based on silicon nano-posts (see figure \ref{fig:Huygens}(f)) have shown efficiencies above 95\% \cite{Arbabi_2015}.

Coming to the end of this subsection, we have discussed and exemplified how the different categories of metasurfaces can be employed to realize compact (i.e., subwavelength-thin) versions of some of the most widely used optical components. The list of functionalities is, however, far greater and, in principle, only limited by your imagination and potential relevance. This point will become more evident in the following sections, but of other flat components, not explicitly discussed in the preceeding text, we mention axicons for the generation of non-diffracting Bessel beams \cite{Pfeiffer_2013,Aieta_2012,Lin_2014} and magnetic mirrors for maximizing the electric field at the surface \cite{Zhijie_2016,Esfandyarpour_2014, Liu_2014, Headland_2015} or for creation of low-profile planar microwave antennas \cite{Feresidis_2005}.

\subsection{Broadband absorbers}
The inherent Ohmic losses associated with plasmonic nanostructures and metasurfaces are often considered detrimental to device performance. However, the unavoidable absorption of light may in certain applications be a distinct advantage, particularly within the field of photovoltaics. For example, plasmonic absorption may reduce the size of the semiconductor absorbing layer in solar cells \cite{Atwater_2010}, or harvest solar energy below the band gap of the semiconductor by injection of energetic (so-called, hot) electrons, generated by non-radiative decay of plasmons in metal nanostructures, into the conduction band of the semiconductor at the metal-semiconductor interface \cite{Knight_2011,Linic_2011,Clavero_2014}. Also, the technology of solar thermophotovoltaics, in which the solar spectrum is first absorbed and then converted to narrowband thermal emission at the band edge of the photovoltaic cell \cite{Rephaeli_2009,Lenert_2014}, can greatly benefit from plasmonic configurations that either absorb light in a broad or narrow wavelength range. Finally, it is worth noting that absorbing structures also find applications within camouflage and stealth technology \cite{Fante_1988,Kern_2003}.

When considering absorbing configurations, one simple, yet very powerful, approach is to utilize multilayer systems that incorporate nanometer-thin lossy films in order to function as Gires-Tournois resonators \cite{Kats_2013,Park_2014,Ding_2015_3}. These systems can be realized without the use of lithography, thus extending large areas, while the strong index contrast between the surrounding medium and the lossy layer(s) makes them insensitive to the polarization and angle of the incident light. Nevertheless, since multilayer systems are homogeneous in the plane and absorption occurs due to propagation in the layers, these configurations cannot be considered as metasurfaces and will not be considered further in this review. Metasurface absorbers, on the other hand, are typically characterized by unit cells featuring single- and multi-resonant design, which translates into narrowband and broadband absorption, respectively. In order to keep the present section relatively compact, we limit the remaining discussion to few representative examples of broadband absorption by metasurfaces, hence disregarding work on narrowband absorbers that, otherwise, are interesting with respect to sensing applications \cite{Tittl_2011} and thermal emitters \cite{Mason_2011,Abbas_2011}. For a more detailed discussion on light absorption by nanostructured materials, we refer to dedicated review papers \cite{Cui_2014,FeiGuo_2014,Radi_2015}.

\subsubsection{Broadband absorbers using multi-resonances}
Intuitively, metasurfaces based broadband absorbers can be realized by properly designed unit cells featuring multi-resonances. One example is depicted in figure \ref{fig:Absorbers}(a), where the broadband absorption is achieved by mixing multiple resonances via carving the top thin metal film of an MIM multilayer stack into the format of the so-called crossed trapezoid arrays \cite{Aydin_2011}. Due to the crossed symmetric arrangement, the fabricated absorber yields broadband and polarization-independent resonant light absorption over the entire visible spectrum ($400-700$\,nm) with an average measured absorption of 0.71, shown in figure \ref{fig:Absorbers}(b). Compared with the crossed-grating structures, the measured absorption is much broader and flatter, which is ascribed to the multiple adjacent plasmonic resonances within the visible wavelength range supporting by the unit cell. Besides designing particular structure featuring multiple resonances, an alternative is to integrate multi-sized resonators parallel in the same plane within a supercell \cite{liu2011taming,cui2011thin,nielsen2012efficient}. It should be noted that the feature sizes of the constituent resonators should have relatively small differences between each other, thus several adjacent narrow absorption bands can mix into one broad absorption band; otherwise, one can only obtain multiband absorbers when significant differences exist in size between neighboring resonators. For this type of horizontal super cell integration, the number of resonances supported within the same unit cell is limited due to the mode competition and finite absorption cross section of each resonator, thus limiting the absorption efficiency. Another way to expand the absorption band is to vertically stack different-sized resonators together within the same unit cell size \cite{ye2010omnidirectional,ding2012ultra,cui2012ultrabroadband,zhu2014ultra,he2014light,ding2014ultrabroadband}. Comparatively, vertical integration of the resonators does not have a limitation on the number of integrated resonators. Hence, the absorption bandwidth of the constructed absorber can be much broader. Particularly, when the thickness of each dielectric and metallic slab is extremely subwavelength, the metal/dielectric multilayer system can be regarded as hyperbolic metamaterials which supports slow-wave modes \cite{cui2012ultrabroadband,he2014light,ding2014ultrabroadband}. As a final comment, it should be mentioned that highly-lossy plasmonic materials can replace the commonly used gold/silver to realize broadband absorption with simple MIM configurations \cite{LiWei2014Absorber,dingSR2016broadband}.

\subsubsection{Nonresonant broadband absorbers}
\begin{figure}[tb]
	\centering
		\includegraphics[width=8cm]{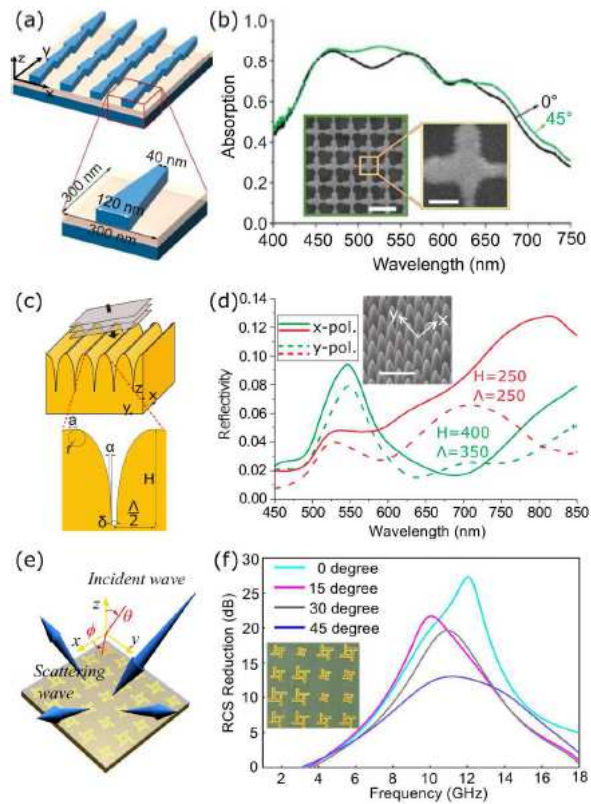}
	\caption{(a) Sketch of plasmonic meta-reflectarray consisting of trapezoidal-shaped 100\,nm thick silver meta-atoms atop a 60\,nm thin glass spacer and 100\,nm silver film. The metasurface is designed for broadband absorption of $x$-polarized incident light in the visible regime. (b) Measured absorption as a function of wavelength for $x$- and diagonal (i.e., $0^\circ$ and $45^\circ$) polarization when the unit cell of the meta-reflectarray consists of two crossed trapezoidal meta-atoms. Inset displays SEM image of the fabricated metasurface. (c) Schematic of the considered configuration for adiabatic focusing of GSP modes, showing the orientation of a groove array and incident plane wave. (d) Reflectivity spectra of $x$-(solid lines) and $y$-polarized (dashed lines) light for crossed convex grooves of different depths $H$ and periods $\Lambda$. (e) Schematic of a diffuse metasurface composed of windmill shaped particles with random reflection phases illuminated by microwave waves, showing the occurrence of electromagnetic diffusion in the upper half-space. (f) The measured RCS reduction for various incident angles with horizontal polarization. The inset shows the optical image of part of the fabricated sample. (a,b) Adapted with permission from \cite{Aydin_2011}; Copyright 2011 Nature Publishing Group. (c,d) Adapted with permission from \cite{Sondergaard_2012}; Copyright 2012 Nature Publishing Group. (e,f) Adapted with permission from \cite{wang2014broadband}; Copyright 2014 Nature Publishing Group.}
	\label{fig:Absorbers}
\end{figure}

Though the multi-resonant metasurfaces are good candidates in realizing broadband absorption, the bandwidth is still necessarily limited owing to the resonant nature of surface plasmon excitations involved. Therefore it is more desirable to have nonresonant absorption. Nonresonant (broadband) absorption has been demonstrated with plasmonic composite metamaterials involving a nanostructured metallic layer embedded in dielectric layers \cite{kravets2008plasmonic,kravets2010plasmonic}, and optical analogues of black holes based on metamaterials \cite{narimanov2009optical,genov2009mimicking} and GSP modes \cite{nerkararyan2011plasmonic}. However, the involved metamaterials are often suffering from complicated three-dimensional (3D) structure and consequently differently in fabrication. Recent progress is related to nonresonant light absorption with pure metal metasurfaces, which utilizes ultra-sharp convex metal grooves via adiabatic nanofocusing of GSP modes excited by scattering off subwavelength-sized wedges \cite{Sondergaard_2012,beermann2013plasmonic,sondergaard2013theoretical}. Figure \ref{fig:Absorbers} (c) conveniently illustrates the 1D array of grooves in gold for nonresonant absorption by adiabatic focusing of GSP modes \cite{Sondergaard_2012}. For subwavelength array periods and $p$-polarized incident light (the electric field is perpendicular to the groove direction), GSP modes will be excited and propagate inside the grooves towards the bottom without any reflection once the regime of adiabatic nanofocusing is realized throughout the groove due to small variations of the GSP propagation constant. To remove the polarization-selective absorption of 1D groove arrays, 2D arrays of crossed grooves were implemented with focused ion-beam (FIB) milling, which have relatively low structural anisotropy (see the inset of figure \ref{fig:Absorbers}(d)). The fabricated 2D groove arrays show a remarkable suppression of gold reflectivity below 13\% for both polarizations in the whole investigated wavelength range of $450-850$\,nm (figure \ref{fig:Absorbers}(d)). This reflectivity signifies an average absorption of 96\% of unpolarized light within this wavelength range in the absence of transmission for the 350-nm-period arrays of 400-nm-deep grooves. Moreover, this approach can naturally be exploited with other metals, for example, with nickel that, apart from being much cheaper than gold and having a much higher melting temperature, features stronger plasmon absorption, promising significantly lower reflectivity levels that can be fabricated with more practical techniques, such as reactive-ion etching.

\subsubsection{Broadband absorbers with random metasurfaces}
In the aforementioned broadband absorbers, individual meta-atom effectively converts the incident wave into heat, thus the reflected energy could be significantly reduced. However, the generated heat within the subwavelength structure inevitably limits the applications, for instance, hiding the objects from radars. Diffuse reflection or low scattering is another useful approach to this aim, since it diffuses the scattered energy and, thereby, suppresses the reflection, which can be regarded as an alternative to realize broadband absorption. A simple example is illustrated in figure \ref{fig:Absorbers}(e), where a low-scattering metasurface operating in the microwave range is consisting of windmill-shaped units with randomly distributed reflection-phase \cite{wang2014broadband}. Because of the random reflection phases, the reflected waves from each part of the metasurface will destructively interfere in all directions of the upper half-space, thereby creating a low-reflection metasurface. In order to redistribute the scattering energy into all the directions, a large phase range is usually required, while the magnitude of the reflection efficient should approach unity, which is different from the typical absorbing metasurface where each meta-atom is required to reflect as low energy as possible. The metasurface shows excellent broadband properties, where the radar cross section (RCS) reduction is over 10\,dB from 7\,GHz to 14\,GHz, as shown in figure \ref{fig:Absorbers}(f). Additionally, the low RCS properties are still maintained with the increased incident angle. Following this concept, a variety of random phase metasurfaces, mostly at microwave frequencies, have been demonstrated to accomplish low-reflection \cite{cui2014coding,su2016ultra}; even at terahertz range the broadband and wide-angle diffusion can be experimentally maintained \cite{dong2015terahertz,Gao_2015}. Very recently, a GSP-based random-phase metasurface at a wavelength of 800\,nm was demonstrated by Anders and co-workers \cite{pors2016random}, which convincingly shows the diffuse scattering of reflected light, with statistics obeying the theoretical predictions.

\subsection{High-resolution color printing}
Colors are significantly important for the perception and identification of our surroundings due to their abilities in carrying rich information. In principle, colors are mainly produced by the light scattering or partial absorption of materials. In nature, there are numerous species whose colors arise from the micro/nanostructure patterns of their skins, which selectively diffract, reflect and scatter the light. Inspired by the natural colors, renewed attention has been drawn to the so-called structural colors with well-designed micro/nanostructures for their great promise in a wide variety of fields such as light modulators, light-emissive diodes, imaging sensor, and colored display devices \cite{ko2000colorhigh,catrysse2004color,chung2012flexible}. Recently, artificial surfaces, on which the colors are generated via a resonant interaction between light and subwavelength nanostructures, have emerged as metasurfaces for the realization of structural colors \cite{xu2011structural,Yinghong_2015,kristensen2016plasmonic}. In the following section, we will discuss recent advances in high-resolution color printing by metasurfaces, and several kinds of representative metasurfaces that have excellent quality or special functionalities are presented. To keep the present section compact, we are mainly focused on 2D high-resolution color printing, hence disregarding some work on 1D grating colors \cite{xu2010nc,zeng2013ultrathin} and multilayer systems \cite{Kats_2013,lee2015angular,li2015colorlarge}.

\subsubsection{Static color printing}
\begin{figure*}[tb]
	\centering
		\includegraphics[width=8cm]{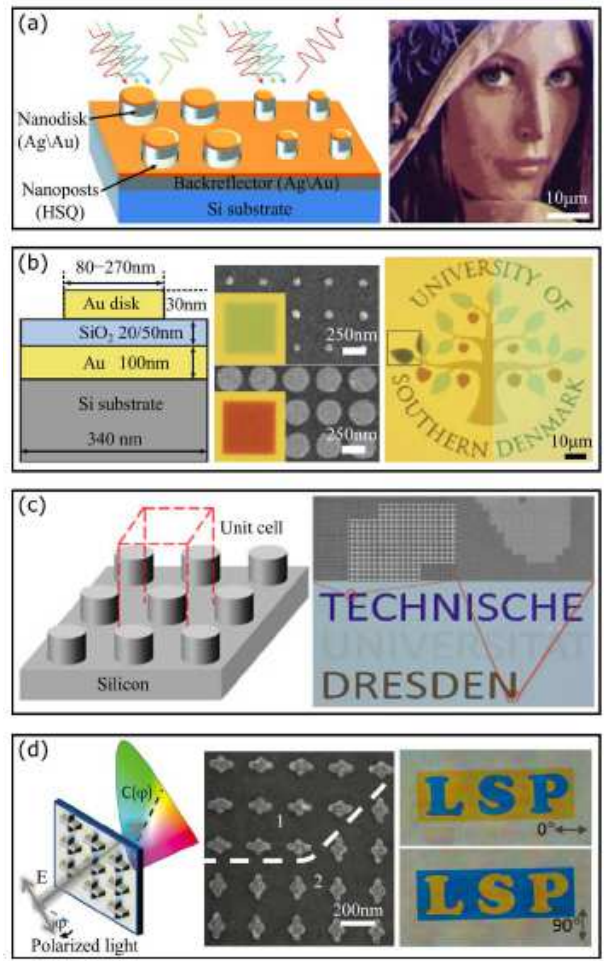}
	\caption{High-resolution color printing. (a) Individual plasmonic pixels made of a metal disk (gold or silver) raised above a holey metal back-reflector by HSQ nanoposts. The resulting print of a photorealistic test image that goes under the name of Lena is shown. (b) Gold nanodisk arrays supported by a continuous silica layer and by a gold metallic film acting as a back-reflector used to represent the logo of Southern Denmark University. (c) Silicon nanodisk arrays used for full-color image printing patterns realized using NIP. (d) Polarization-dependent color printing with anisotropic cross nanoantennas. (a) Adapted with permission from \cite{duan2012color}; Copyright 2012 Nature Publishing Group. (b) Adapted with permission from \cite{alex2014color}; Copyright 2014 American Chemical Society. (c) Adapted with permission from \cite{yang2015diele}; Copyright 2015 AIP Publishing Society. (d) Adapted with permission from \cite{ellenbogen2012chromatic}; Copyright 2012 American Chemical Society.}
	\label{fig:Color}
\end{figure*}

The color printing with subwavelength resolution requires the juxtaposition of individual color pixels of dimensions smaller than the wavelength: that is, between 400 and 700\,nm. However, industrial techniques such as inkjet and laserjet methods are limited at sub-10,000 dots per inch (DPI) because of their micro-sized ink spots and accuracy in the positioning of multiple color pigments. By contrast, resonant metasurfaces enable subwavelength color elements due to strong scattering and absorption. In particular, plasmonic metasurfaces colors can even be printed with a resolution that beats the diffraction limit due to deep-subwavelength light localization \cite{gramotnev2010plasmonics}. For example, hybridized nanodisk and nanohole arrays can be used to obtain high-resolution images \cite{duan2012color,tan2014plasmonic,clausen2014plasmonic,james2016plasmonic}. Specifically, such hybridized plasmonic structures contain a metallic nanodisk suspended on top of a dielectric pillar and a complementary metallic hole as a back reflector. Thus the propagating surface plasmons supported by the holes couple with the localized surface plasmon resonances (LSPRs) in the disks. In this way, a high-resolution color printing technology at the diffraction limit was reported, by depositing a thin layer of Au/Ag on hydrogen silsesquioxane (HSQ) pillars atop a silicon substrate \cite{duan2012color}.
With properly selected disk sizes, an image of Lena was produced, shown in figure \ref{fig:Color}(a). Remarkably, the resulting image closely reproduces the details of the original image down to single-pixel elements, thus enabling color printing at a resolution of $\sim100,000$ DPI. Aluminum has also been adopted as constituent material to enrich the color gamut \cite{tan2014plasmonic,clausen2014plasmonic,james2016plasmonic}, and to realize large-scale high-resolution color printing by using nano-printing technology (NPT) \cite{clausen2014plasmonic}.

High-resolution color printing can also be achieved using GSP-based metasurfaces. As an example, figure \ref{fig:Color}(b) shows a subwavelength plasmonic color orienting in a configuration with gold nanodisks coupled to a continuous gold film  \cite{alex2014color}. By tuning the size of nanodisks and the thickness of the dielectric layer, the reflected color can be tuned. As such, the optical image of the University of Southern Denmark logo reveals a high-quality color print in terms of high contrast, high color uniformity, and high reproduction fidelity due to correct single-pixel color reproduction. Compared to other plasmonic color nanostructures, the GSP-based color printing is insensitive to the incident angle, which is valuable in practical applications. Moreover, while the colors of most plasmonic nanostructures are sensitive to the surroundings, the GSP nanostructure can endure a cover of transparent dielectric overlay without significantly influencing colors, which is a direct result of the GSP field distribution that is almost entirely contained within the MIM structure itself. In the right panel of figure \ref{fig:Color}(b), an optical microscope image is shown to compare the colors of uncovered (left) and PMMA covered (right) parts of the color printing, suggesting that the overall color image remains except for a slight contrast variation. Therefore, this GSP color device can be protected with a transparent dielectric overlay without destroying the colors, providing the sample with the chemical and mechanical stability necessary for use in exposed ambient color printing applications such as, for example, security certificates. Similarly, Aluminum-based GSP structures have also been considered in this context \cite{miyata2016full}. In the context of printing new colors on a prefabricated GSP color substrate, recent work has demonstrated ultrahigh-resolution printing with an unprecedented resolution of 127,000 DPI over large areas using a laser to induce morphology changes at the single-resonator level \cite{zhu2015plasmonic}. All primary colors can be printed with a speed of 1\,ns per pixel, and power consumption down to 0.3\,nJ per pixel.

As complementary to conventional subwavelength color printing by plasmonic metasurfaces, dielectric metasurfaces may be tailored to produce colors with reduced material loss \cite{proust2016all,kuznetsov2016optically}. One example of dielectric color printing is depicted in figure \ref{fig:Color}(c), where each unit cell consists of silicon nanodisk patterned on top of silicon substrate \cite{yang2015diele}. These silicon nanodisks were able to support HE$_{1m}$ leaky modes, which depended on the diameter of the nanodisks, resulting in wavelength-dependent reflection spectra, enabling multicolor generation. In addition, the electric field of HE$_{1m}$ is mainly concentrated around the nanodisks and shows minor dependency on the gap, suggesting that the single nanodisk can act as an individual pixel. Moreover, such silicon nanodisk patterns are quite compatible with existing CMOS technology and could be produced in large-scale using NPT, superior to the present e-beam lithography (EBL) processing suffering from low throughput and high cost. The nano-imprinted sample shows distinct colors with high quality (see the right panel of figure \ref{fig:Color}(c)). While dielectric metasurface color printing is really promising, significant challenges remain to be addressed for practical use. For example, silicon has very strong absorption in the visible light range especially when the wavelength $\lambda$ is smaller than 500\,nm. Thus the silicon metasurfaces face severe challenges in producing distinct color impressions, especially for the blue and purple colors. To address this, TiO$_2$ has been selected to replace silicon, and consequently, distinct colors covering the entire visible spectral range have been demonstrated very recently \cite{Sun2017dielectric}.

Polarization state is an intrinsic property of light, which is uncorrelated with its intensity and wavelength, thereby providing another degree of freedom to manipulate color information. The polarization-dependent color printing could double the information capacity by recording the color data in different polarization states. Different from the polarization-insensitive color printing using circular-symmetric meta-atoms, polarization-dependent color printing can be realized with asymmetric meta-atoms, such as anisotropic nanocrosses \cite{ellenbogen2012chromatic} and elliptical nanodisks \cite{goh2014three}, which respond dissimilarly to different polarized incident lights. Figure \ref{fig:Color}(d) displays an example of polarization-dependent color printing based on subwavelength-spaced anisotropic cross-shaped aluminum nanoantennas, which support two LSPRs for the vertical and horizontal polarized light, rendering different colors. Each anisotropic nanocross could work a yellow color filter for incident polarization along its short arms while being a blue filter for incident polarization along the long arms. With the controlled position and rotating angle of the nanocrosses, active polarized images have been achieved (right panel of figure \ref{fig:Color}(d)). When the light is polarized along one of the principal axis, a brilliant yellow and blue LSP image emerges that can be color switched by rotating the polarization of the incident light by $90^\circ$. In analogy to cross-shaped nanoantennas, asymmetric cross-shaped nanoapertures can also create polarization-switchable images \cite{li2016dual}.

\begin{figure*}[tb]
	\centering
		\includegraphics[width=8cm]{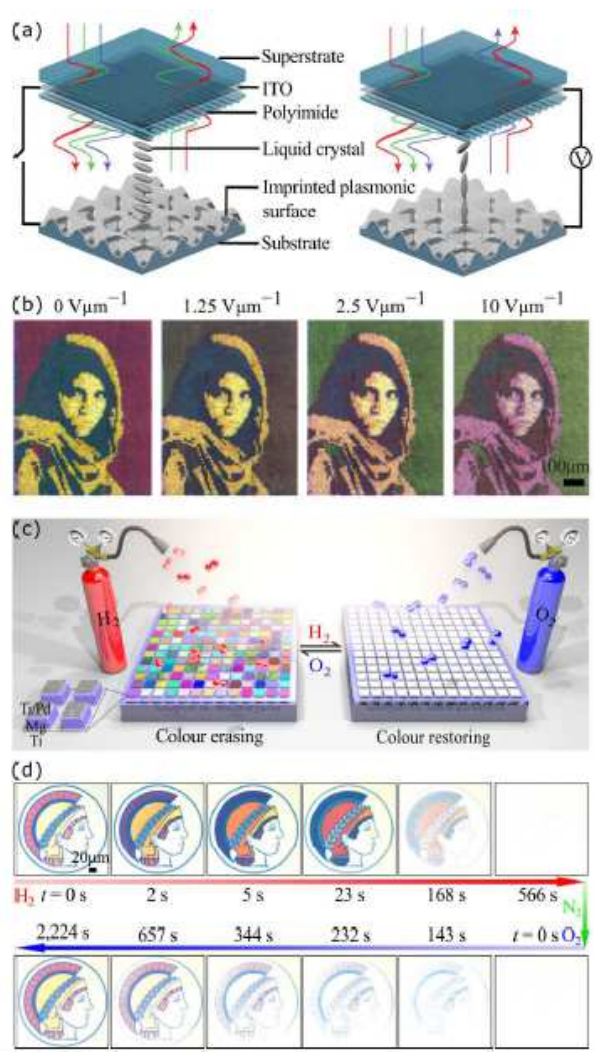}
	\caption{Reconfigurable color printing. (a) The color appearance can be modulated dynamically by incorporating LCs in the near-field of the plasmonic resonances. (b) The resulting nanoimprinted plasmonic color surface can be tuned by applying different voltages. (c) Schematic of the plasmonic metasurface composed of hydrogen-responsive magnesium nanoparticles interacting with incident light. (d) Optical micrographs of the Minerva logo during hydrogenation and dehydrogenation for color erasing and restoring. (a,b) Adapted with permission from \cite{franklin2015polarization}; Copyright 2015 Nature Publishing Group. (c-d) Adapted with permission from \cite{duan2017dynamic}; Copyright 2017 Nature Publishing Group.}
	\label{fig:ColorTunable}
\end{figure*}
\subsubsection{Dynamically reconfigurable color printing}
The previous discussions are mainly focused on static colors. However, it is more desirable to have dynamically reconfigurable colors \cite{liu2012light}. A good example of reconfigurable colors is shown in figure \ref{fig:ColorTunable}(a), where a specially designed nanostructured plasmonic surface is in conjunction with high birefringent liquid crystals (LCs) \cite{franklin2015polarization}. Therefore the plasmonic field can be modulated by the LCs, which is in turn controlled by an external applied voltage, resulting in a large range of color tenability. In combination with imprinted structures of varying periods, a full range of colors spanning the entire visible spectrum has been achieved. Once a particular image has been printed, different color representations can be rendered by applying an external bias, paving the way towards dynamically reconfigure colors for reflective displays, shown in figure \ref{fig:ColorTunable}(b). Dynamic color tuning was also accomplished through controlled hydrogenation and dehydrogenation of the constituent magnesium nanoparticles, which serve as dynamic pixels \cite{duan2017dynamic}. Figure \ref{fig:ColorTunable}(c) schematically illustrates the working principle of this dynamic plasmonic color display, where the plasmonic metasurface is composed of hydrogen-responsive magnesium nanoparticles. The dynamic behavior is enabled by the reversible chemical reaction of Mg in response to hydrogen, that is, a metal-to-dielectric transition to form magnesium hydride (MgH$_2$). The performance of the dynamic display with Minerva logo is presented in figure \ref{fig:ColorTunable}(d). When the hydrogen is loaded, the Minerva logo undergoes dynamic color changes and a series of abrupt color alterations take place within 23\,s. For example, yellow turns red, red turns blue, and blue turns white. After a few minutes, all the colors disappear, revealing a white surface instead of Minerva. By replacing the hydrogen stream with a stream of oxygen, this process is reversible, that is, the logo is gradually restored to its original state, which takes 2,224\,s to complete. Because of the unique hydrogenation/dehydrogenation kinetics afforded by magnesium nanoparticles, this dynamic display has excellent reversibility and durability. Other approaches to dynamically control the colors include using electrochromic polymers \cite{xu2016high} and bimetallic nanodots arrays \cite{wang2016mechanical}.

\subsection{Holograms}
Holography, first developed in 1948 by Denis Gabor \cite{gabor1948new}, is a revolutionary 3D imaging technique to reconstruct the 3D features of an object by scattering an incident coherent source from an optically recorded interference pattern (hologram). Owing to the computer-generated holography (CGH) \cite{gerchberg1972practical}, holograms have gained numerous applications, such as wavefront shaping and optical manipulation. However, the thickness of conventional holograms usually remains comparable to the wavelength since the phase modulation relies on the accumulated phase through propagation. Worth still, their smallest achievable pixel size is usually tens of the wavelengths for visible light, which suffers from issues of high-order diffraction and twin-image. Metasurfaces enable exceptional control over the light with surface-confined planar components, offering the fascinating possibility of engineering the local amplitude, phase and polarization at will, and thus are a good platform to realize all types of computer-generated holograms. In this section, metasurface-based holograms (metaholograms) are classified in terms of the method of phase realization: resonant phase and geometric phase metaholograms.

\subsubsection{Resonant phase metaholograms}
\begin{figure*}[tb]
	\centering
		\includegraphics[width=14.0cm]{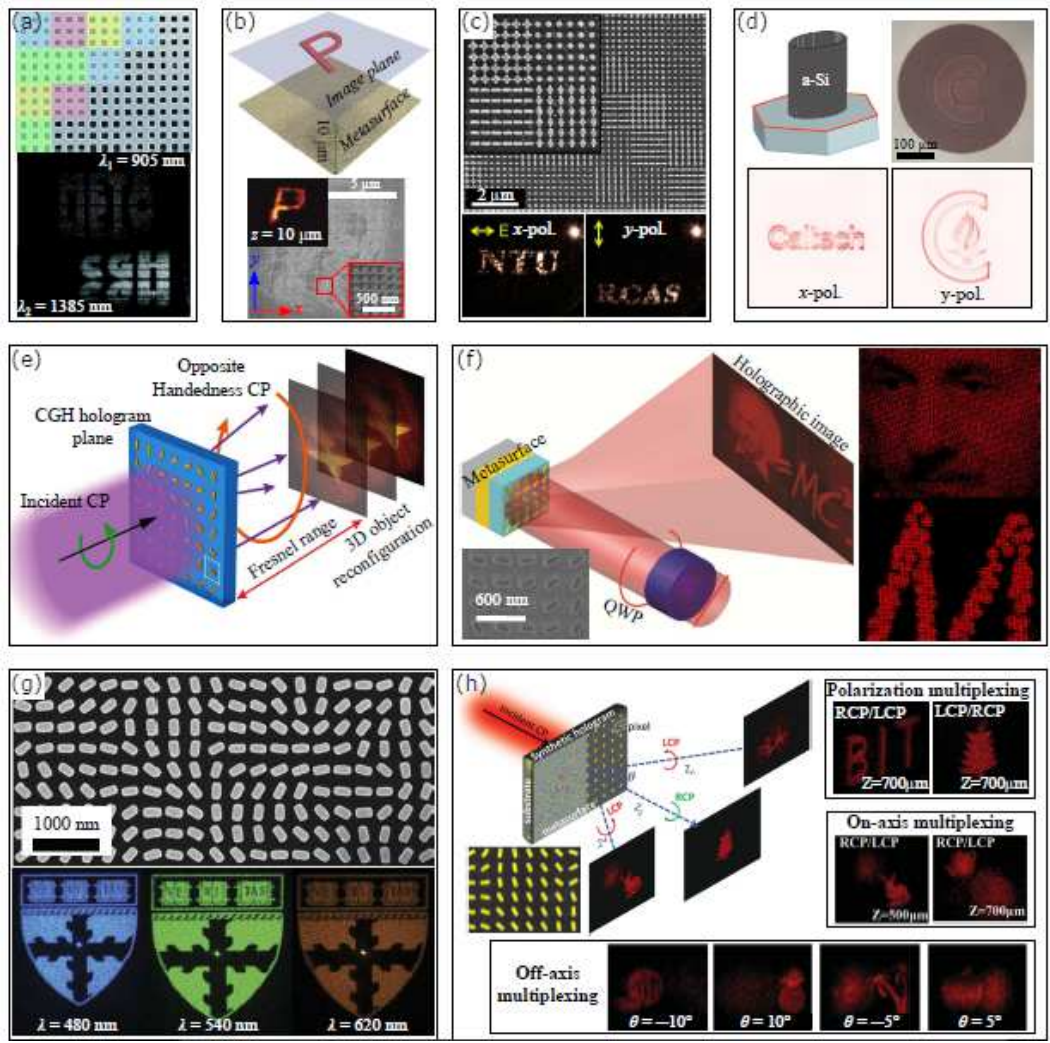}
	\caption{{\scriptsize Metaholograms. (a) Dual-wavelength binary holograms consisting of rectangular nanoapertures. (b) Visible metasurface holography realized by V-shaped nanoapertures. (c) Polarization-multiplexing holograms based on meta-reflectarrays for linear polarization states. (d) Polarization-switchable phase hologram that generates two arbitrary patterns for x- and y-polarized light based on silicon nanoposts. (e) 3D holography consisting of subwavelength metallic nanorods with spatially varying orientations generates an on-axis jet image upon normal incidence of CP light. (f) Highly-efficient reflective holograms employing geometric phase and MIM configuration. (g) High-performance dielectric holograms for red, green, and blue wavelengths with record absolute efficiency employing TiO$_2$. (h) Hybrid multiplexing holograms with geometric metasurface for normal CP incident light in transmission. (a) Adapted with permission from \cite{walther2012spatial}; Copyright 2012 Wiley-VCH. (b) Adapted with permission from \cite{ni2013metasurface}; Copyright 2013 Nature Publishing Group. (c) Adapted with permission from \cite{chen2013high}; Copyright 2013 American Chemical Society. (d) Adapted with permission from \cite{Arbabi_2015}; Copyright 2015 Nature Publishing Group. (e) Adapted with permission from \cite{huang2013three}; Copyright 2013 Nature Publishing Group. (f) Adapted with permission from \cite{Zheng_2015}; Copyright 2015 Nature Publishing Group. (g) Adapted with permission from \cite{Devlin_2016}; Copyright 2016 National Academy of Sciences. (h) Adapted with permission from \cite{huang2015broadband}; Copyright 2015 Wiley-VCH.}}
	\label{fig:Holograms}
\end{figure*}

Binary resonant phase metaholograms have first been proposed by using layered nanostructures \cite{walther2012spatial,larouche2012infrared}. Figure \ref{fig:Holograms}(a) shows a fishnet metasurface consisting of an MIM layer stack with rectangular holes, which is able to control simultaneously the amplitude and phase of the transmitted field, yielding huge flexibility in transmission hologram design \cite{walther2012spatial}. By carefully engineering the dispersion of apertures, coding of two different words `META' and `CGH' has been demonstrated at different wavelengths, $\lambda_1=905$\,nm and $\lambda_2=1385$\,nm, respectively. However, due to the binary nature of the hologram, there is always a twin image and unwanted zero-order diffractions. It is worthy to point out that more complicated multicolor holograms can be achieved by extending the coding methods. One simple, yet very powerful, approach is to combine multiple elements featuring totally different resonant wavelengths, thus different holographic images can be reconstructed by addressing the incident wavelength \cite{montelongo2014plasmonic,huang2015aluminum}. Nevertheless, as a consequence of minimizing the crosstalk between different color images, the efficiency of wavelength multiplexing holograms is inevitably low, usually less than 1\% \cite{huang2015aluminum}.
As a realization of advanced holograms, figure \ref{fig:Holograms}(b) displays an ultrathin (30\,nm) planar metasurface hologram at a wavelength of 676\,nm, consisting of an array of V-shaped nanoapertures, which could generate discrete eight-level phase distribution with a two-level amplitude modulation \cite{ni2013metasurface}. Due to the subwavelength-scale control of the phase and suppressing the undesired co-polarized light in the cross-polarized images, ultra-thin, ultra-small holograms can be recorded, providing unprecedented spatial resolution, low noise, and high precision of the reconstructed image. Since all the orders of diffraction fully contribute to the holographic image, an overall efficiency of $\sim$10\% is achieved, which is about one order of magnitude greater than the existing metamaterial holograms \cite{walther2012spatial,larouche2012infrared}. It is worth noting that major limiting factors of efficiency here are Ohmic losses and low polarization conversion efficiency (see section \ref{sec:Vmetasurface}).

As demonstrated in section \ref{sec:RMs}, meta-reflectarrays play an important role in realizing high-efficiency metasurfaces. As such, meta-reflectarray is a good candidate to realize efficient holograms. Here, we would like to highlight a meta-reflectarray that generates reflective hologram at visible wavelengths for linear polarization states (figure \ref{fig:Holograms}(c)) \cite{chen2013high}. By utilizing the polarization-sensitive response of the gold cross nanoantennas, simultaneously encoding two holograms on the same metasurface has been realized, with each image reconstructed by an incident beam with polarization perpendicular to the other. The functionality of polarization controlled dual images was verified experimentally, where the projected patterns transitioned from `NTU' to `RCAS' with the polarization rotated from $x$- to $y$-direction. The measured efficiency at $\lambda=780$\,nm reaches 18\% at $15^\circ$ incidence and drops down to 8.5\% when the angle is increased to $45^\circ$. The efficiency for meta-reflectarray hologram can be further improved to $\sim$ 50\% at a near-infrared wavelength of 1550\,nm, resulting from the decreased metal loss \cite{yifat2014highly}.

Though the efficiency of holograms can be increased to a certain extent by meta-reflectarrays, it is still far from satisfactory, since the Ohmic losses of metals pose a severe issue, particularly in the visible frequency range. Recently, all-dielectric metasurface holograms have got a rapid development due to their ability in addressing the efficiency issues and realizing Huygens' metasurfaces \cite{Arbabi_2015,Chong_2016,wang2016grayscale}. In a recent paper from Faraon's group, a polarization-switchable phase hologram that generates two distinct patterns for $x$- and $y$-polarized light has been demonstrated by a dielectric metasurface \cite{Arbabi_2015}, where elliptical amorphous silicon nanoposts provide complete control of polarization and phase in transmission (figure \ref{fig:Holograms}(d)). In this case, the word `Caltech' is displayed for input $x$ polarization and an icon is displayed for input $y$ polarization. Additionally, the hologram features measured efficiencies of 84\% and 91\% for $x$- and $y$-polarized incident light at the near-infrared wavelength of 915\,nm. A similar highly-efficient polarization-insensitive metaholograms based on silicon has been realized by Want \emph{et al} \cite{wang2016grayscale}, which produce grayscale high-resolution images and transmit over 90\% of light with a diffraction efficiency over 99\% at a 1600\,nm wavelength. It should be noted that the design approach of dielectric metahologram is applicable to other wavelengths and, in particular, to the visible wavelengths, with properly selected materials with high refractive index, for example, TiO$_2$.

\subsubsection{Geometric phase metaholograms}
In the metaholograms mentioned above, phase variations are achieved with geometry-dependent resonance, thereby resulting in the limited working bandwidth due to the dissimilar dispersion of the meta-atoms resonance. Geometric phase metaholograms, on the other hand, gain the required phase on the cross-polarized light with identical geometric parameters but spatially varying orientations. In this sense, geometric phase metaholograms are dispersionless in terms of the phase. However, it should be noted that conversion efficiency may be strongly wavelength-dependent, as it depends solely upon the designed meta-atoms (see section \ref{sec:GMs}). Because of the linear dependence of geometric phase on the orientation angle of individual meta-atoms, the engineering realization of multilevel phase holograms has been greatly simplified \cite{huang2013three,Zheng_2015,scheuer2015holography}. As an example of geometric phase metaholograms, figure \ref{fig:Holograms}(e) schematically illustrates the 3D CGH images reconstruction using ultrathin plasmonic metasurfaces made of subwavelength metallic nanorods with spatially varying orientations \cite{huang2013three}. The reconstructed image was observed with an optical microscope, and by scanning the image plane position truly 3D images were demonstrated. Due to subwavelength meta-atoms for phase encoding, metasurfaces, therefore, provide a method to increase the field of view for digital holograms, which was evaluated as $\pm40^\circ$. Additionally, the dispersionless nature of our metasurface can result in broadband operation without sacrificing image quality.

To increase the efficiency, a reflectarray metasurface that introduced 16-level geometric phases was demonstrated to create holographic images \cite{Zheng_2015,scheuer2015holography}, where the MIM structure was optimized as a broadband half-wave plate maintaining high polarization conversion, shown in figure \ref{fig:Holograms}(f). Remarkably, this metahologram has a diffraction efficiency of 80\% at 825\,nm and a broad bandwidth between 630\,nm and 1050\,nm with a high window efficiency larger than 50\%, ascribed to the antenna-orientation controlled geometric phase profile combined with the reflectarrays for achieving high polarization conversion efficiency. It should be emphasized that this approach can be readily extended from phase-only to amplitude controlled holograms, simply by changing the size of the nanorods, thereby generating more complex holograms.

As an alternative approach to realizing highly-efficient geometric phase metaholograms, particularly relevant for the visible spectrum, figure \ref{fig:Holograms}(g) shows a high-performance dielectric hologram for red, green, and blue wavelengths with record absolute efficiency ($>78\%$) \cite{Devlin_2016}. As the basis of dielectric metasurfaces, a common material, TiO$_2$, was fabricated based on atomic layer deposition, thereby creating highly anisotropic nanostructures with spatially controlled orientations. The Harvard logo with high-resolution fine features can be seen in the reconstructed images across the visible spectrum for the hologram with a design wavelength of 480\,nm since the geometric phase is a wavelength-independent effect. Besides TiO$_2$, silicon is also a good platform for all-dielectric geometric phase metaholograms \cite{khorasaninejad2016broadband,wang2016visible,huang2016silicon}.

Similar to the resonant phase metaholograms, holography multiplexing can be easily accomplished with geometric metasurfaces by integrating complex multiplexing methods into a single synthetic technique, including polarization \cite{khorasaninejad2016broadband,huang2016silicon,wen2015helicity,huang2015broadband}, position \cite{huang2016silicon,huang2015broadband}, and angle \cite{huang2015broadband}. Figure \ref{fig:Holograms}(h) illustrates a good example of hybrid multiplexing with plasmonic metasurface, which functions as synthetic holograms for normal incident CP light in transmission \cite{huang2015broadband}. Various images have been simultaneously reconstructed at different polarization channels, with different sets of reconstruction parameters, such as positions and off-axis angles, verifying the concepts of solely circular polarization multiplexing and multiple hybrid multiplexing schemes. As a final comment, we emphasize that wavelength multiplexing with geometric metasurface is also possible \cite{wang2016visible,Lie1601102,Sajid2017}, as the conversion efficiency of geometric metasurface is still wavelength-dependent. However, one should carefully optimize each meta-atom to feature resonance peak, which corresponds to a certain wavelength. Therefore, only one kind of meta-atoms is predominantly activated to induce the geometric phase for one wavelength.

From the above discussions, it is evident that metasurface is somehow an ideal platform to realize all types of holograms with subwavelength thickness, high-resolution, low-noise, high precision, and flexibility. Moreover, geometric phase metahologram is more robust against fabrication tolerances and variation of material properties due to the much simpler structure geometry and the geometric nature of the phase. For a more detailed discussion on metasurface holograms and associated applications, we refer to a dedicated progress report \cite{Genevet_2015}.

\subsection{Polarimeters}
Polarimeters, which enable direct measurement of the state of polarization (SOP), have found significant applications in many areas of science and technology, ranging from astronomy \cite{Sterzik_2012} and medical diagnostics \cite{Gupta_2005} to remote sensing \cite{Tyo_2006}, since the SOP carries crucial information about the composition and structure of materials interrogated with light. Despite all scientific and technological potential, polarimetry is still very challenging to experimentally realize as the SOP characterization requires conventionally six intensity measurements to determine the Stokes parameters \cite{Bohren}. Typically, the SOP is probed by utilizing a set of properly arranged polarization elements, for example, polarizers and waveplates, which are consecutively placed in the light path in front of a detector. In this way, the Stokes parameters that uniquely define the SOP are determined by measuring the light flux transmitting through these polarization components. Consequently, polarimeters based on conventional discrete optical components amount to bulky, expensive and complicated optical systems that go against the general trend of integration and miniaturization in photonics. During recent years, SOP detection using metasurfaces has attracted increasing interest due to their design flexibility and compactness.
\begin{figure*}[tb]
	\centering
		\includegraphics[width=8.0cm]{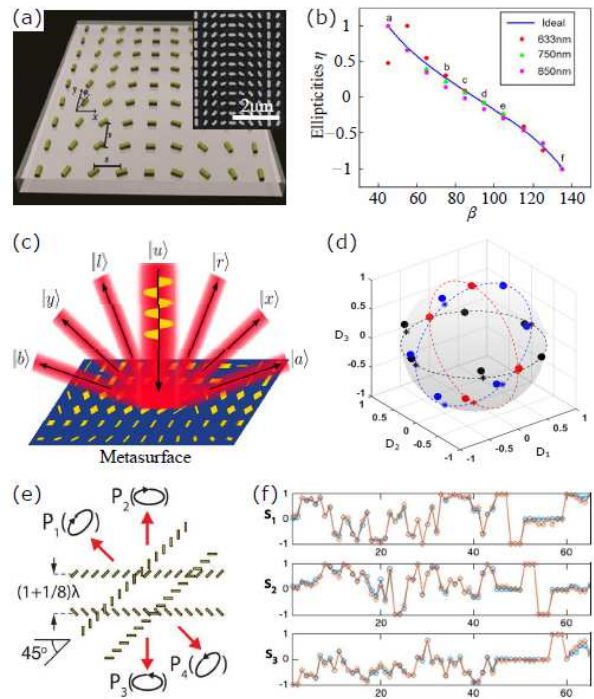}
	\caption{Metasurface polarimeters. (a) Schematic illustration of the phase gradient metasurface. Inset: SEM image of the fabricated metasurface on an ITO coated glass substrate. (b) Experimentally obtained ellipticity $\eta$ versus the incident polarization a function of $\beta$. (c) Illustration of the metagrating’s working principle. (d) Measured diffraction contrasts (denoted by filled circles) for polarization states along the main axes of the Poincar\'e sphere (indicated by asterisks) at 800 nm wavelength. (e) Polarization-selective directional scattering of four elliptical polarization states by two pairs of rows superimposed at a $45^\circ$ relative angle. (f) Measurement of the state of polarization of arbitrarily selected polarizations using the commercial polarimeter (blue) and the metasurface polarimeter (orange) at 1550\,nm wavelength. (a,b) Adapted with permission from \cite{Wen2015oe}; Copyright 2015 Optical Society of America. (c,d) Adapted with permission from \cite{Pors15optica}; Copyright 2015 Optical Society of America. (e,f) Adapted with permission from \cite{BalthasarMueller16optica}; Copyright 2016 Optical Society of America.}
	\label{fig:Polarimeters}
\end{figure*}
\subsubsection{Polarimeters}
In early approaches, metasurfaces together with conventional optical elements, such as polarizers and retardation waveplates \cite{bomzon2001spatial}, or the effect of a polarization-dependent transmission of light through six carefully designed nano-apertures in metal films \cite{Farzaneh2012meas}, were designed for the purposes of polarimetry. Likewise, different types of metasurfaces were proposed to determine certain aspects of the SOP, like the degree of circular polarization \cite{Shaltout2015optica,Wen2015oe}. One example is depicted in  figure \ref{fig:Polarimeters}(a) where an ultrathin (40\,nm) geometric metasurface has been proposed and demonstrated to measure the ellipticity and handedness of the polarized light. Due to the spin-selected opposite slope of constant phase gradient, the decomposed RCP and LCP beam are steered in two directions. By measuring the intensities of the refracted light spots, the ellipticity and handedness of various incident polarization states were characterized at a range of wavelengths and used to determine the polarization information of the incident beam (figure \ref{fig:Polarimeters}(b)). It is worth noting that the information of the incident SOP is incomplete as it is not possible to probe the polarization azimuth angle. To fully characterize the SOP, an extra polarizer is needed.

Recently, on-chip metasurface-only polarimeters have been proposed to simultaneously determine all polarization states, which includes meta-reflectarrays \cite{Pors15optica}, waveguide \cite{Pors16PhysRevApplied} and nanoaperture configurations \cite{BalthasarMueller16optica}. Figure \ref{fig:Polarimeters}(c) illustrates the basic working principle in which an arbitrary polarized incident beam is diffracted into six predesigned directions corresponding to different polarization states by a so-called metagrating, therefore allowing one to easily analyze an arbitrary state of light polarization by conducting simultaneous (i.e., parallel) measurements of the correspondent diffraction intensities that reveal immediately the Stokes parameters of the polarization state \cite{Pors15optica}. Specifically, the metagrating is composed of three interweaved metasurfaces where each metasurface functions as a polarization splitter for a certain polarization basis. The associated diffraction contrasts obtained by averaging three successive measurements at a wavelength of 800\,nm replicate reasonably well the Poincar\'e sphere (figure \ref{fig:Polarimeters}(d)), with points covering all octants of the 3D parameter space and the two-norm deviation between Stokes parameters and experimental diffraction contrasts being around $\sim0.1$. It should be noted that the designed metasurfaces have a rather broadband response, featuring experimentally diffraction contrasts that closely represent Stokes parameters in the wavelength range of $750-850$\,nm. Following this concept, a waveguide metacoupler was proposed, which facilitates normal incident light to launch in-plane photonic-waveguide modes propagating in six predefined directions with the coupling efficiencies providing a direct measure of the incident SOP \cite{Pors16PhysRevApplied}. In the later work \cite{BalthasarMueller16optica}, an ultracompact in-line polarimeter has been demonstrated by using a 2D metasurface covered with a thin array of subwavelength metallic antennas embedded in a polymer film (figure \ref{fig:Polarimeters}(e)). Based on the measured polarization-selective directional scattering in four directions, the SOP was obtained after the calibration experiment. The measurements of several arbitrarily selected polarizations using the commercial polarimeter and the metasurface polarimeter at a wavelength of 1550\,nm are given in figure \ref{fig:Polarimeters}(f), demonstrating the excellent agreement. As a final comment, we note that the degree of polarization (DOP) cannot be measured in the present four-output design. For DOP measurement, more complex antenna array design is needed.
\begin{figure*}[tbh]
	\centering
		\includegraphics[width=8.0cm]{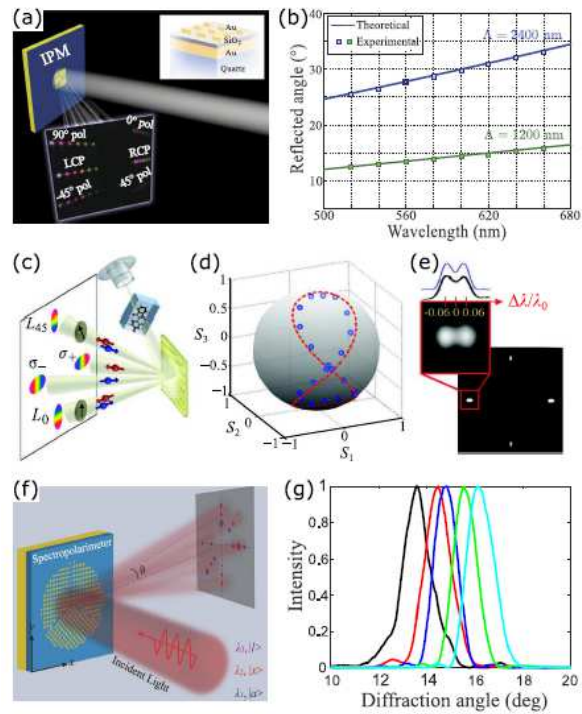}
	\caption{Metasurface spectropolarimeters. (a)  Illustration of the integrated plasmonic metasurface device. Inset: experimental image obtained for input $-45^\circ$ linearly polarized light with six spectral components. (b) Theoretical predictions and experimentally observed spectral dispersion for circular detection channels. (c) Schematic setup of the spectropolarimeter. The SPM is illuminated by a continuum light passing through a cuvette with chemical solvent, then four beams of intensities $I_{\delta+}$, $I_{\delta-}$, $I_{L45}$, and $I_{L0}$ are reflected toward a charge-coupled device camera. (d) Predicted (red dashed curve) and measured (blue circles) polarization states, depicted on a Poincar\'e sphere. (e) Measured far-field intensities for elliptical polarization at two spectral lines (with wavelengths of 740 and 780\,nm) and (inset) the corresponding resolving power (black line) and calculation (blue line) of the 50-$\mu$m diameter SPM. (f) Illustration of the GSP metasurface based beam-size-invariant spectropolarimeter. (g) Normalized measured far-field intensity profile for different wavelengths of the $|x\rangle$ channel. (a,b) Adapted with permission from \cite{Chen16nonotec}; Copyright 2016 IOP Publishing. (c-e) Adapted with permission from \cite{Maguidaaf16science}; Copyright 2016 American Association for the Advancement of Science. (f,g) Adapted with permission from \cite{ Ding2017beam}; Copyright 2017 American Chemical Society.}
	\label{fig:Spectropolarimeters}
\end{figure*}
\subsubsection{Spectropolarimeters}
Besides SOP measurement, spectral analysis is often required. Therefore, spectropolarimeters, which enable simultaneous measurements of the spectrum and SOP, have received much attention during recent years due to their superior capabilities in combining the uncorrected information channels of intensity, wavelength and SOP. Related developments have revealed that segmented \cite{Chen16nonotec,Ding2017beam} and interleaved \cite{Maguidaaf16science, maguid2017multifunctional} metasurfaces can be designed to conduct spectropolarimetry with simultaneous characterization of the SOP and spectrum of optical waves. Figure \ref{fig:Spectropolarimeters}(a) shows a spectropolarimetric metadevice that steers different input polarization and spectral components into distinct directions \cite{Chen16nonotec}. The metadevice comprises six GSP metasurfaces arranged in a $2\times3$ array corresponding to horizontal ($0^\circ$), vertical ($90^\circ$), $\pm45^\circ$, RCP and LCP analyzers. Upon excitation by a probe beam with an arbitrary polarization state, the metadevice generates six intensity peaks in far-field, arising from diffraction of each metasurface. The polarization response of the spectropolarimeter at a given wavelength was determined with careful calibration, that is, analyzing the relationship between the intensity of individual peaks and the incident SOPs. The measured angular dispersion for the LCP and RCP channels are 0.053 $^\circ$/\,nm and 0.024 $^\circ$/\,nm, respectively (Figure \ref{fig:Spectropolarimeters}(b)), revealing the potential of spectral measurement with spectral resolutions up to $\sim0.3$\,nm if this metadevice is inserted into a typical spectrometer setup. However, the rectangular configuration is oriented toward the spectropolarimetry of plane waves, so that the incident light should cover the whole area of the metasurfaces with the same light intensity over all metasurface elements in order to ensure faithful comparison of the corresponding (to particular Stokes parameters) diffraction orders.

By using center-symmetrical configuration, the spectropolarimeters are more compatible with the circular laser beam with a Gaussian profile, decreasing the footprint if the spectral resolution is maintained. This property is realized in both the interleaved \cite{Maguidaaf16science,maguid2017multifunctional} and segmented \cite{Ding2017beam} metasurfaces. An interleaved spectropolarimeter metasurface (SPM), shown in figure \ref{fig:Spectropolarimeters}(c), has enabled simultaneous characterization of the SOP and spectrum in reflection, which is composed of three linear phase profiles associated with different nanoantenna subarrays formed by the random interspersing. Figure \ref{fig:Spectropolarimeters}(d) shows the measured and calculated Stokes parameters on a Poincar\'e sphere for an analyzed beam impinging the SPM at a wavelength of 760\,nm with different polarizations. In addition to the excellent capability of polarization probe, the SPM shows good spectral resolving power (figure \ref{fig:Spectropolarimeters}(e)), which was measured to be $\lambda/\Delta\lambda \approx 13$ when the diameter is 50 $\mu$m. Similarly, this interleaved center-symmetrical approach can be applied with dielectric metasurfaces for spectropolarimetry \cite{maguid2017multifunctional}. Here, it should be emphasized that additional calibration experiment is needed since only four channels are designed; otherwise, the Stokes parameters cannot be determined.

To increase the detection robustness, a segmented plasmonic spectropolarimeter, featuring the self-calibrating nature, has been demonstrated \cite{Ding2017beam}. Figure \ref{fig:Spectropolarimeters}(f) schematically illustrates the GSP metasurface based spectropolarimeter, which is consisting of three gap-plasmon phase-gradient metasurfaces that occupy $120^\circ$ circular sectors each. This center-symmetrical configuration would diffract any normally incident (and centered) beam with a circular cross-section to six predesigned directions, whose polar angles are proportional to the light wavelength, while contrasts in the corresponding diffraction intensities would provide a direct measure of the incident polarization state through retrieval of the associated Stokes parameters. We would like to stress that uneven illumination of the three metasurfaces caused by misalignment will not affect the retrieved Stokes parameters and no calibration is needed since the Stokes parameters are only related to the relative diffraction contrasts for three polarization bases which have the self-calibrating nature. The proof-of-concept 96-$\mu$m-diameter spectropolarimeter operating in the wavelength range of $750-950$\,nm exhibits excellent polarization sensitivity with beam-size-invariant property. Additionally, the experimentally measured angular dispersion $\Delta\theta/\Delta\lambda$ for $|x\rangle$ channel is 0.0133 $^\circ$/\,nm, corresponding to a measured spectral resolving power of 15.2 if $0.7^\circ$ is taken as the actual minimum resolvable angular difference from the Rayleigh criterion (figure \ref{fig:Spectropolarimeters}(g)). It is worth noting that the spectral resolving power can be further improved with proper design, for instance, using superdispersive off-axis metalenses that simultaneously focus and disperse light of different wavelengths \cite{Khorasaninejad2017Spectroscopy,Alexander2017spectrometer}.

\subsection{Surface waves couplers}
Besides the unprecedented capability of metasurfaces in manipulating the propagating waves (PWs) in free space, another exciting feature of metasurfaces is the ability to control surface waves (SWs), such as surface plasmon polaritons (SPPs) and their low-frequency counterpart, that is, spoof SPPs on artificial surfaces \cite{pendry2004spoof,maier2007plasmonics}. In the present section, we will review some of the research progress on using metasurfaces to unidirectionally generate SWs from freely PWs.
\subsubsection{Unidirectional SWs couplers}
\begin{figure*}[tbh]
	\centering
		\includegraphics[width=12.0cm]{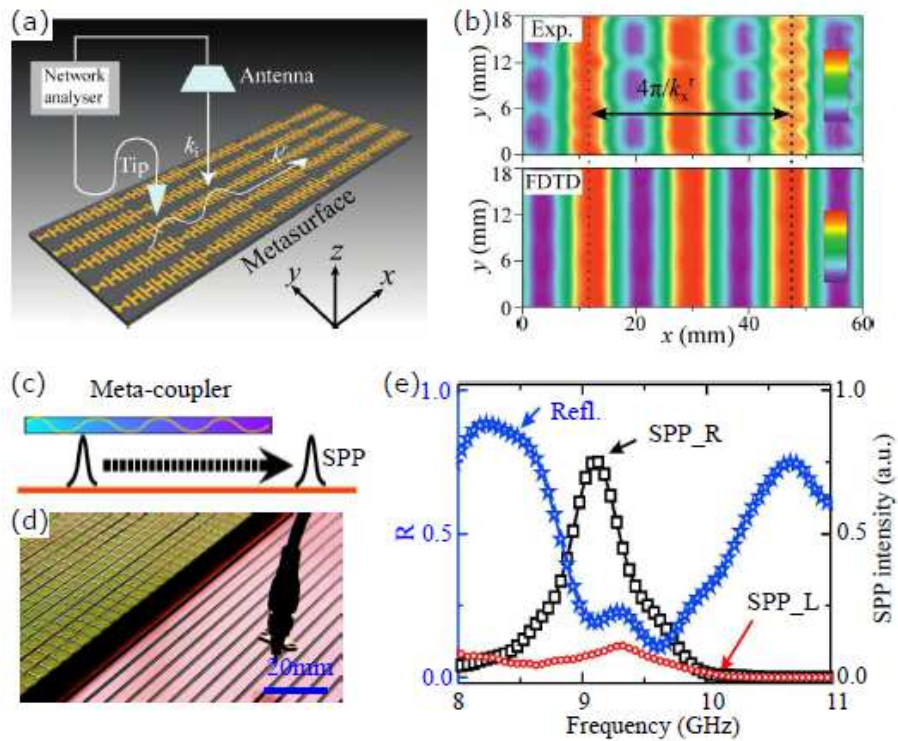}
	\caption{Metasurface SWs couplers. (a) Schematic picture describing the near-field scanning technique. (b) E$_z$ distributions (with phase information included) on part of the $\xi=1.14k_0$ meta-surface under illumination of a normally incident $x$-polarized wave, obtained by near-field scanning measurements (top) and simulations (bottom). (c) Configuration of the SPP meta-coupler: a transparent gradient metasurface is placed at a certain distance above the target plasmonic metal. (d) Image of part of fabricated sample and experimental setup where monopole antenna is used to probe the spoof SPPs field distribution. (e) Measured spectra of the integrated reflection (blue stars, left axis) and intensities of the excited spoof SPPs (right axis) flowing to the right side (black squares) and left side (red circles). (a,b) Adapted with permission from \cite{Sun_2012}; Copyright 2012 Nature Publishing Group. (c-e) Adapted with permission from \cite{sun2016high}; Copyright 2016 Changchun Institute of Optics, Fine Mechanics and Physics.}
	\label{fig:SWsD}
\end{figure*}

Conventionally, prisms or gratings can be used for unidirectional SWs generation with obliquely incident beams \cite{maier2007plasmonics}. However, this implementation has a stringent requirement of the oblique incident angle and excitation position, limiting its practical applications. Additionally, prisms are too bulky and not suitable for integrated plasmonic devices. Metasurfaces, on another hand, can achieve arbitrary phase gradient, or an in-plane effective wave vector that is matched with the wave vector of SWs, thereby allowing for conversion between a propagating mode in free space and a guided mode \cite{Sun_2012,wang2012coupler,epl2013coupler}. Figure \ref{fig:SWsD}(a) displays a reflective gradient metasurface that can convert a PW into a driven SW bounded on its surface with nearly 100\% efficiency \cite{Sun_2012}. Due to the unidirectional reflection-phase gradient, asymmetric coupling between PW and SW, and consequently unidirectional SWs launching has been realized (figure \ref{fig:SWsD}(b)). Distinct from the prism or grating couplers based on resonant coupling between PW and SW, the momentum mismatch between PW and SW is compensated by the reflection-phase gradient, and a nearly perfect PW-SW conversion can happen for any incidence angle larger than a critical value, confirming the robustness and flexibility of this meta-coupler. However, the generated SW is not an eigenmode of the phase-gradient metasurface because of its spatial inhomogeneity. Therefore the conversion efficiency decreases significantly when the size of incident beam increases, arising from the significant scattering caused by inter-supercell discontinuities before coupling to another bounded mode of other system \cite{epl2013coupler}.

To further increase the coupling efficiency, a new SPP meta-coupler has been demonstrated \cite{sun2016high}. Figure \ref{fig:SWsD}(c) illustrates the working principle of the high-efficiency SPP coupler consisting of a transparent gradient metasurface placed at a certain distance above the target plasmonic metal: The incident wave is first converted into a driven SW bound on the metasurface and then resonantly coupled to the eigenmode (i.e., SPP wave in optical frequencies) on the plasmonic metal. Based on this new configuration, the nonnegligible issues \cite{epl2013coupler} that severely affect the coupling efficiency can be resolved, thus leading to a theoretical efficiency up to 94\% as predicted by model calculations. As a practical realization, a realistic device operating in the microwave regime has been fabricated (figure \ref{fig:SWsD}(d)). Both near-field and far-field experiments demonstrate that the designed meta-coupler exhibits a spoof-SPP conversion efficiency of $\sim73$\% (figure \ref{fig:SWsD}(e)), which is much higher than those of all other available devices operating in this frequency domain. In particular, the efficiency is insensitive to the size of the incident beam.

\subsubsection{Polarization-controlled SPPs couplers}
Though the aforementioned metasurfaces couplers show high performance when operating with SWs excitation, the direction of SWs generated is predefined without any tunability. Additionally, SPPs are essentially transverse magnetic waves with the magnetic field oriented perpendicular to the propagation plane, a very important feature that dictates the polarization sensitivity of the SPPs excitation efficiency by free propagating radiation. Thus light power carried by the orthogonal polarization is usually lost.

\begin{figure*}[tbh]
	\centering
		\includegraphics[width=16.0cm]{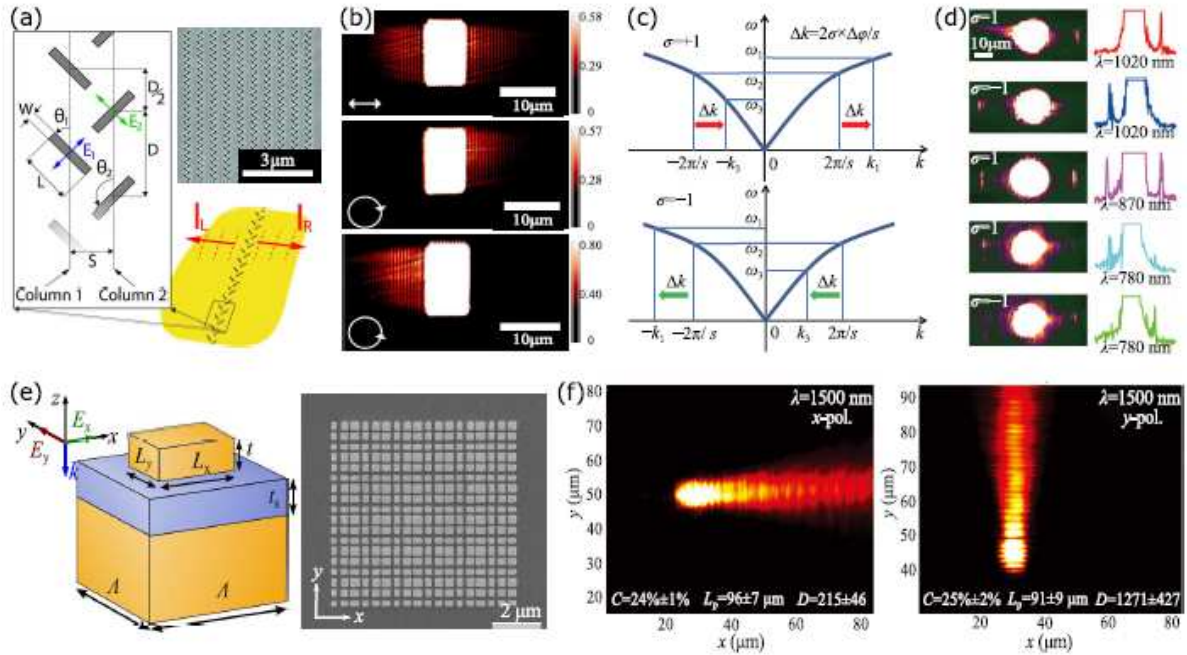}
	\caption{Polarization-controlled SPPs couplers. (a) Closely spaced subwavelength apertures as polarization-selective SPPs plane-wave sources. (b) Near-field images of the structure in (a) under illumination with different polarizations. (c) Dispersion curve of SPPs and the momentum matching condition for ordinary and anomalous diffraction orders. (d) Experiment demonstration of selective unidirectional SPP generation. (e) Sketch of a basic GSP unit cell, and SEM image of fabricated 2D-periodic polarization-sensitive SPP coupler composed of $6 \times 6$ super cells. (f) Recorded LRM images for a free-space wavelength of 1500\,nm with the coupling efficiency $C$, propagation length $L_p$, and directivity $D$ indicated in each image. (a,b) Adapted with permission from \cite{Lin2013coupler}; Copyright 2013 American Association for the Advancement of Science. (c,d) Adapted with permission from \cite{Huang2013coupler}; Copyright 2013 Changchun Institute of Optics, Fine Mechanics and Physics. (e,f) Adapted with permission from \cite{Anders2014coupler}; Copyright 2014 Changchun Institute of Optics, Fine Mechanics and Physics.}
	\label{fig:SWsT}
\end{figure*}

Very recently, polarization-controlled tunable directional SPPs coupling has been demonstrated using arrays of elongated rectangular apertures in an otherwise opaque metal film, so that the direction of SPPs excitation was switched by the handedness of a CP incident beam \cite{Lin2013coupler,miroshnichenko2013polarization,yang2014broadband,Huang2013coupler,Shitrit2013Rashba}. Figure \ref{fig:SWsT}(a) and \ref{fig:SWsT}(b) displays the spin-dependent directional SPPs coupler by Capasso's group \cite{Lin2013coupler}. An elongated rectangular shape of each aperture in a metal film selectively interacts with incident light that is polarized perpendicular to it, acting as an equivalent electric dipole and giving rise to SPPs. If an array of such apertures is arranged in a column with a spacing that is smaller than the SPPs wavelength, the launched SPPs are plane waves that propagate perpendicularly away toward either side of the column. When two columns with apertures oriented at $90^\circ$ with respect to each other are placed in close proximity, that is, a quarter of the SPPs wavelength, the SPPs interference becomes independent of the direction of linear polarized light but dependent on the helicity of CP light (figure \ref{fig:SWsT}(a)). The top right panel of figure \ref{fig:SWsT}(a) illustrates the SEM image of a polarization-controlled SPPs coupler with multiple column pairs. Figure \ref{fig:SWsT}(b) shows the near-field intensity distribution of the device by different polarization states. For circular polarization, the propagating direction of the SPPs waves is opposite for different handedness. In contrast, the launched SPPs have equal intensity toward either side of the coupler when the incident light is linearly polarized. In addition, this metasurface coupler can be bent into a circle to achieve spin-dependent focus. Near-field interference can also be induced by a CP dipole, thereby allowing unidirectional propagation of SPPs in a symmetric structure excited by CP light \cite{rodriguez2013near}.

As an alternative, geometric metasurface can provide a spin-sensitive phase gradient, enabling polarization-controlled SPPs launching \cite{Huang2013coupler,Shitrit2013Rashba}. For this geometric metasurface \cite{Huang2013coupler}, the lattice constant between neighboring apertures $s$ and the step of the rotation angle $\Delta\phi$ offer the necessary phase matching condition for a beam to excite SPPs at normal incidence, as illustrated in figure \ref{fig:SWsT}(c). A unidirectional propagating SPPs can only be excited for an incident spin $\sigma$ by momentum matching: $k_{SPP}=2\sigma\Delta\phi/s+2m\pi/s$, where $m$ is the chosen direction order. This asymmetric matching due to spin-sensitive geometric phase enables unidirectional SPPs excitation. For example, when $\sigma=+1$, the phase matching condition is shifted to a higher frequency $\omega_1$ for SPPs propagating along $+x$ direction, and to a lower frequency $\omega_3$ for propagating along $-x$ direction, which is ascribed to the extra positive in-plane momentum arising from the phase gradient. As shown in figure \ref{fig:SWsT}(d), the propagation direction of the SPPs can be dictated to the opposite directions when the circular polarization of the incident beam is switched at $\lambda=1020$\,nm and $\lambda=780$\,nm. In contrast, the SPPs are excited equally along both directions at $\lambda=870$\,nm, due to the fact that the momentum conservation for both directions is matched simultaneously for the two first-order ordinary refracted beams. Such a concept can also lead to spin-controlled multidirectional SPPs excitation by using inverse asymmetric 2D kagome lattice metasurface \cite{shitrit2013spin}.

Note that, in these nanoaperture configurations, the SPPs excitation involves both the transmission through narrow apertures and the coupling of the transmitted radiation to SPPs, so that the overall coupling efficiency is very low. This can be improved by optimizing the array configuration and employing GSP structures \cite{liu2012compact,Anders2014coupler,acsph2016coupler}. Here, we would like to highlight an efficient unidirectional polarization-controlled SPPs coupler by using GSP-based metasurfaces \cite{Anders2014coupler}. In particular, arrays of GSP resonators that would produce two independent orthogonal phase gradients upon reflection in two respective linear polarizations of incident radiation have been proposed, so that the incident radiation (with arbitrary polarization) can efficiently (up to 40\%) be converted into SPPs propagating in orthogonal directions dictated by the phase gradients. The basic GSP unit cell is consisting of a gold nanobrick atop a 50-nm-thick SiO$_2$ spacer layers covering 80-nm-thick gold films supported by glass substrates, shown in figure \ref{fig:SWsT}(e). The GSP-based coupling arrays have been fabricated for operation at telecom wavelengths, and figure \ref{fig:SWsT}(f) displays the recorded images from leakage radiation microscopy (LRM) for a free-space wavelength of 1500\,nm, featuring the coupling efficiency of $\sim25$\% for either of two linear polarizations and the directivity of SPP excitation exceeding 100. Furthermore, this concept was applied to realize a wavelength-sized meta-scatterer for efficient and polarization sensitive SPPs excitation, suggesting the great potential for ultracompact plasmonics circuits.

\subsection{Nonlinear metasurface}
The aforementioned exotic phenomena and unprecedented applications of metasurfaces are relying on the linear light-matter interaction. Recently metasurfaces with tailorable nonlinear optical properties have offered new degrees of freedom in designing photonic devices with remarkable performance. The key impact of metasurfaces on nonlinear responses is the unique ability to localize electromagnetic fields in nanoscale volumes, permitting control of the properties of light at dimensions much smaller than its wavelength. In this way, nonlinear response is governed to a large extend, boosting dramatically the magnitude of available nonlinearities. At the same time, metasurfaces significantly relax the phase matching condition, which should be strictly satisfied in conventional bulky materials, since their thickness is reduced to the scale of subwavelength. Moreover, various parameters of the nonlinear optical response such as intensity, phase, and polarization can be effectively controlled by changing the shape anisotropy and geometry of a particular metasurface.
\begin{figure*}[tbh]
	\centering
		\includegraphics[width=14.0cm]{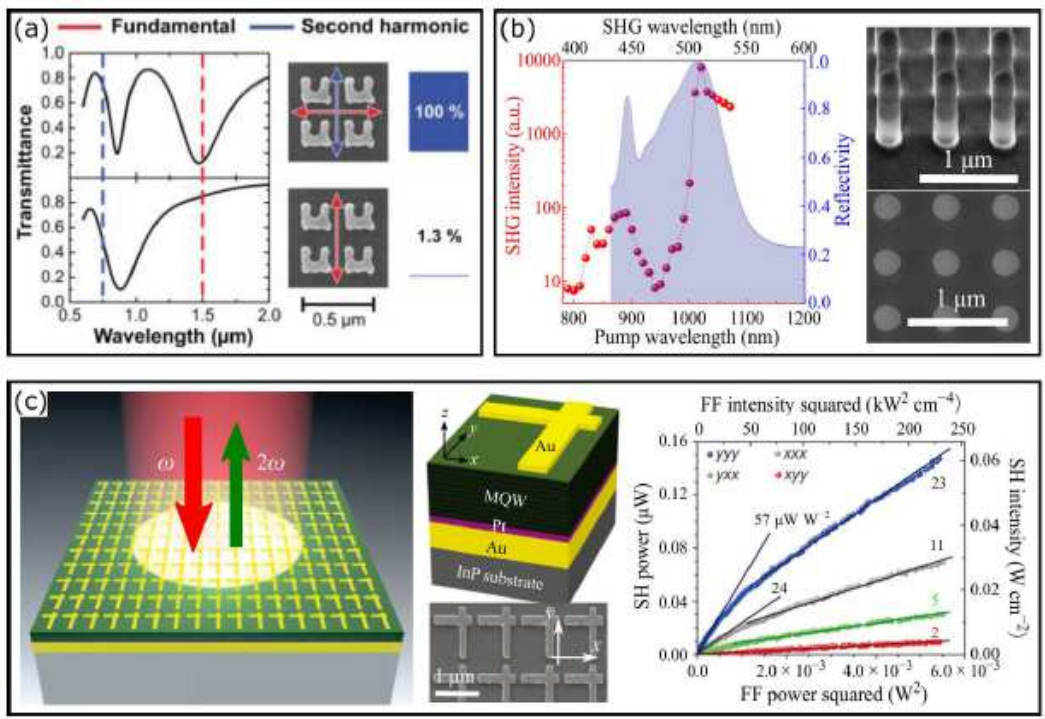}
	\caption{Nonlinear metasufaces. (a) SHG with noncentrosymmetric SRRs for different pump polarizations (red arrows). The blue bars highlight the corresponding measured SHG signal strengths, normalized to 100\% for the strongest SHG signal obtained from the fundamental magnetic resonance. (b) Resonantly enhanced SHG using GaAs all-dielectric metasurfaces. (c) Metal/MQWs hybrid metasurface exhibiting giant nonlinear response. (a) Adapted with permission from \cite{klein2006second}; Copyright 2006 American Association for the Advancement of Science. (b) Adapted with permission from \cite{liu2016GaAs}; Copyright 2016 American Chemical Society. (c) Adapted with permission from \cite{lee2014giant}; Copyright 2014 Nature Publishing Group.}
	\label{fig:Nonlinear}
\end{figure*}
\subsubsection{Nonlinear metasurfaces}
Since noble metals such as gold and silver exhibit strong second- and third-order nonlinear susceptibilities, $\chi^{(2)}$ and $\chi^{(3)}$, respectively, plasmonic metasurfaces can be engineered to tailor the nonlinear optical response with properly designed meta-atoms. The combination of the strong field enhancement obtained with plasmonic metasurfaces and the intrinsic nonlinearities of metals have readily resulted in efficient nonlinear optical processes, including second harmonic generation (SHG) \cite{klein2006second,butet2010optical,zhang2011three,husu2012metamaterials}, third harmonic generation (THG) \cite{lippitz2005third,hentschel2012quantitative,navarro2012broad}, or four-wave mixing (FWM) \cite{palomba2012optical,harutyunyan2012enhancing,suchowski2013phase}, underlining their great potential to design advanced nonlinear plasmonic devices. Many meta-atoms with symmetry considerations based on selection rules \cite{shen1984principles} have been designed, including SRRs \cite{klein2006second,ciraci2012origin,o2015predicting}, metallic hole arrays \cite{van2006strong,xu2007second}, fishnet structures \cite{shorokhov2016ultrafast}, gold T-dimers \cite{canfield2007local}, MIM nanodisks \cite{kruk2015enhanced}, L-shaped nanoparticles \cite{husu2012metamaterials,czaplicki2011dipole},  exacerbating specific features of the nonlinear response. For example, single layer metallic SRR arrays are efficient to enhance the SHG signal when magnetic dipole resonances are excited, as compared with purely electric dipole resonances, shown in figure \ref{fig:Nonlinear} (a). In such noncentrosymmetric SRRs, the exciting field and generated nonlinear surface currents couple with the bright plasmonic modes, leading to efficient excitation \cite{klein2006second}.

The fairly large absorption of metals within plasmonic metasurfaces ultimately affects their applications in nonlinear optics as the generated high-order harmonic suffers strong attenuations during propagation. Additionally, the patterned metallic nanostructures have low melting point, thus they may be damaged under strong laser illumination. An exciting alternative to circumvent the issue of losses and heat consists in the use of all-dielectric metasurfaces (see section \ref{sec:BuildingBlocks}), in particular, semiconductor metasurfaces, with low absorption, high refractive index, and strong nonlinear susceptibilities. Highly-efficient THG has been demonstrated with silicon meta-atoms by utilizing the magnetic Mie resonance \cite{shcherbakov2014enhanced} and Fano-resonance \cite{yang2015nonlinear}. Besides silicon, germanium nanodisks have also been explored to enhance THG excited at the anapole mode \cite{grinblat2016enhanced}. However, due to the centrosymmetric crystal structure of silicon, second-order nonlinear optical phenomena were not observed in silicon-based metasurfaces. Very recently, resonantly enhanced SHG using dielectric metasurface was demonstrated by Liu \emph{et al}, which is made from gallium arsenide (GaAs) that possesses large intrinsic second-order nonlinearity (figure \ref{fig:Nonlinear}(b)) \cite{liu2016GaAs}. By using arrays of cylindrical resonators, SHG enhancement factor as large as $10^4$ relative to unpatterned GaAs has been observed. Moreover, the measured nonlinear conversion efficiency is $\sim2\times10^{-5}$ with a pump intensity of $\sim3.4$ GW/cm$^2$.

The aforementioned nonlinearities in metasurfaces have been mostly realized by exploiting the natural nonlinear response of metals or dielectric materials. However, the associated optical nonlinearities are far too small to produce significant nonlinear conversion efficiency and compete with conventional bulky nonlinear components for pump intensities below the materials damage threshold. By coupling electromagnetic modes in plasmonic metasurfaces with intersubband transitions of multiple-quantum-well (MQW) semiconductor heterostructures, nonlinear metasurfaces with giant SHG response have been experimentally demonstrated in the mid-infrared range \cite{lee2014giant}, shown in figure \ref{fig:Nonlinear}(c). In addition to achieving strong field enhancement at both the fundamental and second harmonic (SH) frequencies, the plasmonic metasurfaces tailored the near-field polarizations, thereby converting giant nonlinear susceptibility of MQW heterostructures, which is intrinsically polarized normal to the surface \cite{rosencher1996quantum}, into any in-plane element of the nonlinear susceptibility tensor of the metasurface. This meta-atom/MQW hybrid metasurfaces could achieve a nonlinear conversion efficiency of $\sim2\times10^{-6}$ with a low pumping intensity of only 15\,kW\,cm$^{-2}$, corresponding to an effective second-order nonlinear susceptibility $\chi^{(2)}$ of $\sim5\times10^{4}$\,pm\,V$^{-1}$ (figure\ref{fig:Nonlinear}(c)). Following this concept, a larger susceptibility $\chi^{(2)}$ of $\sim2.5\times10^{5}$\,pm\,V$^{-1}$ was demonstrated by a hybrid nonlinear metasurface composed of SRR arrays and MQWs where the SRRs can enhance the pump and SH signals simultaneously \cite{campione2014second}. Furthermore, deeply subwavelength ($\sim\lambda/20$) metal-semiconductor nanocavities were introduced, which not only convert $z$-polarized MQW nonlinear susceptibility into the transverse plane but provide further enhancement to the nonlinear response\cite{lee2015aom,nookala2016ultrathin}. Combined with innovations in the MQW design, this approach allowed to produce a record-high second-order nonlinear optical response of $1.2\times10^{6}$\,pm\,V$^{-1}$ at $\lambda=10$\,$\mu$m, which is about 3$-$5 orders of magnitude higher than that of traditional nonlinear materials and nonlinear plasmonic metasurfaces, and the experimentally achieved absolute conversion efficiency is up to 0.075\% using pumping intensities of only 15\,kW\,cm$^{-2}$ \cite{lee2015aom}. It should be noted although the effective nonlinear susceptibility is significantly enhanced by orders of magnitude compared to traditional nonlinear materials; the absolute conversion efficiency is still low.
\begin{figure*}[tbh]
	\centering
		\includegraphics[width=12.0cm]{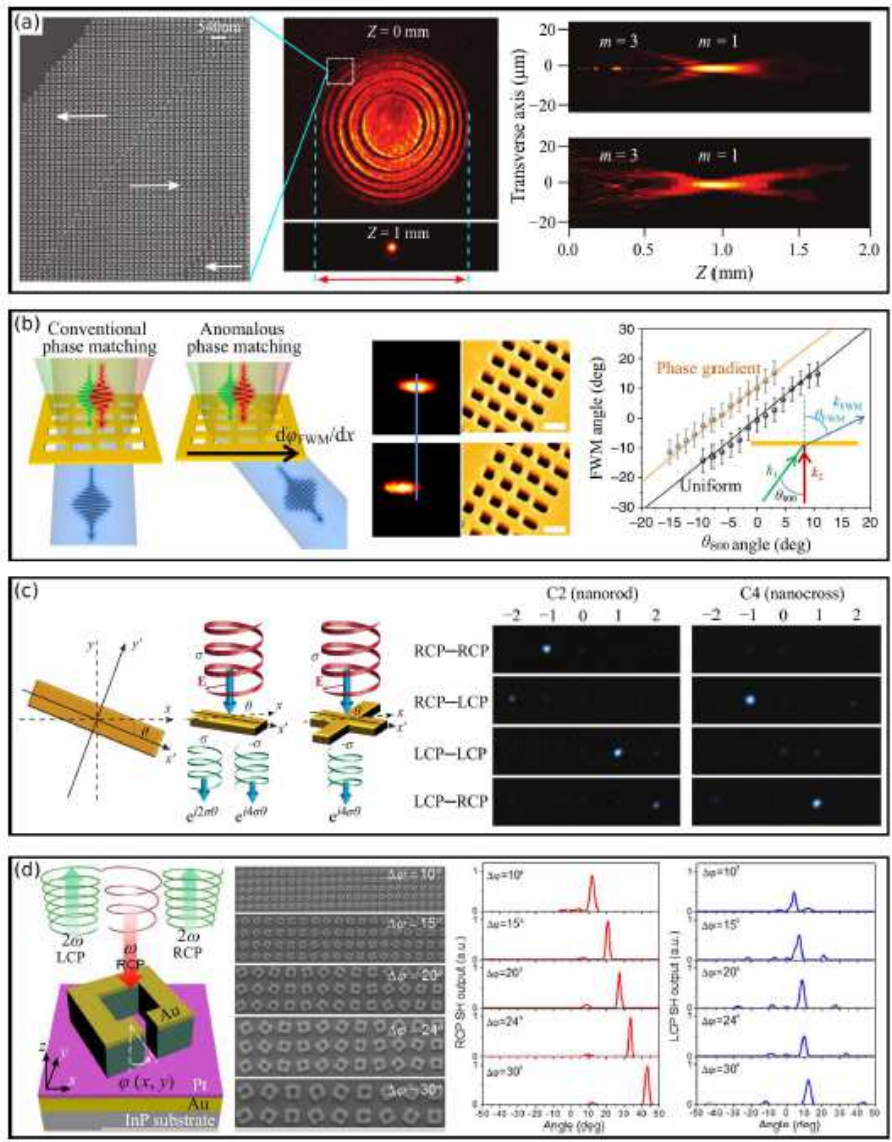}
	\caption{{\scriptsize Nonlinear metasufaces with phase control. (a) Nonlinear Fresnel zone plate. Left panel: SEM image of a portion of the sample showing mirror inversion of the SRRs in adjacent zones; Mid-panel: recorded normalized images of SH; Right panel: simulation (top) and experimental results (bottom) of transverse focusing of SH ($m$ denotes focusing order). (b) Phase control of FWM in plasmonic metasurfaces. Left panel: illustration of the anomalous phase-matching condition; Mid-panel: CCD images of FWM signals from uniform (top), and phase gradient (bottom) structures; Right panel: angle dependence of the phase-matching angle for the uniform and phase-gradient metasurfaces. (c) THG from nonlinear geometric metasurfaces with continuous phase control. (d) Metal/MQWs hybrid metasurface exhibiting a geometric phase for SHG. Left panel: schematic of metasurface unit cell; Mid-panel: fabricated gradient SRR arrays with differing angular steps $\Delta\phi$; Right panel: far-field profiles of RCP and LCP SH output. (a) Adapted with permission from \cite{segal2015controlling}; Copyright 2015 Nature Publishing Group. (b) Adapted with permission from \cite{almeida2016subwavelength}; Copyright 2016 Nature Publishing Group. (c) Adapted with permission from \cite{li2015continuous}; Copyright 2015 Nature Publishing Group. (d) Adapted with permission from \cite{nookala2016ultrathin}; Copyright 2016 Optical Society of America.}}
	\label{fig:NonlinearPhase}
\end{figure*}
\subsubsection{Nonlinear metasurfaces with phase control}
Besides the enhanced nonlinear responses, metasurfaces, particularly phase-gradient metasurfaces, may show even greater control of the local phase, amplitude and polarization response of high harmonic signal, thus achieving advanced nonlinear functionalities beyond the nonlinear metasurfaces with uniform meta-atoms. As a simple demonstration, nonlinear radiation steering has been demonstrated by nonlinear metamaterial-based photonic crystals (NLMPCs) (figure \ref{fig:NonlinearPhase}(a)) \cite{segal2015controlling}. By simply reversing the orientation of the SRRs comprising the NLMPCs with a period, periodic inversion of the effective $\chi^{(2)}$ can be achieved, inducing a $\pi$ phase shift of the local SHG radiation for linearly polarized light. Based on the generated nonlinear binary phase, the SHG signal can be directed from NLMPCs with controlled emission angle. Moreover, an NLMPC-based ultrathin nonlinear Fresnel zone plate (FZP) was obtained, which focuses the SH signal to a spot of 7 $\mu$m with large SH intensity enhancement. Similarly, the nonlinear binary phase was used to radiate SH signals into different directions depending on their polarization states by using SRRs/MQWs hybrid nonlinear metasurfaces \cite{wolf2015phased}. Moreover, a binary nonlinear metasurface consisting of SRR meta-atoms has been used to realize SH Airy and vortices beams by manipulating both the phase and amplitude of the quadratic nonlinear coefficient locally \cite{keren2015nonlinear}.
Such binary nonlinear beam-steering devices only use two phase steps, which inevitably introduce undesired diffraction effects. For more complex nonlinear beam manipulation and shaping, a continuous and spatially varying phase manipulation is needed. Figure \ref{fig:NonlinearPhase}(b) illustrates the full nonlinear phase control for FWM with linearly polarized light in plasmonic metasurfaces, which is achieved by introducing a spatially varying phase response of a metallic metasurface consisting of subwavelength nonlinear nanoapertures designed specifically for the nonlinear signal \cite{almeida2016subwavelength}. Specifically, the nonlinear phase of FMW signals can be continuously tuned from 0 to 2$\pi$ through adjusting the geometry of nanoapertures. For such interfaces, a new, anomalous nonlinear phase-matching condition, that is, the nonlinear analogue of the generalized Snell's law (middle and right panels of figure \ref{fig:NonlinearPhase}(b)), has been derived, that differs from the conventional phase-matching schemes in nonlinear optics. In addition, ultrathin frequency-converting lenses with tight focusing were demonstrated. Following this strategy, nonlinear multilayer metasurface holograms have been demonstrated, with a background free image formed at the third harmonic of the illuminating beam \cite{almeida2016nonlinear}.

Though continuous nonlinear phase control over the 2$\pi$ range can be achieved with linearly polarized light, the phase and amplitude of high-order signal are strongly depending upon the size and shape of each meta-atom, which may affect the nonlinear response since the giant nonlinear effects are very sensitive to the variations in the local resonances of meta-atoms. As an alternative, geometric phase (or Pancharatnam-Berry phase) approach is an ideal tool to realize full phase control with a nearly uniform nonlinear response for CP light, based on the suitably designed nonlinear meta-atoms with gradually changed orientations \cite{li2015continuous,tymchenko2015gradient}. Inspired by the concept of spin-rotation coupling, nonlinear geometric metasurfaces have been demonstrated, featuring homogeneous linear optical properties but spatially varying effective nonlinear polarizability with continuously controllable phase (figure \ref{fig:NonlinearPhase}(c)) \cite{li2015continuous}. For a CP fundamental wave propagating along the rotational axis of a meta-atom, the nonlinear polarizability can be expressed as $\alpha^{n\omega}_ {\theta,\sigma,\sigma}\propto e^{(n-1)i\theta\sigma}$ and $\alpha^{n\omega}_ {\theta,-\sigma,\sigma}\propto e^{(n+1)i\theta\sigma}$, thus introducing geometric phase $(n-1)\theta\sigma$ or $(n+1)\theta\sigma$ into the nonlinear polarizabilities of $n$th harmonic generation with the same or opposite circular polarization to that of the fundamental mode, where $\sigma=\pm1$ represents the LCP or RCP light and $\theta$ is the orientation of meta-atom. Therefore the spin-induced geometric nonlinear phase, combined with selection rules of harmonic generation for CP light \cite{konishi2014polarization,chen2014symmetry}, enables complete control over the propagation of harmonic generation signals. As shown in figure \ref{fig:NonlinearPhase}(c), meta-atoms with two-fold symmetry (C2) diffracts THG signals with RCP and LCP states to the first and second diffraction orders, respectively, while the meta-atoms with four-fold symmetry (C4) diffracts the opposite CP THG signal only to the first diffraction order direction. Moreover, by combining the concept of nonlinear geometric metasurfaces with CGH technique, spin and wavelength multiplexed nonlinear metasurface holography was demonstrated \cite{ye2016spin}.

Nonlinear geometric metasurfaces have also been demonstrated by using hybrid SRRs/MQWs system, which simultaneously provides nonlinear efficiencies that are many orders of magnitude larger than those in other nonlinear setups, and, at the same time, is capable of controlling the local phase of the nonlinear signal with high precision and subwavelength resolution. Based on this new platform, the nonlinear wavefront can be controlled at will, enabling beam steering, focusing and polarization manipulation \cite{nookala2016ultrathin,tymchenko2015gradient}. Figure \ref{fig:NonlinearPhase}(d) depicts the experimental realization of the hybrid nonlinear geometric metasurfaces for SH steering, which are consisting of spatially arranged subwavelength metal-semiconductor resonators with an angular rotation step of $\Delta\phi$ between adjacent unit cells along one direction \cite{nookala2016ultrathin}. For a RCP incident beam, RCP and LCP SH beams will be generated toward two different directions $\theta_{R(R)}=arcsin[(3\Delta\phi/360^\circ)\lambda_{2\omega}/d]$ and $\theta_{L(R)}=arcsin[(\Delta\phi/360^\circ)\lambda_{2\omega}/d]$ respectively, where $\lambda_{2\omega}$ is the SH wavelength and $d$ is the period of the super cell. As is seen in the right panel of figure \ref{fig:NonlinearPhase}(d), experimental results fully confirm these predictions, steering most of the RCP and LCP signals away from the normal in specified directions.

Coming to the end of this subsection, we have presented a brief overview of the field of nonlinear metasurfaces and discussed how the metasurfaces can be employed to realize various functionalities with remarkable improvement. For a more detailed discussion on the nonlinear response of nanostructures, we refer to some review papers \cite{kauranen2012nonlinear,Minovich_2015,butet2015optical}.

\section{Conclusions and Outlook}
Metasurfaces, especially gradient metasurface, have becoming a rapidly growing field of research due to their exceptional capabilities of realizing novel electromagnetic properties and functionalities. In this paper, we have reviewed the development of gradient metasurfaces by introducing their fundamental concept, classification, physical realization, and a few of representative applications. As a rapidly developing area of research which is expanding very fast every day, to review all aspects of metasurfaces mentioned in the available literature is impossible and hardly instructive. There are still many other aspects not included here, such as parity-time metasurfaces \cite{Lawrence2014,Monticone2016}, ultrathin invisibility cloaks \cite{Ni2015cloak}, photonic spin Hall effects \cite{Yin2013spin}, mathematical operation \cite{Silva2014Mathematical,Pors2015Analog}, etc.

Owing to the unprecedented control over light with surface-confined planar components, metasurfaces are expected to extend to a much broader horizon beyond what we have discussed in this review article, offering fascinating possibilities of very dense integration and miniaturization in photonic/plasmonic devices. We think that metasurfaces could have a significant impact on the following promising areas which still remain largely unexplored.
\begin{enumerate}
  \item{\emph{Multifunctional Metasurfaces.} The up-to-date metasurface design is typically focused on single, on-demand light manipulation functionality, which is not compatible with the desired goal of multifunctional flat optics. Metasurfaces that facilitate efficient integration of multiple diversified functionalities into one single ultrathin nanoscale device have become an emerging research area. Very recently, Hasman and colleagues have proposed a generic approach to realize multifunctional metasurfaces via the synthesis of the shared-aperture antenna array and geometric phase concepts \cite{Veksler2015Multiple}. By randomly mixing of optical nanoantenna subarrays, where each subarray provides a different phase function in a spin-dependent manner, multiple wavefronts with different functionalities can be achieved within a single shared aperture, without reducing the numerical aperture of each sub-element \cite{Maguidaaf16science,maguid2017multifunctional,lin2016photonic}. Xu et al recently demonstrated high-efficiency bifunctional devices to achieved distinct functionalities in the microwave regime based on anisotropic meta-atoms with polarization-dependent phase responses \cite{cai2017}.}
  \item{\emph{Dynamically reconfigurable metasurfaces.} Adding tunability and reconfigurability into metasurfaces is highly desirable, as it extends their exotic passive properties and enables dynamic spatial light modulation over an ultrathin surface \cite{Zheludev_2012}. Reconfigurable metasurfaces can be achieved in the aspect of structure, that is, by altering the shape of individual meta-atoms, or by manipulating the near-field interactions between them, which can be accomplished through flexible substrates made of polydimethylsiloxane (PDMS) \cite{kamali2016decoupling} and microelectromechanical systems (MEMS) \cite{ou2013electromechanically}. Besides the structure aspect, a more general strategy of realizing tunable metasurfaces is to hybridize metasurfaces with active functional materials, including semiconductors \cite{chen2008experimental}, graphene \cite{yao2014electrically,emani2015graphene}, phase-change materials \cite{wuttig2007phase,wang2016optically}, transparent conducting oxides (TCOs) \cite{Park2017ITO}, and liquid crystals \cite{sautter2015active}. In particular, graphene-based metasurfaces have realized the ultimate reconfigurability and multiple functionalities in the in the mid-infrared and THz frequency ranges \cite{emani2015graphene}, resulting from the extraordinary optoelectronic properties and largely tunable carrier density of graphene. In addition, phase-change materials have been extensively utilized in optical data-storage systems and photonic devices due to their outstanding switchable optical properties by thermal, laser or electrical current pulses with controlled duration and intensity \cite{wuttig2007phase}. For example, germanium-antimony-tellurium (GST) chalcogenide glass has recently been used to demonstrate all-optical, non-volatile, metasurface switch \cite{wang2016optically}.}
  \item{\emph{Quantum Metasurfaces.} There is an explosively growing interest in combining the existing quantum systems with metasurfaces, creating the so-called metasurfaces, which expands the properties of metasurfaces towards the quantum regime \cite{tame2013quantum}. More specifically, by integrating non-classical light sources such as single emitters with properly designed optical meta-atoms, metasurfaces are thus endowed with quantum information processing capabilities. Lon{\v{c}}ar and co-works are pioneering the research on enhanced single-photon emission by color centers in diamond/meta-atoms hybrid systems \cite{choy2011enhanced,Sipahigil2016}. As an emerging area of research, quantum metasurfaces opens up a new frontier, which will considerably advance our understanding of the fundamental physics associated with light-mater interaction within quantum regime, and facilitate the realization of novel quantum photonic devices \cite{noginov2009demonstration,oulton2009plasmon,akimov2007generation,falk2009near}, including nanoscale lasers, single photon/plasmon sources, and quantum plasmonic circuitry.}
\end{enumerate}

\section*{Acknowledgements}
This work was funded by the European Research Council (the PLAQNAP project, Grant 341054) and the University of Southern Denmark (SDU2020 funding).

\clearpage
\section*{References}
\bibliography{References}

\end{document}